\documentclass[preprint]{aastex}

\usepackage{amsmath,amssymb}

\usepackage{color}
\definecolor{black}{rgb}{0,0,0}
\definecolor{red}{rgb}{1.0,0,0}

\begin{document}

\title{The Spectral Energy Distribution of Self-gravitating Interstellar Clouds\\I. Spheres}
\shorttitle{The SED of Self-gravitating Interstellar Clouds, I.}
\shortauthors{Fischera \& Dopita}

\author{J\"org Fischera\footnote{Current address: CITA, University of Toronto, 60 St. George Street, ON M5S3H8, Canada}\ \& Michael A. Dopita}
\affil{Research School of Astronomy \& Astrophysics, Institute of Advanced Studies\\The Australian National University, Cotter Road, Weston Creek, ACT 2611 Australia}
\email{fischera@cita.utoronto.ca,mad@mso.anu.edu.au}

\begin{abstract}
	We derive the spectral energy distribution (SED) of dusty, isothermal, self gravitating, stable and spherical clouds externally heated by the ambient interstellar radiation field. For a given radiation field and dust properties, the radiative transfer problem is determined by the pressure of the surrounding medium and the cloud mass expressed as a fraction of the maximum stable cloud mass above which the clouds become gravitational unstable. 
	To solve the radiative transfer problem a ray-tracing code is used to accurately derive the light distribution inside the cloud. This code considers both non isotropic scattering on dust grains and multiple scattering events. The dust properties inside the clouds are assumed to be the same as in the diffuse interstellar medium in our galaxy. We analyse the effect of the pressure, the critical mass fraction, and the ISRF on the SED and present brightness profiles in the visible, the IR/FIR and the submm/mm regime with the focus on the scattered emission and the thermal emission from PAH-molecules and dust grains.
\end{abstract}


\newpage

\section{Introduction}

{Self-gravitating clouds are an essential constituent of the in-homogeneous interstellar medium. They are either regions of ongoing or, if the clouds become cold enough to be gravitational unstable, future star-formation. They are important for the chemistry in the ISM through the formation of molecules, in particular molecular hydrogen, and they play probably an essential role for the life-cycle of interstellar dust grains since in dense environments grains are thought to grow to larger sizes through a variety of physical processes. 
Those grain processes not only alter the mean optical grain properties and therefore the extinction curve, most importantly at ultraviolet and far-ultraviolet wavelengths, but will probably also lead to a different integrated grain surface area. Because molecules form on dust grains and their formation rate is controlled by the strength of the photo dissociation flux inside the cloud these grain processes will have direct consequences for the formation of molecules.
Our knowledge of the dust properties in clouds is furthermore crucial for our understanding of the origin of the interstellar extinction curve as parts of the dense material will be redistributed into the diffuse ISM. This will enrich the gas with large grains with a composition possibly rather different to grains initially formed around cold stars, in planetary nebulae, or in the metal rich part of the ejected material of supernova explosions.}

The spectral energy distribution (SED) of self-gravitating clouds might have an important contribution to the global SED of galaxies, in particular at far infrared and submm wavelength, thanks to the large population of cool dust which they are expected to contain.

The origin of self-gravitating clouds is likely related to turbulent motion. A turbulent medium would not only naturally produce a distribution of cloud masses as observed for the ISM \citep{Elmegreen2002} but is also possibly directly linked to the initial mass function of the stars \citep{Padoan1997,Padoan2002, Bate2003}. An isothermal turbulent medium leads to a multi-fractal structure where large clouds actually consist of distributions of clouds of different sizes. As shown by simulations \citep{Vazquez1994,Padoan1997,Passot1998} and verified analytically \citep{Nordlund1999}, the probability distribution of the local density in an isothermal turbulent medium can be well described by a simple log-normal density distribution. In the presence of gravity this distribution is squeezed by increasing the number of higher densities while the number of low density values decreases \citep{Nordlund1999}. This happens as clouds become more compact due to gravitational pressure. If the gravitational forces are strong enough to hold the cloud together against the turbulent motions the density enhancements become stable. If the gravitational pressure surpasses the maximum thermal and, possibly,  magnetic pressure support, then they will collapse and will form stars. 

Here, we consider stable, isothermal, self-gravitating clouds assumed to be in pressure equilibrium with their surrounding medium.
To be able to infer the dust properties from observations of the re-emitted and the scattered light and to minimise any complications by radiative transfer effects we shall limit our attention to simple, idealistically spherical, cloud structures. We find them in small molecular clouds, the so called Bok Globules \citep{Bok1947}, and, to somewhat less extend, in translucent clouds, so named as they have attenuation in the range $1<A_{\rm V}< 5$ \citep{Dishoeck1988}. We further will consider only those isolated clouds which are illuminated by the interstellar radiation field (ISRF) which is, as first approximation, assumed to be isotropic.

Clouds often show a rather elongated structure which might be idealized through self-gravitating infinite cylinders. The SED of those clouds is analyzed in a following paper (paper II in this series).

Translucent clouds are the interface between the diffuse and the dense molecular phase of the cold neutral medium. Although the clouds are optically thick they are still transparent enough to allow optical and UV absorption line measurements which enable us to analyse the different processes in dense gas in more detail. Measurements of the extinction curve for example \citep{Cardelli1988,Boulanger1994,Rachford2002} give us essential information about the grain properties in clouds important to infer the conditions and the processes responsible for grain growth. 

Bok Globules are highly optically thick at their centres and appear as dark clouds in front of the distant stars and nebula. These clouds are through their simple, almost spherically shape ideal objects to study the formation of low massive stars and the molecular phase of the ISM. That those clouds are indeed regions of star formation as suggested by \citet{Bok1947} has been revealed by an IRAS survey that showed that $\sim 23\%$ of Bok Globules are associated with a young stellar object with a mass of at least $0.7~M_{\sun}$, the detection limit of the survey, in the cloud centre \citep{Yun1990}.
 
The scattered light from almost spherical globules which are illuminated by an idealised isotropic radiation field has been used not only to infer the properties of the grains as the typical size, the dust albedo and the asymmetry of the scattered light but also to derive the density profile at the rim of the cloud \citep{Mattila1970,FitzGerald1976,Witt1990,Lehtinen1996}.

Molecular line observations show that Bok Globules have in general a rather steep density profile at the outskirts approximately described by a power law $n(r)\propto r^{-2}$ \citep{Arquilla1985}. In the centre of stable clouds the profile becomes rather flat \citep{Dickmann1983,Lehtinen1995}. Qualitatively the density profile of translucent clouds is similar but less steep at the outskirts.

The questions we seek to address in this paper are {what is the spectral energy distribution (SED) of stable self-gravitating clouds and how is the scattered light and the thermal re-emission affected by the density structure of these clouds. 
Although there are several indications that the grain properties in dense clouds are different than in the diffuse interstellar medium we will base our calculations on the mean properties derived for the Milky Way. We will address the effect of different grain properties in the discussion. 

Calculations of the spectral energy distribution from isothermal self-gravitating clouds were presented by \citet{Evans2001} and \citet{Stamatellos2003} but they used a different approach to model the dust emission. We will discuss their models in more detail in the discussion.

Our work may be best compared with the one presented by \citet{Bernard1992}. The model assumptions, however, are different in several respects. In particular it was not their intention to model the scattered light as it is here. Further, the calculation in this paper is based on a physical, although simple, cloud model 
while the results by Bernard et al. were based on a simplified density structure with a flat density profile in the cloud centre and a power law profile at the outer cloud region $n(r)\propto r^{-\beta}$ with $\beta$ chosen to be 0, 1, and 2. For a fixed density at the cloud boundary their SEDs were derived for several values of the central density, the power $\beta$, and the total extinction $A_{\rm V}$ through the cloud centre. In our model of stable clouds the SEDs are determined by only two parameters, the pressure outside of the cloud and the critical mass fraction of the cloud mass relative to the critical mass where the cloud becomes gravitational unstable against gravitational collapse. The calculations of Bernard et al. and ours are based on the mean properties of the dust grains in the ISM.}

The paper is outlined as follows: In section 2 we will discuss the physical properties of the isothermal clouds model. 
In section 3 we describe the dust model used in all calculations and in section 4 the radiative transfer program. In section 5 we will adopt this model to derive the SED and brightness profiles for a number of different parameters. We show the effect of the compactness of the clouds on the emission of PAH molecules and the dust emission. The results are discussed in section 6. A summary of our results is given in section 7.\newline

\section{Model of isothermal clouds}

\label{sectionmodel}
To model the SED of self-gravitating clouds we consider the clouds to be isothermal and to be in pressure equilibrium with the surrounding medium. Particular interest is given for clouds embedded by the WNM of our galaxy. The pressure of this medium, which includes the thermal and magnetic pressure, is taken to be $p_{\rm ext}/k = 2\cdot 10^4~{\rm K/cm^3}$ based on the work from \citet{Boulares1990}. Following \citet{Curry2000} the pressure due to cosmic rays is neglected as they will pervade both the cloud and the ambient medium.
For simplicity the isothermal clouds are taken to be spherical. In addition we neglected the observed complex substructure likely related to turbulent motion inside the clouds. The clouds have a radial density profile which is determined by the equilibrium of the isothermal and gravitational pressure which is obtained by solving the Lane-Emden-Equation. The mathematics structure of such clouds is developed in the appendix to this paper.

We point out that our approach stays valid in the case where the turbulent motion and magnetic pressure inside the cloud can be approximated by an increase of the thermal motion.

The problem of isothermal clouds plays an important role in astrophysics as it gives insights in the inner structure of stars and about the star formation and had therefore been studied in the past \citep{KippenhahnWeigert1990}. Isothermal spheres have furthermore been used to analyse extinction measurements \citep{Alves2001,Kandori2005} or to model the observed infrared emission of dense clouds \citep{Evans2001} often referred to as pre-stellar cores. In general a very good agreement has been found. In the following we will describe some properties that are important for the solution of the radiative transfer problem.

\subsection{The critical cloud mass}

It is well known from the application of the Lane-Emden equation that the clouds can only remain stable against gravitational collapse for masses below a critical value; $M_{\rm cl,max}$. 
This critical mass is given by:
\begin{equation}
	\label{eq_critmass}
 \left(\frac {M_{\rm cl,max}}{M_{\sun}} \right)= 2029 \left(\frac{T_{\rm cl}}{100~{\rm K}}\right)^2\left(\frac{\mu}{1.0}\right)^{-2}
 \left(\frac{p_{\rm ext}/k}{10^4\,{\rm K cm^{-3}}}\right)^{-\frac{1}{2}}
\end{equation}

\subsection{The density profile}

\begin{figure}
  \includegraphics[width=\hsize]{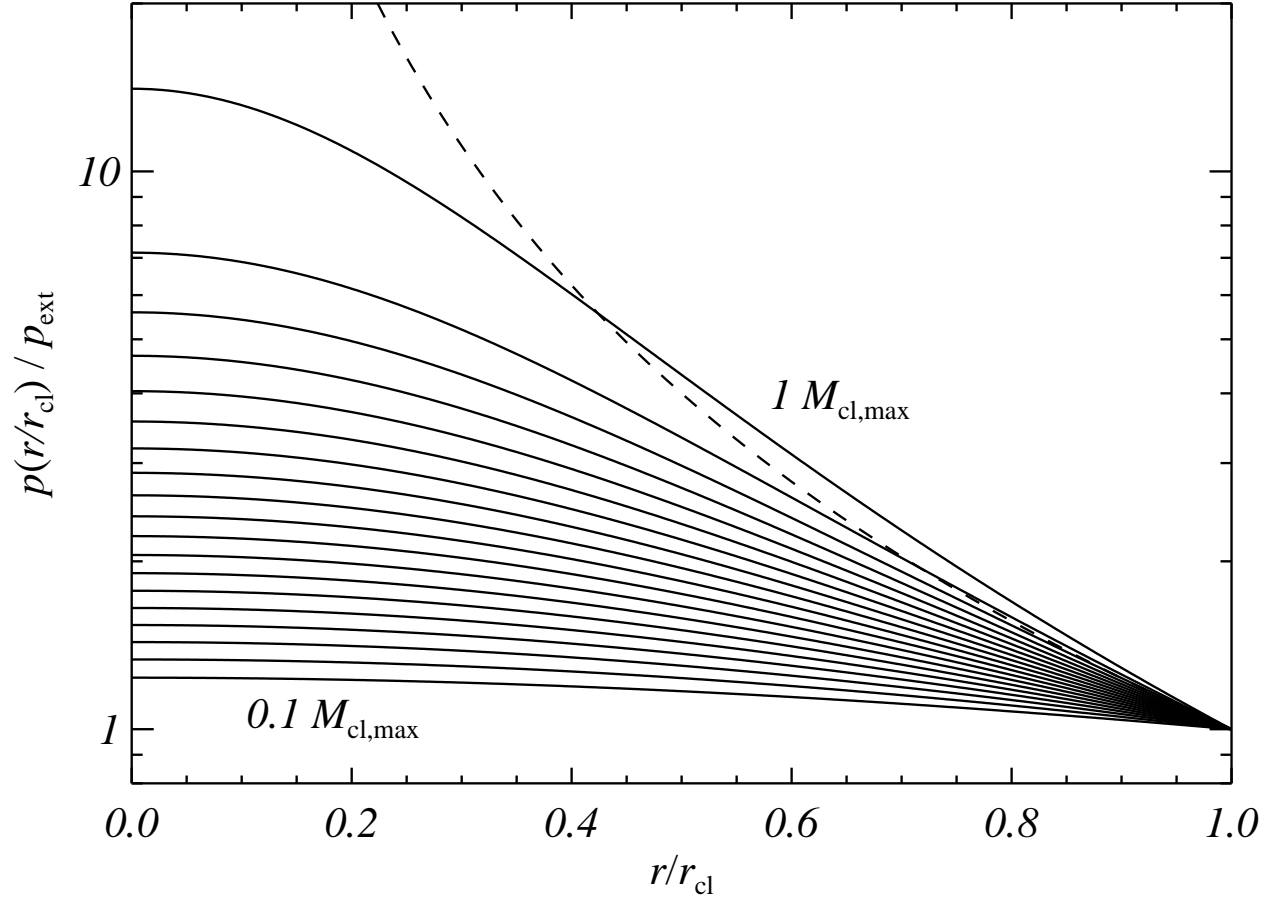}
  \caption{\label{figdensprofile} Pressure profile $p(r/r_{\rm cl})/p_{\rm ext}$ (or density profile) of isothermal clouds. The cloud mass given as fraction of the maximum cloud mass $M_{\rm cl,max}$ of non collapsing clouds is varied from 0.1 to 1 in steps of 0.05. For comparison also a density profile $n(r)\propto r^{-2}$ is shown (dashed line).}
\end{figure}

The density profile of isothermal clouds depends on the critical mass fraction 
$f=M_{\rm cl}/M_{\rm cl,max}$ of the clouds mass $M_{\rm cl}$ and the maximum stable mass $M_{\rm cl,max}$. In general the profile becomes steeper for higher mass fraction $f$ (Fig.~\ref{figdensprofile}). Close to the cloud centre the profile flattens. Apart from clouds close to the maximum cloud mass the density profile outside the central region is less steep than a power law $\rho\propto r^{-2}$.

\begin{figure}
	\includegraphics[width=\hsize]{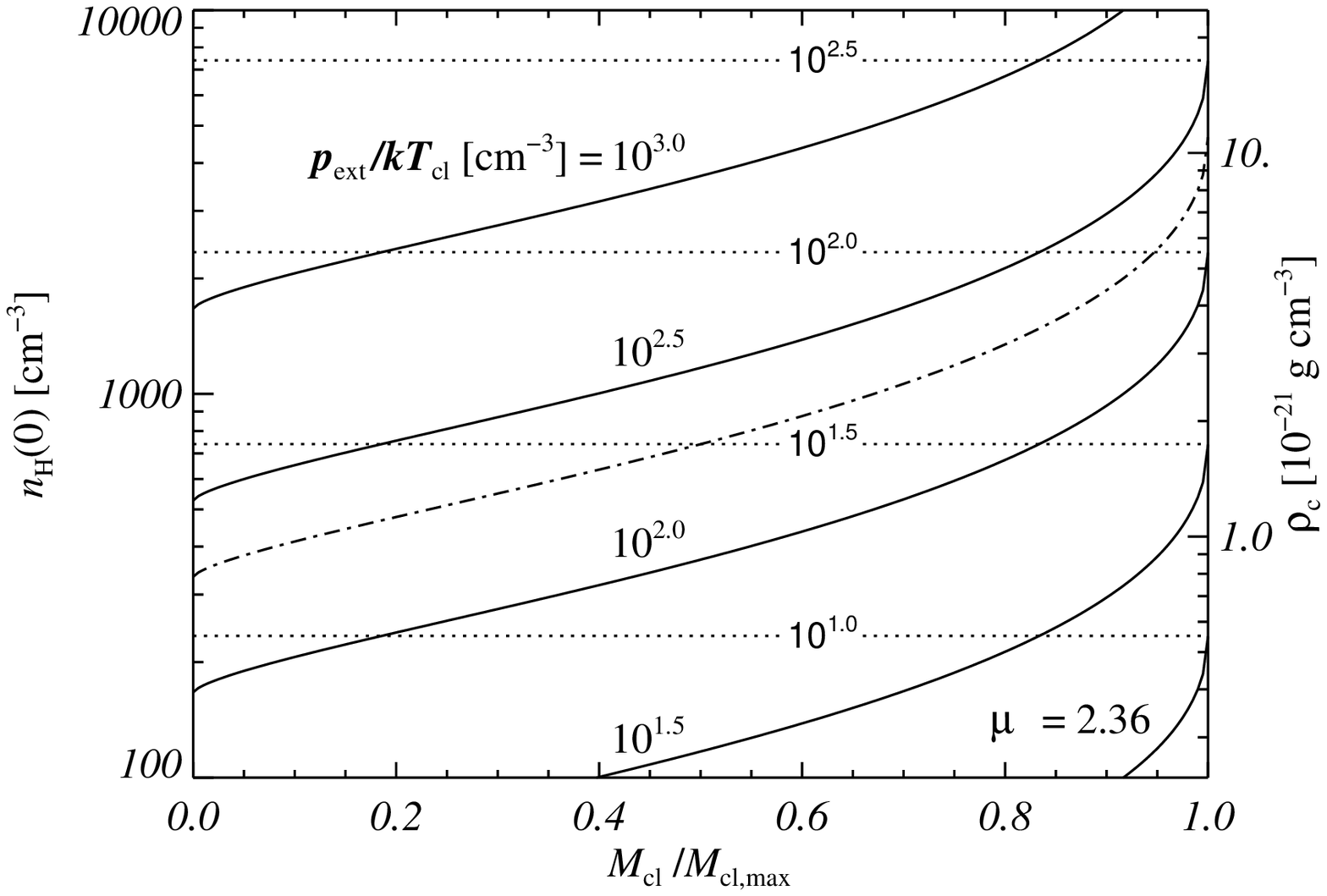}
	\caption{\label{figcentraldensity} Central density in isothermal clouds as function of critical mass fraction $f$ for several values of $p_{\rm ext}/kT_{\rm cl}$. The dashed-dotted curve shows the density values for the parameters assumed in the paper. The maximum central densities for different parameters $p_{\rm ext}/kT_{\rm cl}$ are shown as dotted horizontal lines. The abundances are taken to be solar and the hydrogen is assumed to be completely in form of molecular hydrogen giving a mean molecular weight $\mu\sim 2.36$. }
\end{figure}

For given mass fraction the density increases proportional to $p_{\rm ext}\mu/T_{\rm cl}$. The relationship between the central density and mass fraction $f$, outer pressure $p$, and cloud temperature $T$ is visualised in Fig.~\ref{figcentraldensity} where the mean molecular weight is taken to be 2.36 consistent with a gas with solar abundance where hydrogen is completely molecular.

{A self-gravitating isothermal cloud at the limit of its critical mass against gravitational collapse is known as Bonnor-Ebert sphere \citep{Ebert1955,Bonnor1956}. This is characterised by an over-pressure of 14.04 in the centre relative to the outer medium. As the central density is very sensitive close to the critical mass and decreases strongly toward slightly smaller masses stable clouds should be characterised with over-pressures of less than a factor of 10. A cloud with a mass fraction of $f=0.9$ has for example an over-pressure of less than 6. }

\subsection{The cloud size}

\begin{figure}
	\includegraphics[width=\hsize]{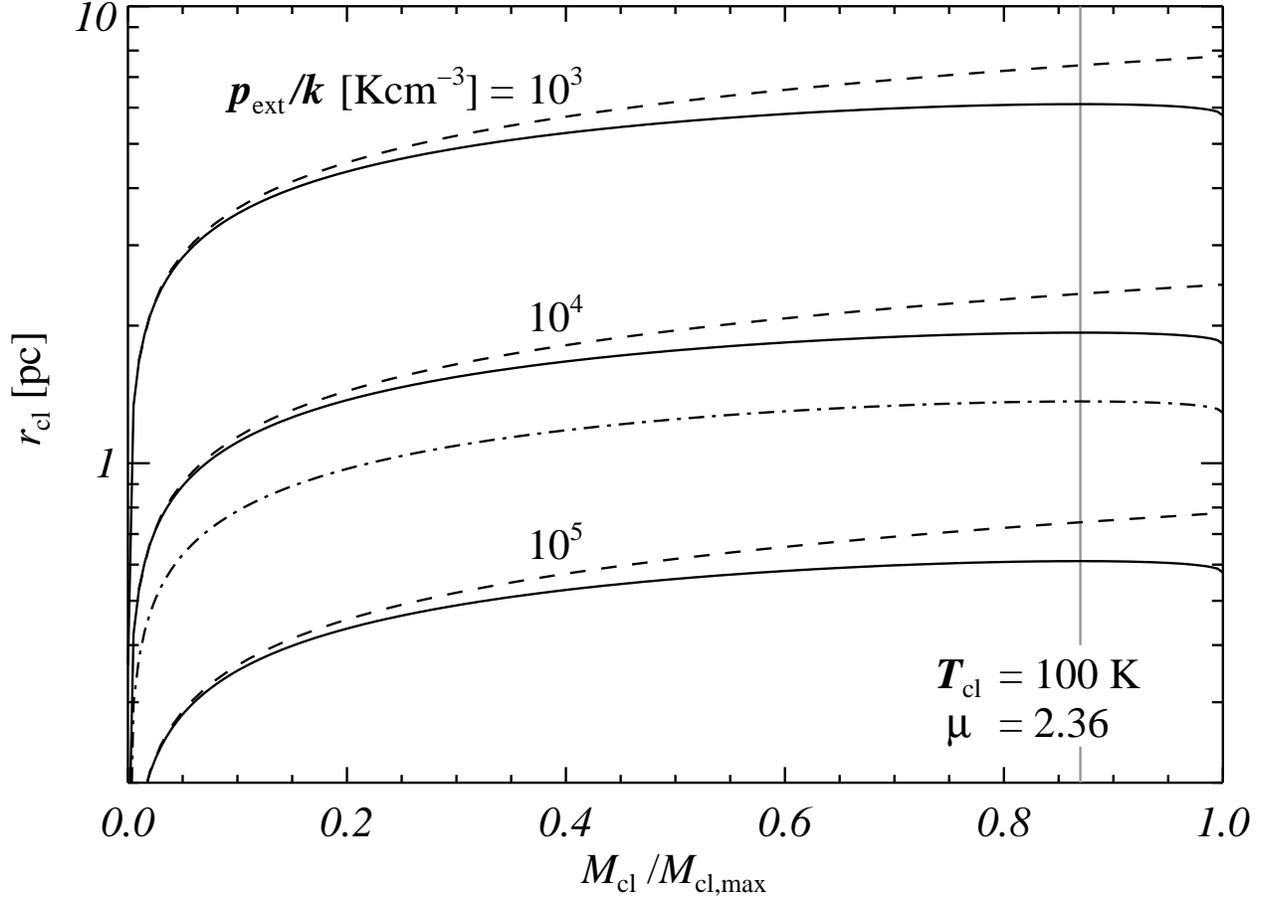}
	\caption{\label{figcloudradius} The cloud size as function of the critical mass fraction $f=M_{\rm cl}/M_{\rm cl,max}$ for various values of the external pressure $p_{\rm ext}/k$ (solid lines). The cloud temperature is taken to be $100~{\rm K}$ and the mean molecular weight $\mu=2.36$. The dashed-dotted curve shows the corresponding cloud sizes for our model 
assumption ($p_{\rm ext}/k=2\cdot 10^4~{\rm K/cm^3}$). For comparison also the sizes of homogeneous filled clouds are shown (dashed lines).  The vertical grey line marks the maximum cloud extension for $f=0.8698$.}
\end{figure}

The cloud size is not a critical value for the radiative transport problem as this is entirely determined (for given dust properties and external radiation field) by the critical mass fraction $f=M_{\rm cl}/M_{\rm cl,max}$ and the outer pressure $p_{\rm ext}$. However, the size determines the total cloud luminosity and is important from an observational point of view. For given critical mass fraction the cloud radius is proportional to $T_{\rm cl}\mu^{-1} p_{\rm ext}^{-1/2}$. A cloud which is ten times as hot is also ten times larger, but it has 100 times larger mass. In Fig.~\ref{figcloudradius} the cloud radius is compared with a simplified model of homogeneous spheres as function of the critical mass fraction $f$. The two curves start to deviate for critical mass fractions larger than $\sim 0.1$ where the density profile becomes significantly steeper with increasing $f$. As more mass is concentrated in the cloud centre the cloud size becomes smaller with respect to the homogeneous sphere. Above $f=0.8698$ the cloud size shrinks for increasing  critical mass fraction. 

\subsection{The column density $N_{\rm H}$}

The most important parameter which determines the radiative transfer problem is the column density towards the cloud centre. For a given outer pressure $p_{\rm ext}/k$ the column density $N_{\rm H}$ of all hydrogen depends only on the ratio $f=M_{\rm cl}/M_{\rm cl, max}$.
For given mass fraction $f$ the column density increases proportional to the square root of the outer pressure: 
\begin{equation}
   N_{\rm H}(f,p_{\rm ext}) = \sqrt{{p_{\rm ext}/p_{\rm ext}'}}\,N_{\rm H}(f,p_{\rm ext}')
\end{equation}
However, the critical mass decreases with pressure. Therefore, in higher pressure regions the clouds are less massive but more compact and more opaque. 

\begin{figure}
  \includegraphics[width=\hsize]{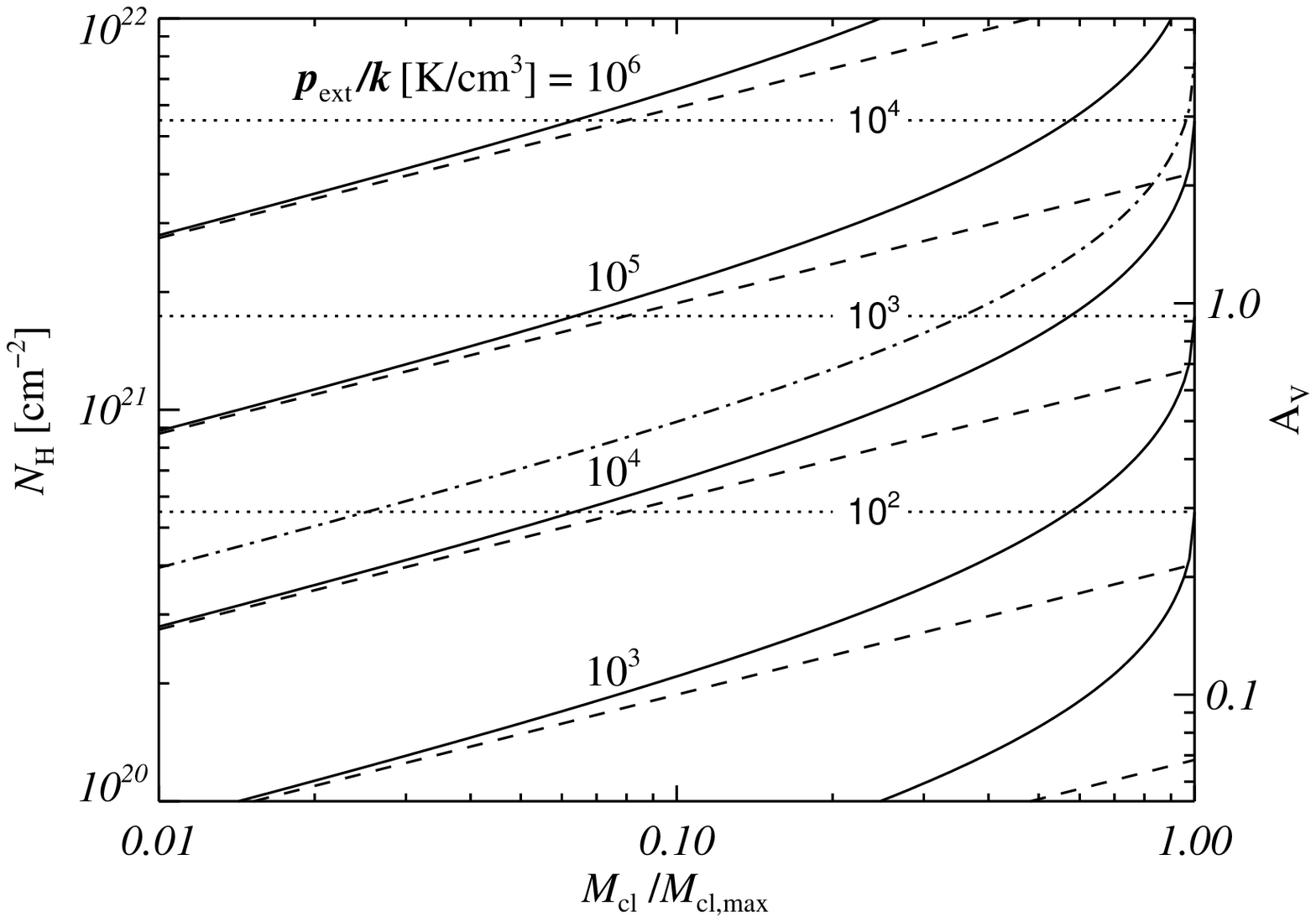}
  \caption{\label{figcolumn} The column density and extinction $A_{\rm V}$ towards the centre of gravitationally stable isothermal clouds as function of their critical mass fraction $f$. The extinction in V is obtained by using a `gas-to-dust ratio' $N_{\rm H}=5.8\,10^{21}\,E(B-V)$ \citep{Bohlin1978} and adopting a value of  $R_{\rm V}=3.1$  for the absolute to relative extinction, derived from the observed mean extinction curve of our galaxy \citep{Fitzpatrick1999}. The column density depends on gas pressure of the ambient medium which is varied over several magnitudes up to $p_{\rm ext}/k=10^6~{\rm K/cm^3}$. The dashed-dotted line corresponds to a pressure of $p_{\rm ext}/k=2\cdot10^4~{\rm K/cm^{3}}$ appropriate for the ISM of our galaxy. For comparison also the column density through homogeneous clouds is shown (dashed lines). The maximum column density is shown as vertical dotted lines labeled with the corresponding pressure $p_{\rm ext}/k$ of the ambient medium.} 
\end{figure}

The variation of column density as function of the critical mass fraction $f$ is shown in Fig.~\ref{figcolumn}. This is again compared with the column density through an homogeneous sphere. For small critical mass fraction both values are the same. Towards higher critical mass fraction the column density through the isothermal clouds increases strongly relative to the homogeneous sphere.
For stable isothermal clouds embedded in the warm neutral medium (WMN) with $p_{\rm ext}/k=2\cdot 10^4~{\rm K/cm^3}$ one expects total column densities no larger than $N_{\rm H}=1.6 \times 10^{22}~{\rm cm^{-2}}$, a value slightly smaller than the column density through molecular clouds ($N_{\rm H}=2N_{\rm H_{2}} \sim 2 \cdot 10^{22}~{\rm cm^{- 2}}$; \citep{Dopita2003}).
Gravitational stable clouds should have total column densities $N_{\rm H} $ of a few $~10^{21}~{\rm cm^{-2}}$, a value which is quite similar to the column densities for translucent clouds \citep{Dishoeck1988,Rachford2002}.

In Fig.~\ref{figcolumn} we also give the attenuation towards the cloud centre in the visual assuming typical dust properties. Clouds with column densities of $10^{21}~{\rm cm^{-2}}$ have an attenuation of $A_{\rm V}\approx 0.5$. The attenuation towards the cloud centre of clouds in the WNM of our galaxy is not larger than 3. Typical clouds in the WNM are therefore marginally optically thick in the optical. Self-gravitating clouds with high critical mass fraction fulfil the criterion of translucent clouds given by \citet{Dishoeck1988}. In higher pressure regions such as in star-burst galaxies or HII-regions these isothermal self-gravitating clouds are expected to be optically thick.

\begin{figure}
	\includegraphics[width=\hsize]{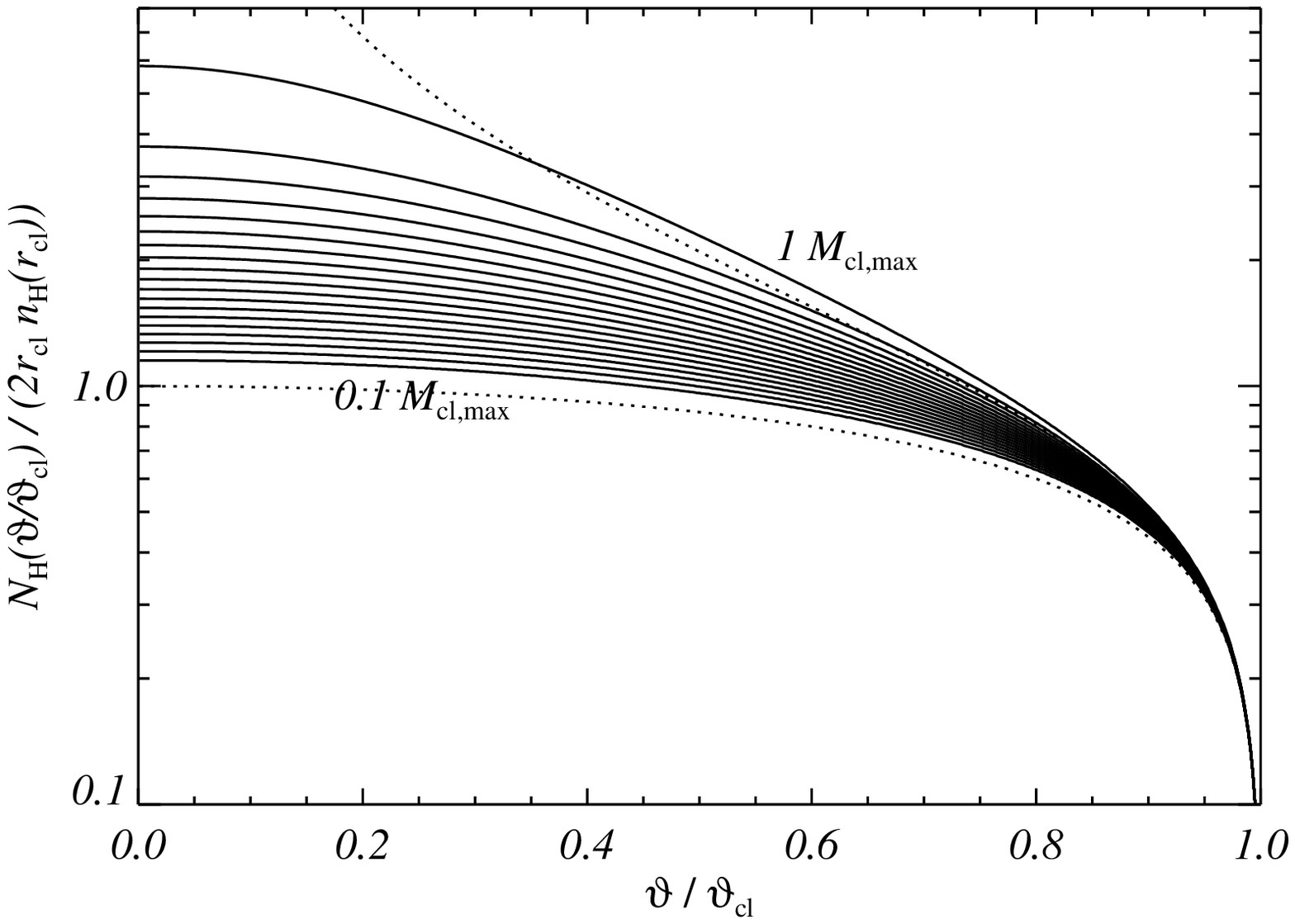}
	\caption{\label{figcolumnprof}Profile of the column densities through distant isothermal clouds ($\vartheta_{\rm cl} = r_{\rm cl}/D\ll 1$, where $r_{\rm cl}$ and $D$ are the cloud radius and the distance to the cloud centre). The column density is normalised to $2r_{\rm cl} n_{\rm H}(\rm cl)$. The profile depends on the critical mass fraction $f=M_{\rm cl}/M_{\rm cl,max}$ which is varied from 0.1 to 1.0 in steps of 0.05. The dotted lines show the profiles for a homogeneous sphere (lowest curve) and a sphere with a steep density profile $\rho\propto r^{-2}$.}
\end{figure}

The steeper radial density profile for high critical mass fraction provides a steeper profile of the column density, as shown in Fig.~\ref{figcolumnprof}. Clouds close to the critical mass against gravitational collapse have column densities through the central region almost 6 times higher than through a homogeneous sphere of the same dimension and outer density. 

In the frame of the model we can classify the different cloud types. Translucent clouds characterized by an $A_V$ from 1 to $\approx 5$ in the WNM with $p_{\rm ext}/k=2\cdot 10^4~{\rm K/cm^3}$ would typically have a mass in the range from 0.1 to 0.9 of the critical mass. Clouds with less than $10\%$ of the critical mass are diffuse and clouds with a mass fraction of more than $\approx 90\%$ are dense clouds or Bok Globules. This shows why Globules often have a rather steep density profile that is at the outskirts approximately a power law with $n_{\rm H}\propto r^{-2}$ while translucent clouds typically have a flatter density profile.

\section{The Dust Model}


In this paper we assume that the dust properties inside the isothermal clouds are the same as the mean properties of the diffuse ISM of our galaxy. For the translucent clouds with $A_V< 3$ this might be a very good approximation. However, we do not expect this to be the case for more compact clouds. In these, grain growth processes probably play an important role \citep{Whittet2001}, and our assumption about the grain properties in those clouds  could well break down. 

The dust is assumed to have a certain composition and size distribution. The number of grains of composition $i$ in the radius interval $a...a+{\rm d}a$ is given by ${\rm d}n(a)={\rm d}a\,n_{\rm H}\,\zeta_i f_i(a)$  where $n_{\rm H}$ is the number density of hydrogen atoms and $\zeta_i$ a constant for which we have chosen the abundance of a key element condensed in grain species $i$. We consider a mixture of silicate, graphite, and iron grains and additional PAH-molecules. The dust grains composed of silicate, graphite, and iron are assumed to have a power law distribution with smooth exponential cut-offs at both the small and large grain sizes:
\begin{equation}
	\ln \tilde f_i(a)=  -k_i\ln a-(a_{i,\rm min}/a)^{m_1}-(a/a_{i,\rm max})^{m_2}.
\end{equation}
For PAH-molecules we assumed a simple log-normal density distribution. 
\begin{equation}
	\ln \tilde f_{\rm pah}(a) =  -\ln a-\frac{1}{2{\sigma^2_{\rm pah}}}{\left(\ln a-\ln a_0\right)^2}
\end{equation}
with $\ln a_0 = \ln \left<a_{\rm pah}\right>-\frac{1}{2}\sigma_{\rm pah}^2$
where $\left<a_{\rm pah}\right>$ and $\sigma_{\rm pah}$ are the mean grain size and the standard deviation. 

The distribution functions $\tilde f_i(a)$ are normalised to the expression
\begin{equation}
	f_i(a) = \frac{{\rm m_{u}}}{(4/3)\pi\int \tilde f_i(a)a^3\,{\rm d}a}\left<\frac{\mu}{\rho}\right>_i\tilde f_i(a)
\end{equation}
to obtain the correct units for $\zeta_i$. ${\rm m_u}$ is the atomic mass unit and $\left<\mu/\rho\right>_i$ the averaged ratio 
of the mean atomic mass $\mu_i$ of the chemical composition and $\rho_i$ the density of the grain.

Following \citet{LiDraine2001} the emission features of the PAH-molecules differ between charged and non charged states. To determine the SED of isothermal clouds we assumed the probability of ionised PAH-molecules as given for the cold-neutral-medium (CNM); see Fig.~7, of \cite{LiDraine2001}. 

\begin{figure*}[htb]
	\includegraphics[width=\hsize]{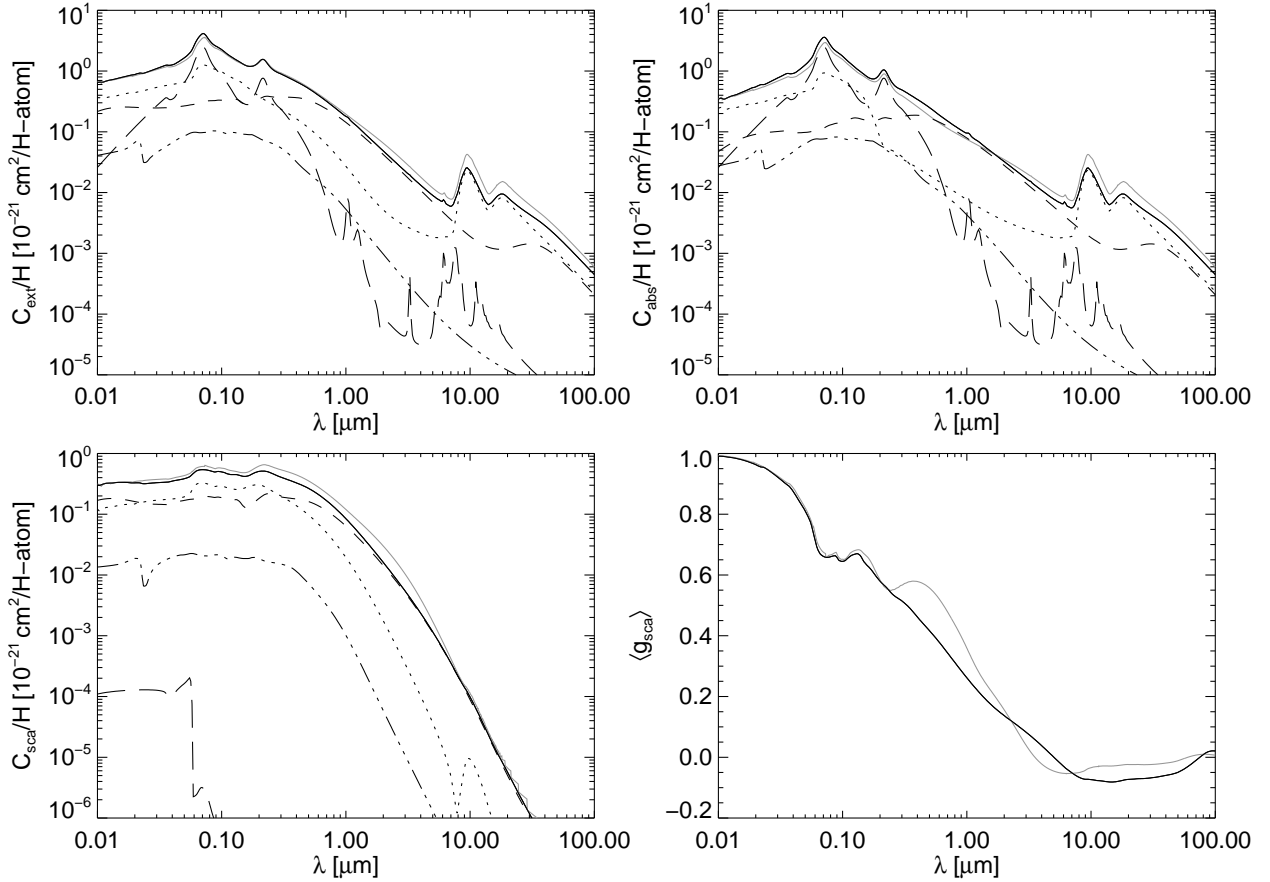}
	\caption{\label{figopt1} Mean optical properties of the dust grains inside the clouds (black solid lines). The data are compared with the mean properties obtained by \citet{Weingartner2001} (grey solid lines). Their values for the extinction, absorption, and  scattering has been corrected by a factor 0.93. The contribution of different dust compositions to the mean values are shown as dashed (graphites), dotted (silicates), long-dashed (PAH-molecules), and dashed-dotted (iron grains) lines.}
\end{figure*}

The parameters for the grain composition and for the size distribution are obtained by fitting simultaneously  the mean extinction of our galaxy (closely consistent with \citet{Fitzpatrick1999}, the diffuse IR-emission measured with DIRBE on board of the COBE satellite \cite{Arendt1998} and the depletion of key elements (C, O, Si, Mg, Fe) from the gas phase \citep{Kimura2003}. 
The depletion is generally obtained by comparing the abundance in the gas phase with the abundances derived for the sun or other stars from photo-spheric using radiative transfer codes. The actual abundance of the gas phase depends on the ionisation fraction of the gas. Kimura provides two depletion factors for an ionisation fraction of 0.25 and 0.45. We adopted the values for 0.45. The depletion of the elements of interest in this work are based on the solar abundances as given by \citet{Holweger2001}. The values are consistent with the ones given by \citet{Zubko2004}. 

The actual composition of silicate grains is not certain but it seems that silicates have generally a low
iron content \citep{Gail1999}. In our model we considered silicates to be a mixture of magnesium rich Olivine (itself a mixture between Fosterite ($\rm Mg_2SiO_4$) and Fayalite ($\rm Fe_2SiO_4$) summarized as $\rm Mg_{2x}Fe_{2(1-x)}SiO_4$) and Quartz ($\rm SiO_2$). For Olivine we used $x=0.7$ consistent with theoretical calculations about the formation of dust grains in the winds of oxygen rich stars \citep{Gail1999}. For simplicity we assume for the different silicate compositions the same optical properties.

The assumed physical grain properties are summarized in table~\ref{tablephysical}.
The optical properties for silicate and graphite grains are taken from \citet{DraineLee1984,LaorDraine1993,Weingartner2001} and the ones for un-ionised and ionised PAH-molecules from \citet{DraineLi2007}. For iron grains we adopted the optical properties given by \citet{Fischera2004}.

In the calculation of the thermal re-emission by dust grains we accurately take into account the stochastic heating of the grains by the ambient radiation field. Large grains maintain a certain temperature determined by the equilibrium of heating and cooling rates. However, in the case of small grains where the thermal energy is typically smaller than the energy of the absorbed photon, the stochastic heating can lead to large temperature fluctuations.

The probability distribution for the temperature of small grains is obtained by adopting the method described by \citet{Voit1991} which is a combination of the numerical integration of \citet{Guhathakurta1989} and a stepwise analytical approximation. 
This method avoids inaccuracies of the pure numerical integration in the limit of small temperature fluctuations in the transition from the behaviour of small to large grains and allows a relatively small
number of integration intervals.

The grain model calculation implies for every grain size and composition equilibrium between absorbed emitted radiation. The energy is
conserved better than $2~\%$ where the largest discrepancy occurs for smallest grain sizes. The individual grain spectra are scaled to have the energy balance fulfilled.

The heat capacities for silicates, graphites and iron grains needed to derive the temperature probability distributions are summarised in \citet{Fischera2000}. For PAH-molecules we used the corrected\footnote{There is a mistake in their recommended formula for the thermal heat,  a two-dimensional  Debye-model with slightly different Debye temperatures relative to the original work by \citet{Krumhansl1953}. In the formula the dimension needs to be in the nominator but appears in the de-nominator instead so that the thermal heat is under-estimated by a factor 4. It appears that the authors were using the correct formula as it would have produced too much emission at short wavelengths.}
formula for the thermal heat given by \citet{LiDraine2001}.

\begin{table}[htb]
	\caption{\label{tablephysical} Physical Grain Parameters}
	\begin{tabular}{llccc}
		\multicolumn{2}{l}{compostion}  & density $\rho_i$& atomic weight $\mu_i$& \\
			& 		& $[{\rm g/cm^3}]$ & $[{\rm AMU}]$\footnote{atomic mass units} \\
		\hline
		graphite 	&		& 2.24 & 12.01 \\
		PAH\tablenotemark{a}	&		& 2.24 & 12.01 \\
		iron 	&		&7.53 & 55.85 \\
		Fosterite 	&(${\rm Mg_2SiO_4}$)& 3.21 & 140.694 \\
		Fayalite 	&($\rm Fe_2SiO_4$) 	& 4.30 & 203.776 \\
		Quartz  	&($\rm SiO_2$)	& 2.2 & 60.08 \\
	\end{tabular}
	\tablenotetext{a}{For PAH-molecules we ignored the additional weight by hydrogen atoms.}
\end{table}

\begin{table}[hbt]
	\caption{\label{tablemodelparameter}
	Grain Model Parameters}
	\begin{tabular}{lcccccc}
		& $\zeta_i$ & $k$ & $a_{\rm min}$ & $a_{\rm max}$ & $m_1$ & $m_2$\\
		& $n_{i}/n_{\rm H}~[10^{-6}]$ & & $[\mu{\rm m}]$ & $[\mu{\rm m}]$ & \\
	\hline
	silicate & 33.22 & 3.5 &0.001 & 0.257 & 3 & 3\\
	graphite & 199.9 & 5.5 & 0.070 & 1.0 & 3 & 3\\
	iron & 11.14 & 3.5 & 0.001& 0.1 & 3 & 3 \\
	\hline
		& $\zeta_i$ & & $\left<a_{\rm pah} \right>$ & $\sigma_{\ln a_{\rm pah}}$ \\
		& $n_{i}/n_{\rm H}~[10^{-6}]$ & & $[{\rm nm}]$ & $[\ln{\rm nm}]$ & \\
	\hline
	PAH & 72.82 & & 0.650 & 0.119 \\ 
	\end{tabular}
\end{table}

The assumed model parameters obtained through a combined fit are summarized in table~\ref{tablemodelparameter}. The fraction of condensed silicon in Olivine is 0.74. The mean optical properties for our grain composition and size distribution are shown in Fig.~\ref{figopt1}. We note the following principal features:
\begin{itemize}
\item The feature in the extinction curve at 2200~\AA{} is entirely explained by the presence of PAH-molecules. 
\item The size distribution of carbon grains (PAH-molecules and graphite grains) is bi-modal. The PAH-molecules have a narrow distribution around 6.5~\AA{} while the graphite grains are restricted to the relatively large grains $a > 0.01~\mu{\rm m}$ which do not show a strong absorption feature at 2200~\AA{}. This result is similar to the one found by \citet{Zubko2004} for a mixture composed of silicates, graphites, and PAH-molecules, and strengthens the association of the 2200~\AA\ absorption feature with PAH molecules, rather than with graphite. 
\item To explain the diffuse emission at $60~\mu{\rm m}$ we introduced iron grains which were also proposed to explain the $60~\mu{\rm m}$-emission measured with IRAS \citep{Chlewicki1988}. The size distribution for iron grains is chosen to produce a maximum flux at $60~\mu\rm m$. A different assumption for iron grains (allowing smaller iron grains) could well modify the shape of the excess and could even reduce the amount of iron necessary to be condensed in iron grains.
\end{itemize}


The model parameters are somewhat different to the ones given by \citet{Weingartner2001} which leads to differences in the mean optical properties. Even though the mean extinction curve is similar our dust-model predicts higher absorption relative to the scattering. In addition the scattered light in the UV is somewhat less forward-scattered. 

\section{The radiative transfer code}

The code used to derive the SED of isothermal clouds will be fully described in a separate paper. Here we summarize its main features:

The code is quite general and can be applied both to clouds that are heated only from outside by an isotropic radiation field or to shells heated both from a central source and an outer radiation field. In the code the light of the sources and the scattered light and the emitted light is treated separately to allow a quantitative analysis of the importance of the different components. Here, we follow the nomenclature used in the code. The light of the internal source and the external source are labelled with index `1' and `4', the scattered light and the emitted light with `2' and `3'.

To solve the radiative transfer problem including re-emission from dust grains a ray-tracing code is used because it allows a realistic determination of the light distribution of the external light $I_{\nu}^{(4)}(r,\vartheta)$, the scattered light $I_{\nu}^{(2)}(r,\vartheta)$, and the emitted light $I_{\nu}^{(3)}(r,\vartheta)$ inside the cloud. Non-isotropic scattering as well as multiple scattering events are both taken accurately into account. To properly account for re-heating by dust grains and scattering of re-emitted photons the problem is solved iteratively. Polarisation is considered to be unimportant in determining the form of the SED and is therefore neglected.

The cloud is divided in $N$ shells of constant density that are each optically thin to the attenuated radiation. Therefore, in case of very compact clouds the column density through the optically thin shells increases inwards. We used a maximum optical thickness relative to the attenuated flux density of $\tau=0.3$. 

To probe the angular distribution of the light inside individual shells we used the mean 
distances $\left<r_i\right>$ of each thin shell $i$ to the cloud centre as the rays' impact parameters (that is, the closest distance achieved by a given ray from the cloud centre). One further ray is added with impact parameter zero. The angular information is obtained for each ray as they cross the thin shells at different angles. The light distribution in shell $i$ is derived by calculating the light along $i+1$ rays. Since rays interior to a shell cross the shell twice they provide brightness information both for light crossing the shell from outside to the inside and for the light crossing the shell from inside to the outside. One ray provides the brightness of the light which passes the shell tangentially. Using this method we obtain for shell $i$ the brightness information at $2i+1$ different angles. 

Absorption and scattering are derived using the averaged optical properties in the shells.
To calculate non-isotropic scattering we made use of the Henyey-Greenstein scattering function \citep{Henyey1941} that is a smooth angular function of the scattering angle and that depends only on the $g$-parameter, the cosine weighted mean of the scattering function of non-polarised light.

The scattering problem is solved by deriving successively the brightness distribution $I_{\nu}^{(2(s))}(r,\vartheta)$ produced by photons scattered $1,2,3,...,s$ times.
We consider the flux of scattering events as unimportant if the fractional increase of the total luminosity of scattered photons in all shells is less than $10^{-3}$. In general the maximum number of scattering events taken into account is less than 10 in all calculations. The iterative process terminates if the increase in luminosity per additional scattering  is also less than $10^{-3}$.

As a main result the radiative transfer code provides the intensities $I^{(n)}_\nu(r_{\rm cl},\vartheta)$ of the attenuated flux ($n=4$), the scattered flux $(n=2)$, and the flux of the dust emission ($n=3$) at the cloud radius $r_{\rm cl}$. In addition it also provides the scattered intensities ($n=2(s)$) of photons scattered exactly $s$ times before leaving the cloud. The flux density of the different components leaving the cloud is given by:
\begin{equation}
	F_{\nu}^{-(n)} = 2\pi\int_0^{\pi/2} {\rm d}\vartheta\,\sin\vartheta \cos\vartheta\,I_\nu^{(n)}(r_{\rm cl},\vartheta)
\end{equation}
where $n=4$, 3, 2 or $2(s)$.
The flux density entering the cloud is given by
\begin{equation}
	F_{\nu}^{+(4)} = \pi J_\nu^{ISRF}
\end{equation}
where $J_\nu^{ISRF}$ is given by equation \ref{eqisrf}.
The corresponding integrated flux has to be equal to the scattered, the re-emitted, and the attenuated light leaving the cloud:
\begin{equation}
	F^{+(4)} = F^{-(2)}+F^{-(3)}+F^{-(4)}.
\end{equation}
We have used this relation to verify our calculations.

\section{Model Calculations and Results}

\begin{deluxetable}{cccccccrcc}
\tablecolumns{10}
	
	\tablecaption{\label{tablecloudparameter}Cloud Parameters\tablenotemark{a}}
	\tablehead{
		\colhead{$M_{\rm cl}$} & \colhead{$p_{\rm ext}/k$} & \colhead{$M_{\rm cl}$} & \colhead{$r_{\rm cl}$}& \colhead{$n_{\rm H}(0)$} & \colhead{$N_{\rm H}$\tablenotemark{b}} & \colhead{$\tau_V$\tablenotemark{b}} & \colhead{$N_{\rm shell}$\tablenotemark{c}} & \colhead{$N_{\rm sca}$\tablenotemark{d}} & \colhead{accuracy\tablenotemark{e}}\\
		\colhead{$[M_{\rm cl,max}]$} & \colhead{$[{\rm K/cm^3}]$} & \colhead{$[M_\odot]$} & \colhead{[pc]} &
		\colhead{$[\rm cm^{-3}]$} & \colhead{$[10^{21} \rm cm^{-2}]$} & \colhead{} &\colhead{} & \colhead{} & \colhead{$[\%]$}
		}
		\startdata
		0.500 & $1\cdot 10^3$ & 574.2 & 5.582 & 36.99 & 0.489 & 0.225 & 5 & 4 & 0.01\\ 
		0.500 & $1\cdot 10^4$ & 181.6 & 1.765 & 369.9 & 1.548 & 0.712 & 14 & 6 & 0.07\\ 
		0.001 & $2\cdot 10^4$ & 0.257 & 0.174 & 336.4 & 0.180 & 0.083 & 2& 4 & -0.17 \\
		0.010 & $2\cdot 10^4$ & 2.568 & 0.373 & 347.8 & 0.395 & 0.182 & 4 & 4 & 0.02\\
		0.100 & $2\cdot 10^4$ & 25.68 & 0.786 & 412.4 & 0.932 & 0.429 & 9 & 5 &  0.09\\
		0.500 & $2\cdot 10^4$ & 128.4 & 1.248 & 739.7 & 2.189 & 1.006 & 20 & 6 & 0.08\\
		1.000 & $2\cdot 10^4$ & 256.8 & 1.290 & 4682. & 7.761 & 3.568 & 70 & 8 & 0.04\\
		0.500 & $1\cdot 10^5$ & 57.42 & 0.558 & 3699. & 4.895 & 2.251 & 44 & 8 & 0.09 \\
		0.500 & $1\cdot 10^6$ & 18.16 & 0.177 & $3.70\cdot 10^4$ & 15.48 & 7.117 & 139 & 9 & 0.11\\
		1.000 & $1\cdot 10^6$ & 36.32 & 0.182  & $2.34\cdot 10^5$ & 54.88 & 25.23 & 353 & 9 & 0.10
		\enddata

	\tablenotetext{a}{The cloud temperature is chosen be 100 K and the mean molecular weight 2.36.}
	\tablenotetext{b}{Column density and optical depth measured to the cloud centre.}
	\tablenotetext{c}{Number of shells used to solve the radiative transfer problem.}
	\tablenotetext{d}{Number of scattering events taken into account.} 
	\tablenotetext{e}{Determined as the total flux (attenuated flux, scattered flux, and re-emitted flux) leaving the cloud relative to the flux entering the cloud.}
\end{deluxetable}

In the following we show the effects of the critical mass fraction $f$ and the pressure $p_{\rm ext}/k$ of the ambient medium of the clouds on the radiative transfer through isothermal clouds. 
The pressure $p_{\rm ext}/k$ is varied from $10^3$ up to $10^6~{\rm K/cm^3}$ and cloud masses from $0.1\%$ up to 100\% of the critical mass. 
The clouds are assumed to be heated by the interstellar radiation field which is adopted from \citet{Mathis1982}. It is given by
\begin{equation}
	\label{eqisrf}
	J_\nu^{ISRF} = \chi\left(J_\nu^{\rm UV}+\sum_{i=2}^4 W_i B_\nu(T_i)\right) + B_\nu(2.7~{\rm K}),
\end{equation}
with the dilution factors $W_2=10^{-14}$, $W_3=10^{-13}$, $W_4=4\,10^{-13}$ and temperatures $T_2=7500~{\rm K}$, $T_3 = 4000~{\rm K}$, and $T_4=3000~{\rm K}$. $J_\nu^{\rm UV}$ is the mean intensity of the UV-radiation in the solar neighbourhood (Table C1, \citet{Mathis1982}). This radiation field has a maximum photon energy of $13.6~{\rm eV}$. Apart of the ISRF of the local neighbourhood ($\chi=1$) we also assumed a 100 times stronger radiation fields  ($\chi=100$).

We assumed a cloud temperature of 100~K. For the gas we adopted both solar abundance of helium ($n_{\rm He}/n_{\rm H}\approx 0.1$) and solar abundance of heavier elements and assumed that hydrogen is completely molecular giving a mean molecular weight of $\mu=2.36$. The mass and the size can be transformed for other conditions using the relations given in section \ref{sectionmodel}. The parameters of the clouds considered in this paper are listed in table~\ref{tablecloudparameter}. In addition we give the accuracy of the SED obtained, the number of shells used to solve the radiative transfer problem, and the number of scattering events taken into account. The energy conservation (see for details in Sect.~\ref{section_sed}) in all calculations is better than $0.2\%$.

First we will examine in some detail how the grain heating and grain temperatures vary radially inside the cloud. Then we will show how the different parameters effect the mean SED and and the brightness profiles at different wavelengths.
Most attention is paid on the situation where clouds are embedded in the WNM ($p_{\rm ext}/k=2\cdot 10^4~{\rm K/cm^3}$) where the mass is chosen to be 0.5 and 1.0 of the critical mass. The results are compared with corresponding calculations for clouds in a much higher pressure region where we have chosen $p_{\rm ext}/k=10^6~{\rm K/cm^3}$.

\subsection{Grain heating}

\label{sectheating}

\begin{figure*}
	\includegraphics[width=0.49\hsize]{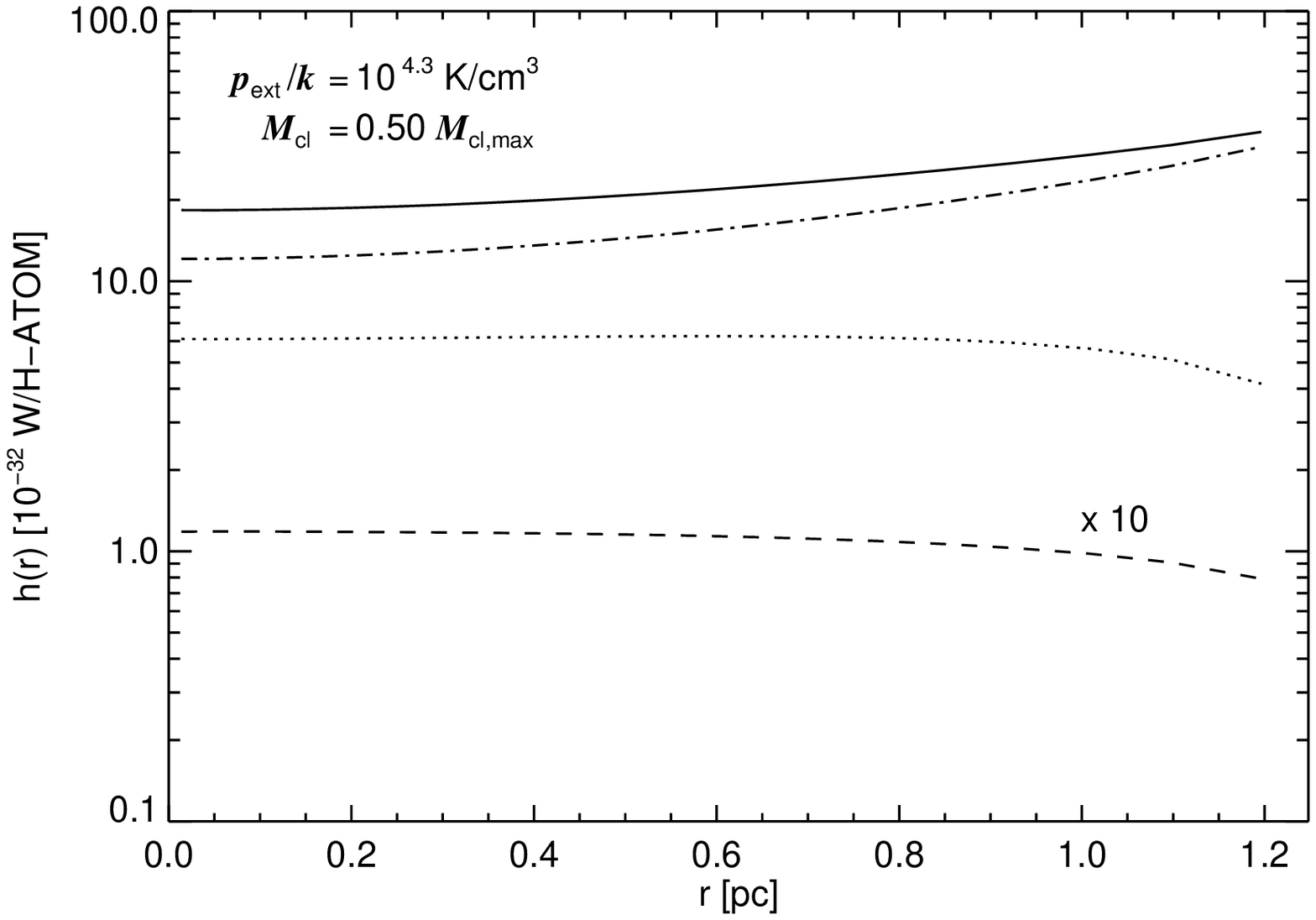}
	\hfill
	\includegraphics[width=0.49\hsize]{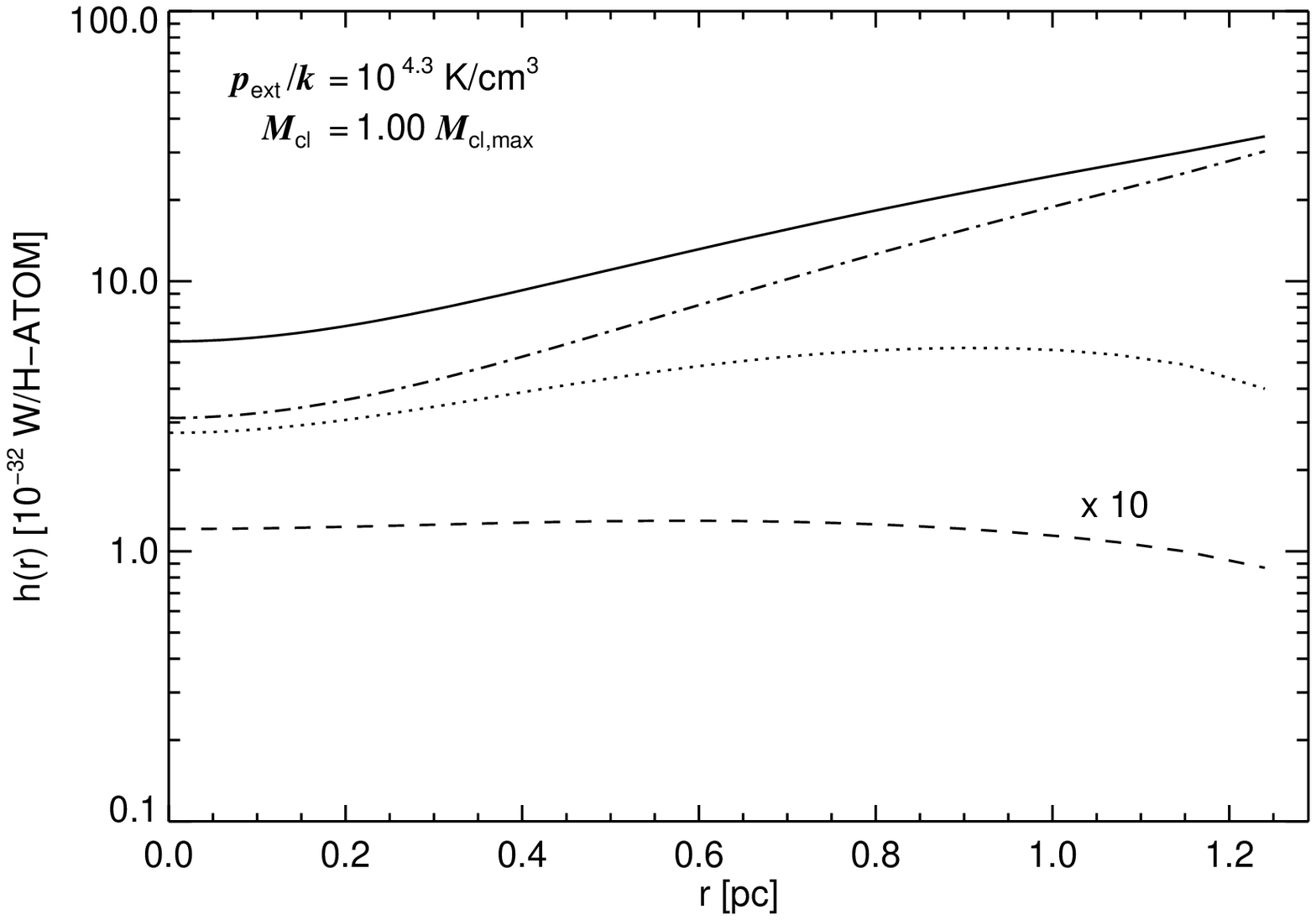}

	\vspace{0.1cm}
	\includegraphics[width=0.49\hsize]{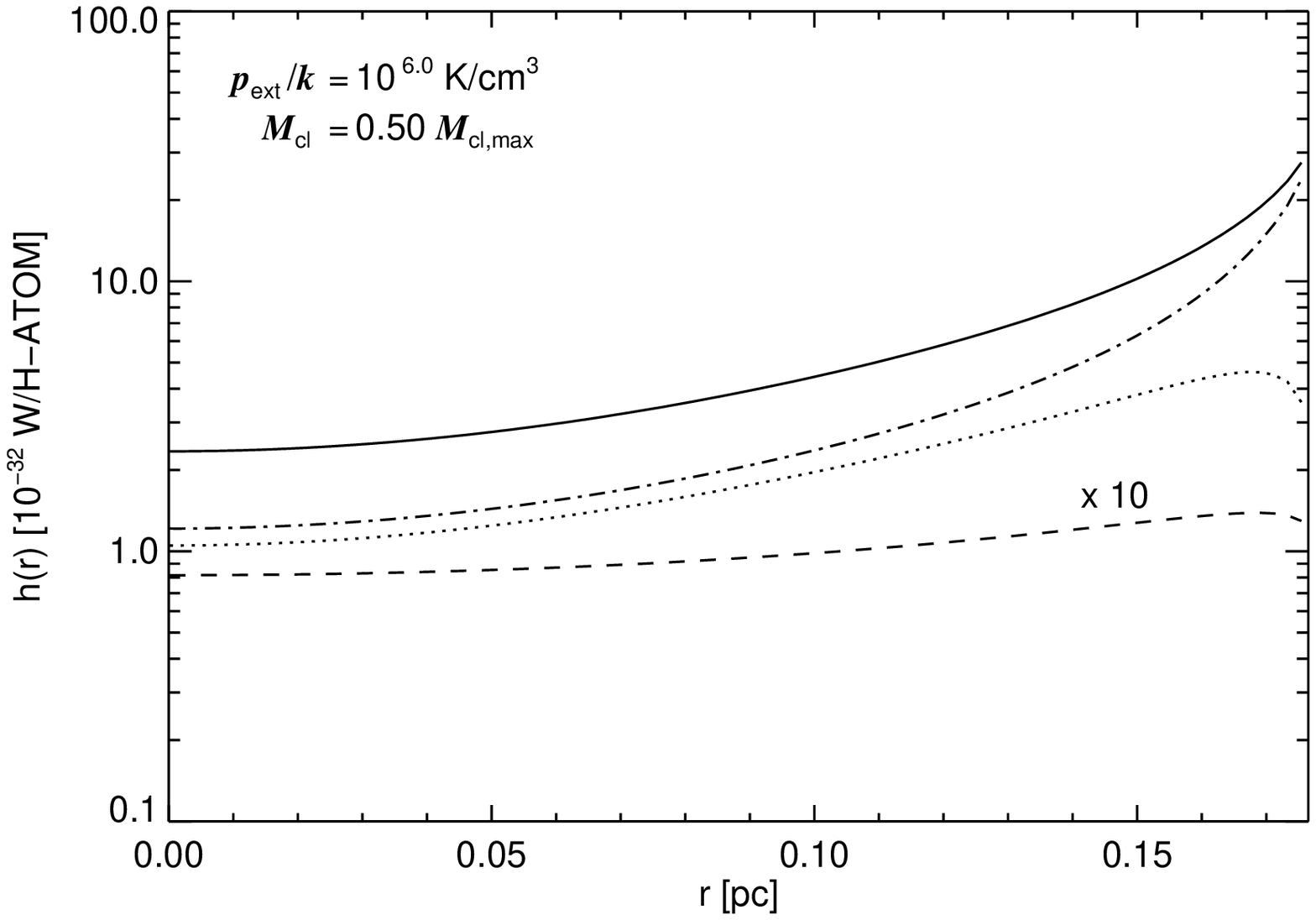}
	\hfill
	\includegraphics[width=0.49\hsize]{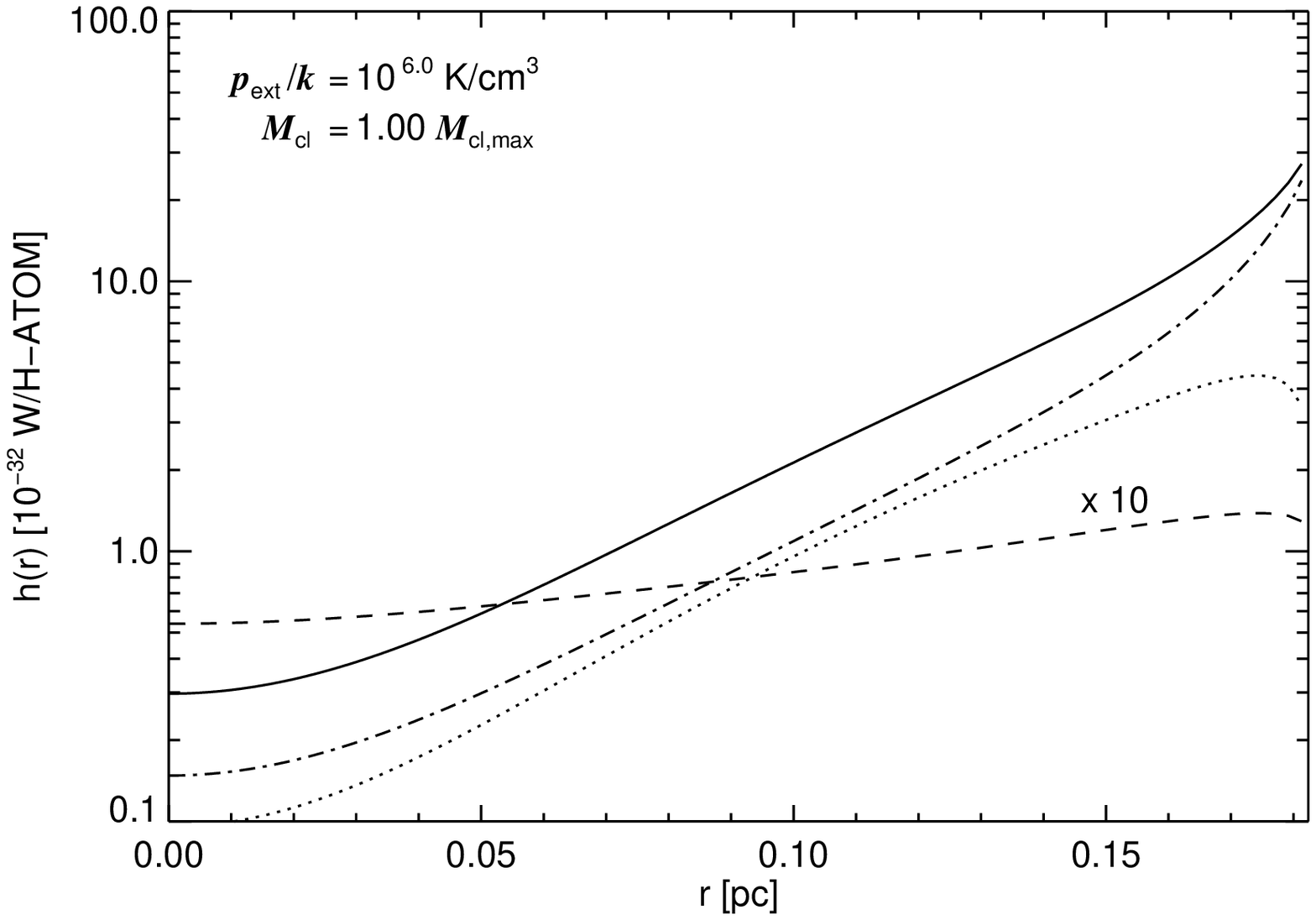}
	\caption{\label{figheating}Heating rate of the grains as function of distance from the cloud centre for different assumptions of the critical mass fraction and the outer pressure. The clouds are heated by the local ISRF ($\chi=1$). The cloud temperature is assumed to be 100 K and the mean molecular weight is taken to be  $\mu=2.36$. The total heating rate is shown as solid line. The heating rates caused by the different components of the radiation, the attenuated external flux, the scattered flux, and the flux of the thermal re-emission, are shown as dotted-dashed, dotted, and a dashed lines, respectively. The heating by the thermal re-emission is scaled by a factor of 10 to allow it to be plotted along with the other heating rates.}
\end{figure*}

To analyse the grain heating in more detail we have separated the heating rate into the three components of the radiation field inside the cloud; heating by the attenuated external flux, heating by the scattered flux, and finally, heating by the flux of the thermal re-emission. The relative contribution to the total heating rate varies as function of the distance towards the cloud centre and depends on the compactness of the clouds and the radial density profile. In Fig.~\ref{figheating} the heating rate is shown as distance from cloud centre for different assumptions about the outer pressure and the critical mass fraction. For those calculations we assumed the local interstellar radiation field with $\chi=1$.

The variation of the heating as function of radius reflects the different density profiles. For example, in case of clouds with the critical mass we see as a strong decrease close to the cloud centre resulting from the steep density profile.
 
As remarked above, stable clouds embedded in an ISM with $p_{\rm ext}/k=2\cdot 10^4~{\rm K/cm^3}$,  are not very optically thick. Clouds with a high critical mass fraction of 0.5 have an optical thickness towards the cloud centre of only $\sim 0.7$ in V. Therefore, the total heating rate in the centre of those clouds is less than a factor of 2 lower than in the outskirts.

The heating is predominantly caused by the attenuated external light. For clouds with small critical mass fraction the scattered light does not make a strong contribution and the heating by thermal emission from grains is negligible. The contribution of the scattered light to the total heating increases with critical mass fraction. In the central regions of clouds close to the gravitational collapse the scattered light is almost as important as the attenuated flux. 

Clouds in the high pressure region are 7 times more optically thick. As a result the total heating decreases by more than a factor of 10 towards the cloud centre. Interestingly we found that the scattered light never dominates the heating of the attenuated flux. The heating by the thermal dust emission is significantly higher but still below the other contributions. While the heating by thermal re-emission is still unimportant in the centre of clouds with $f=0.5$ the contribution increases up to $50\%$ in case of critical stable clouds.

The contribution of the thermal emission depends somewhat on the strength of the external radiation field. If the radiation field is more intense the grains will be warmer, which shifts the radiation of the dust re-emission to shorter wavelengths. Because the absorption efficiency of most grains increases towards shorter wavelengths it therefore becomes more likely that emitted photons are once again absorbed by other dust grains. However, in our models this effect is not very strong. For example, if the radiation field around clouds with $f=0.5$ and $p_{\rm ext}/k=10^6$ is increased by a factor of one hundred the contribution of re-emitted photons is increased by less than a factor of two.

\subsection{The grain temperatures}

\begin{figure*}

  \includegraphics[width=0.49\hsize]{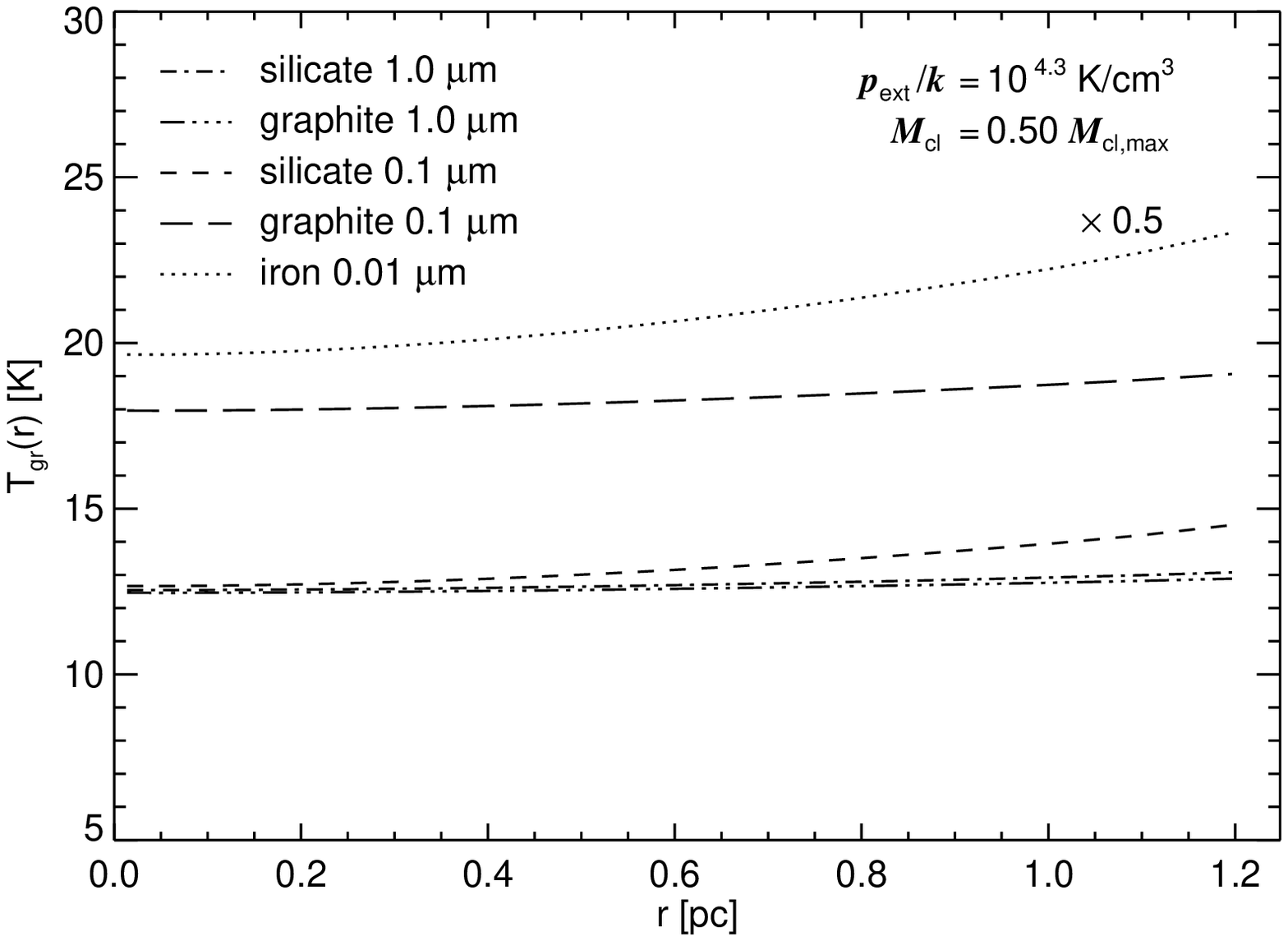}
	\hfill
  \includegraphics[width=0.49\hsize]{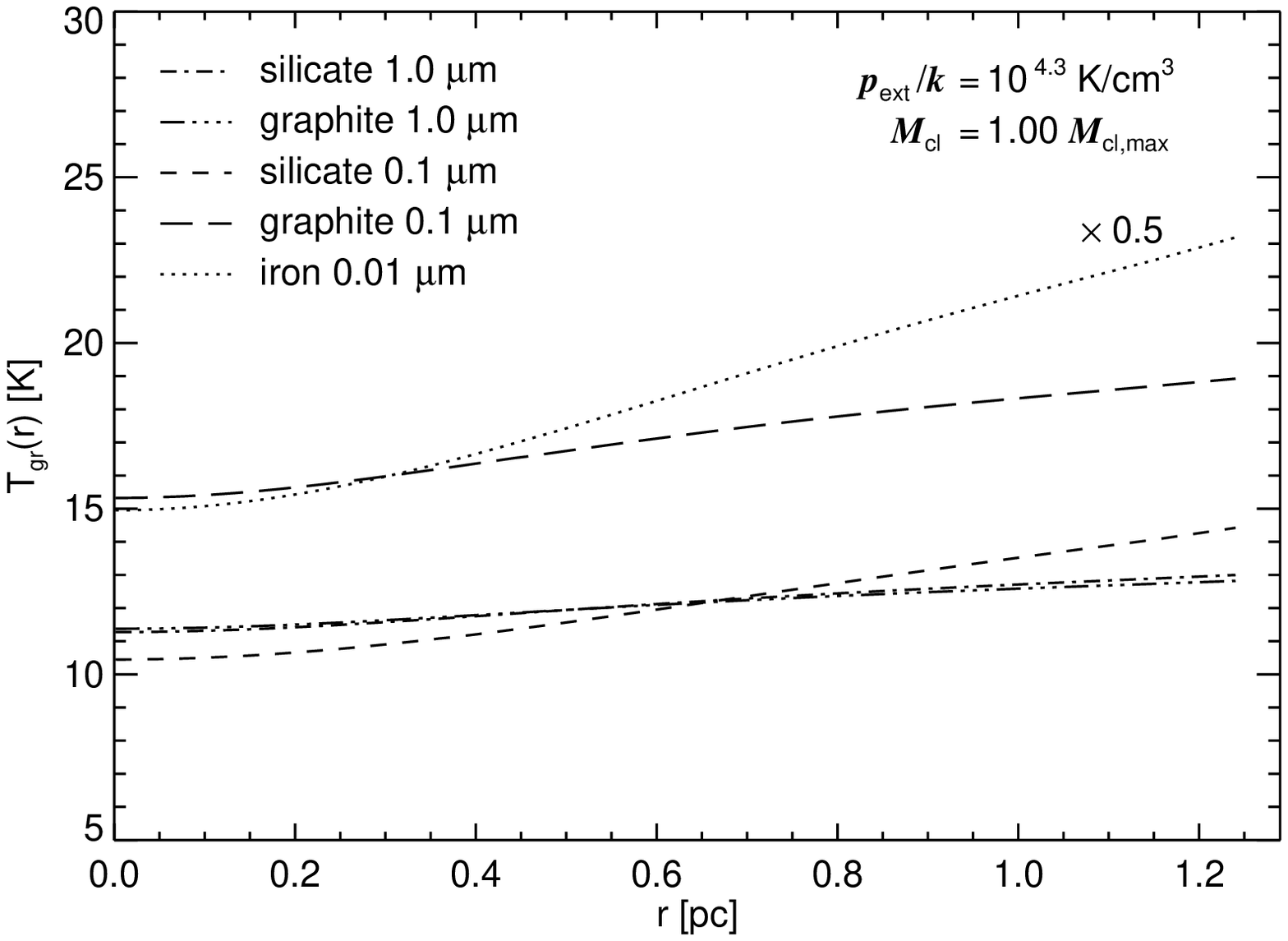}\\

  \includegraphics[width=0.49\hsize]{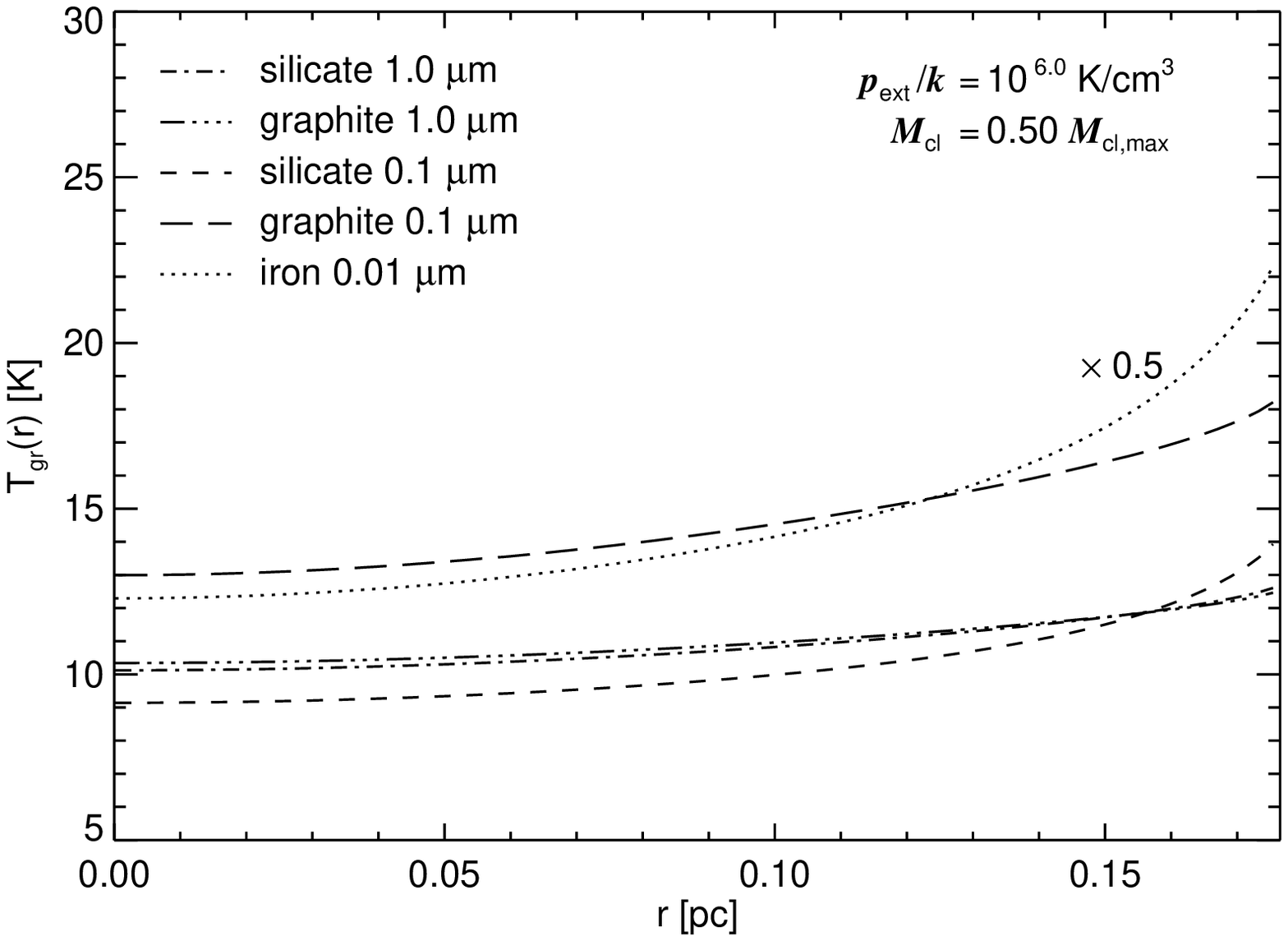}
	\hfill
  \includegraphics[width=0.49\hsize]{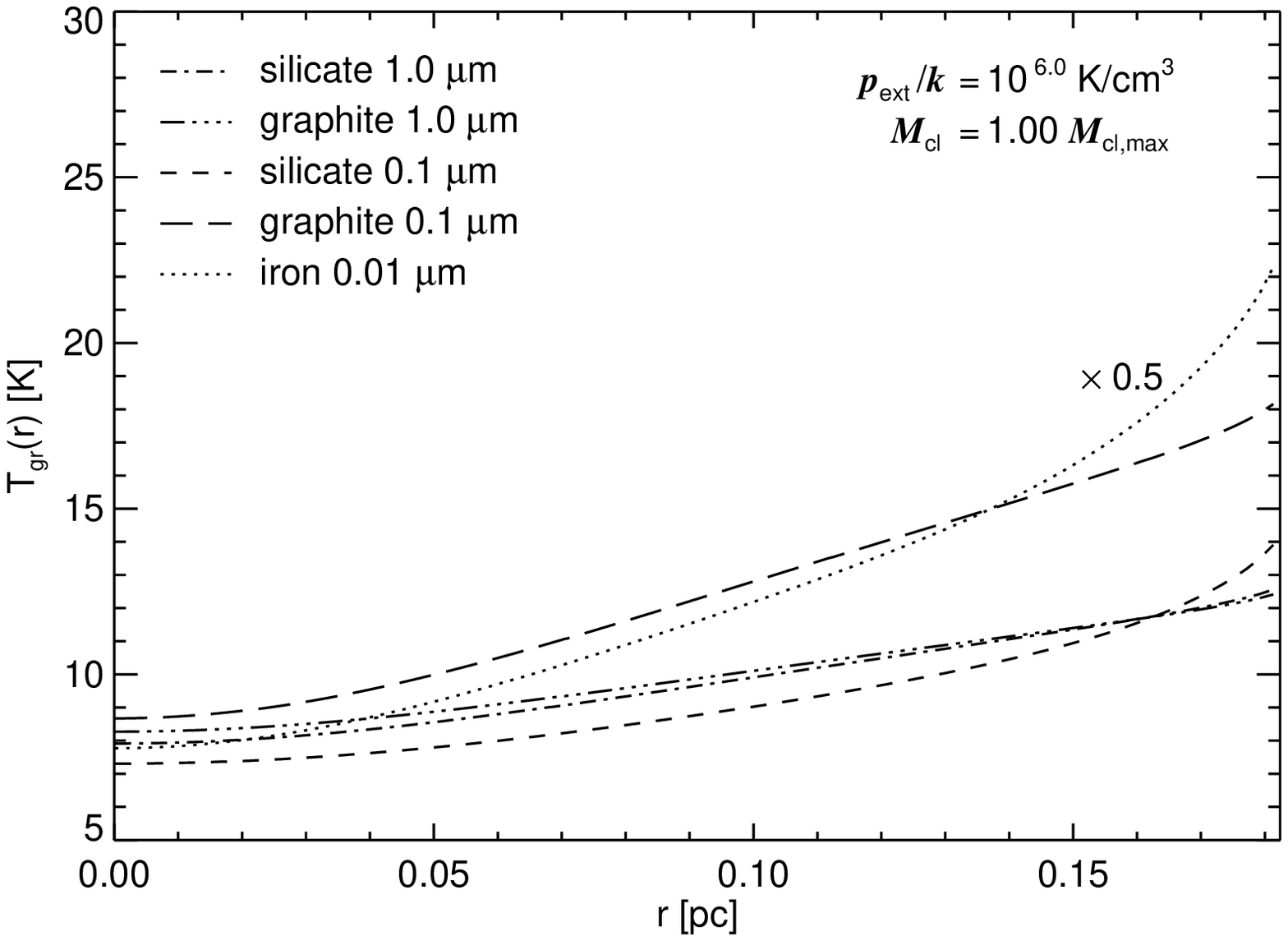}
	\caption{\label{figgraintemp}Grain temperatures of certain grain sizes and compositions inside isothermal clouds as function of distance to the cloud centre.
	 The temperatures of iron grains have been divided by a factor of 2. }
\end{figure*}

The grain temperature distribution inside the clouds determines the spectral shape of the thermal re-emission spectrum. A first indication of likely grain temperatures is given by consideration of the radial variation of the heating rate. From the results presented above we would expect that the grain temperature in isothermal clouds which are embedded by a gas with $p_{\rm ext}/k=2\cdot 10^4$ is not very different to the temperatures of grains which are heated directly by the ISRF, unless the clouds are close to collapse.

In detail, the grain temperature is a function of the grain size and the grain composition as seen in Fig.~\ref{figgraintemp}. Here we have chosen grains of sufficient size so that they do not show a strong temperature fluctuation and can well be described by a single temperature. For iron we have chosen a size of $0.01~\mu{\rm m}$. Because small iron grains do not cool efficiently their temperatures are generally much larger than for other grain types. We have divided their temperature by a factor of two to allow a direct comparison on the plot. One should also be aware that the temperature of iron grains is a strong function of grain size. If the grains are heated by the ISRF the temperature decreases rapidly towards larger sizes, shows a minimum at $0.25~\mu{\rm m}$ and increases slightly to even larger sizes \citep{Fischera2004}. Therefore, the temperature of large iron grains is not necessarily higher than grains of other materials but similar sizes.

The decrease in grain temperature towards the cloud centre reflects firstly the decrease of grain heating. But we also see in Fig.~\ref{figgraintemp} that the temperatures of small grains decrease more rapidly than the larger grains. The strongest radial variation of temperature is seen for the small iron grains. This effect is caused by the fact that the absorption behaviour decreases for wavelengths typically larger than the grain size (or for wavelengths $\lambda > a/2\pi $). Small grains are efficiently heated by UV radiation. Due to the high extinction at those wavelengths the UV-radiation is strongly reduced which leads to a strong decrease of the grain heating of small grains. Large grains are less effected as they are also strongly heated by the optical light.

In case of high optical depths both the UV-radiation and the optical radiation is strongly absorbed at the outskirts of the clouds, so the grains in the central region become predominantly heated by infrared photons. As the wavelengths dependence of the absorption coefficients at those wavelengths are similar for all grains (the absorption coefficient decreases for wavelengths larger than the grain size proportional to $1/\lambda^2$) the grain temperatures become equal. However, in all cases the small iron grains show significantly higher temperatures.

We note that for all stable clouds embedded by a medium with $p_{\rm ext}/k=2\cdot 10^4~{\rm K/cm^3}$ the grain temperatures lie above 10~K. Colder temperatures can only be produced in collapsing clouds if the assumed grain properties are valid for all critical mass fractions.

\subsection{The SEDs of Isothermal Clouds}

\label{section_sed}

\begin{figure*}
	\includegraphics[width=0.49\hsize]{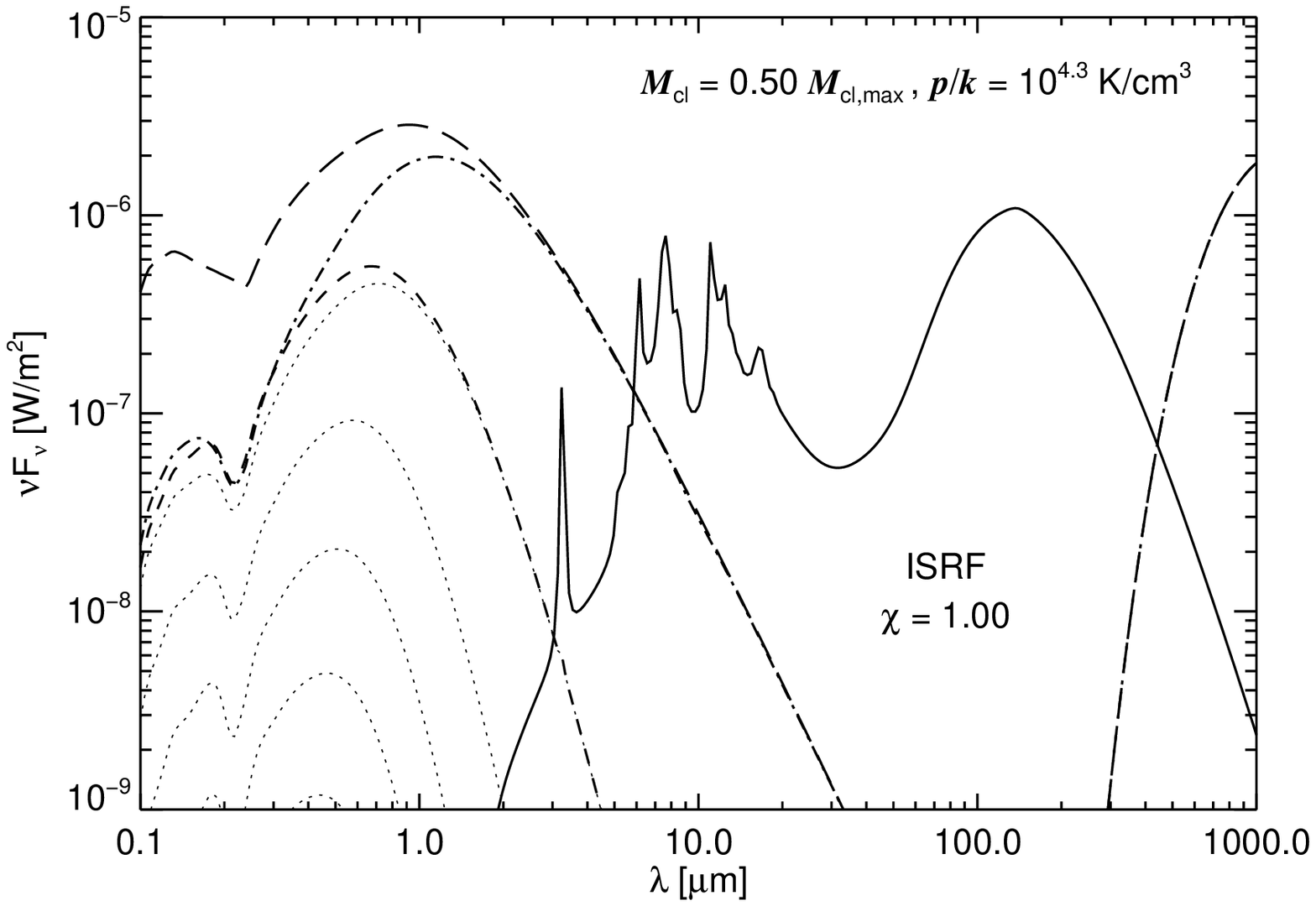}
	\hfill
	\includegraphics[width=0.49\hsize]{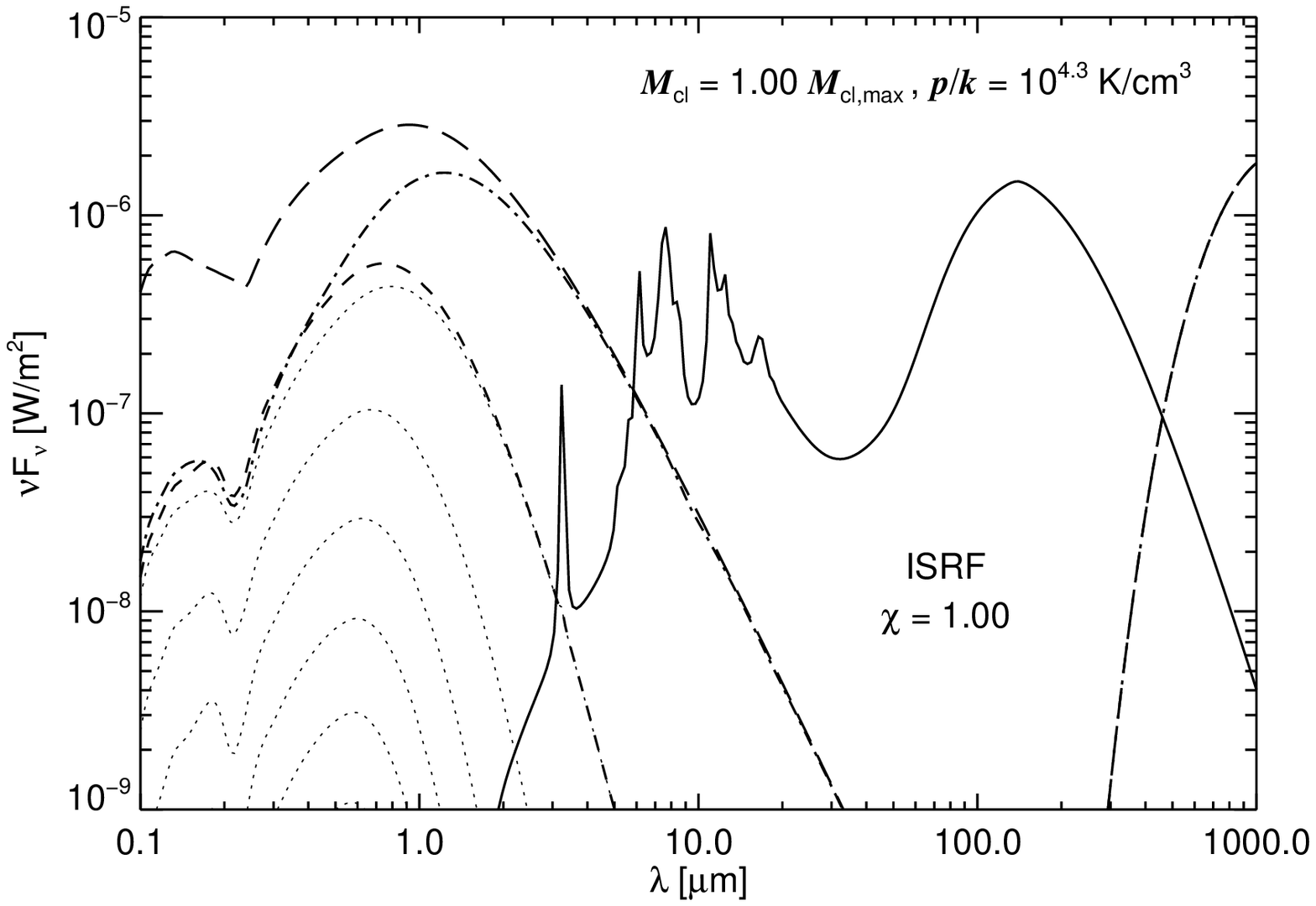}\\

	\includegraphics[width=0.49\hsize]{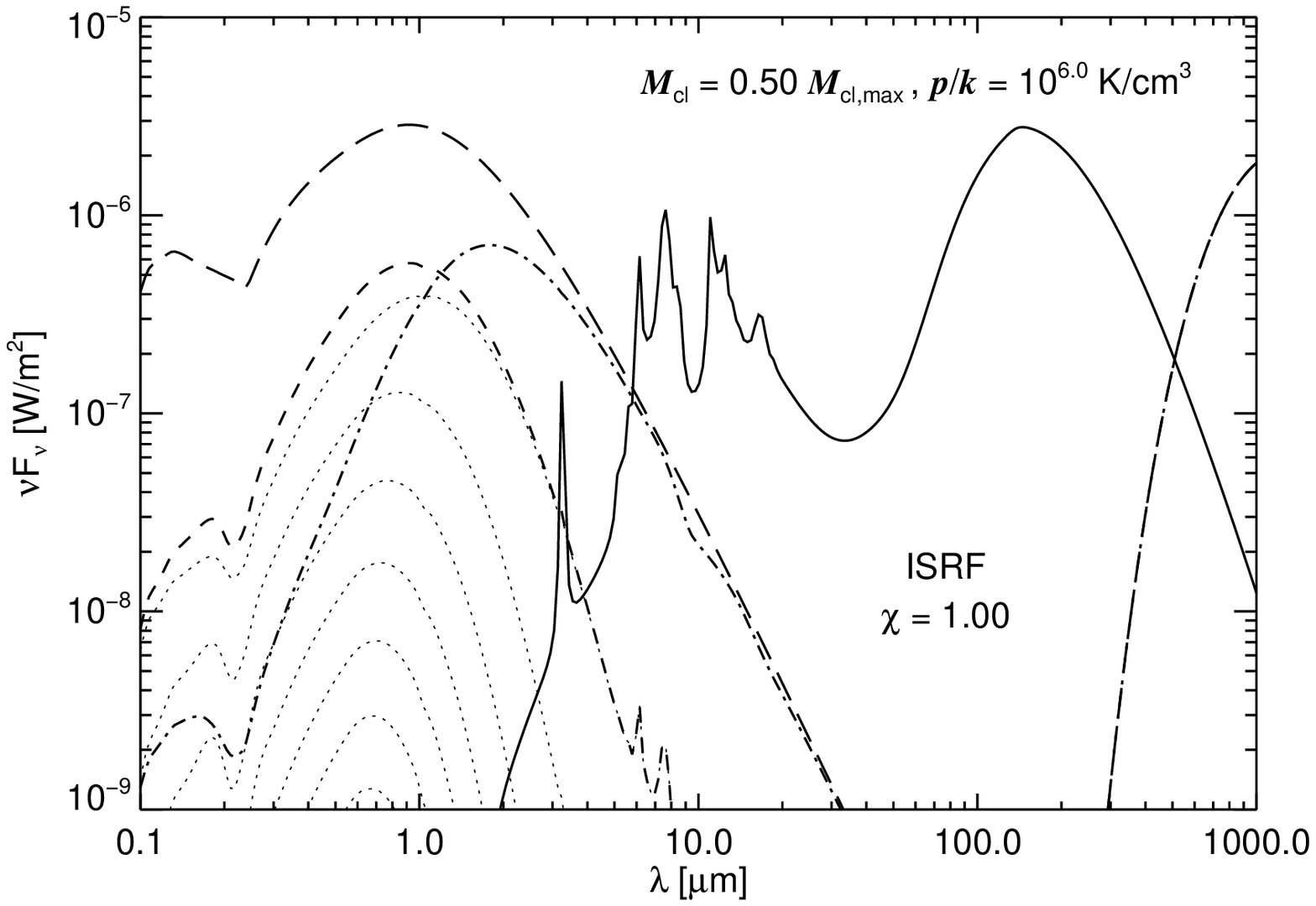}
	\hfill
	\includegraphics[width=0.49\hsize]{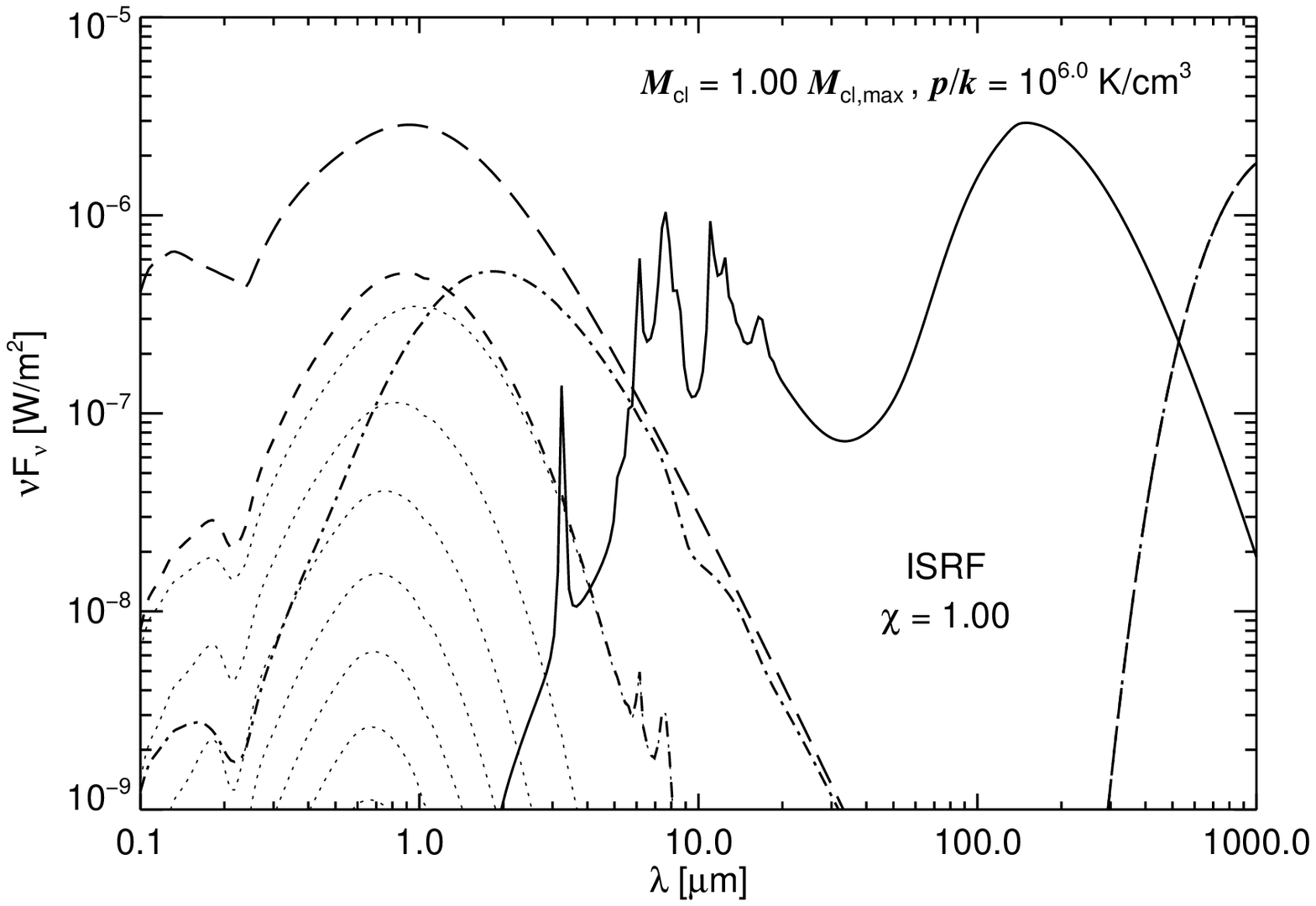}

	\caption{\label{figsed}Spectral energy distribution of isothermal clouds with $f=0.5$ and $f=1.0$ heated by the ISRF appropriate to the local neighbourhood ($\chi=1$). The outer pressure $p_{\rm ext}/k$ is taken to be $2\cdot 10^4$ and $10^6~{\rm K/cm^3}$, respectively. The ISRF entering and heating the cloud is shown as long-dashed curve. The emission at long wavelengths is the cosmic background radiation. The attenuated external emission leaving the cloud is shown as dashed-dotted curve. Shown as short dashed line is the scattered radiation. This is the sum of all the radiation caused by photons scattered $s$ times (dotted curves). The upper dotted curve is the radiation of photons scattered only once, the dotted curve below the radiation of all photons scattered twice and so on. The thermal re-emission spectrum is the solid line.}
\end{figure*}

The SED from isothermal clouds 
heated by the ISRF appropriate to the solar neighbourhood is shown in Fig.~\ref{figsed}.
In the NIR the thermal re-emission spectrum is dominated by the PAH-emission, at wavelengths longer than $\approx 30~\mu{\rm m}$ by emission from dust grains. Because the clouds are more optically thick in higher pressure regions more optical and UV-light is transformed to thermal emission in such regions. 

The scattered light makes an important contribution to the SED in the optical and UV. If we consider clouds with high critical mass fraction embedded in a pressure region $p_{\rm ext}/k = 2\cdot 10^4 ~{\rm K/cm^3}$ roughly half of the flux at wavelengths shorter than $0.5~\mu{\rm m}$ is caused by scattered photons. As will be shown in Sect.~\ref{sectprofsca} the contribution of the scattered light to the total flux varies strongly with the line of sight.  In higher pressure regions scattered light makes the dominant contribution at wavelengths shorter than $\approx 1~\mu{\rm m}$. In addition multiple scattering events become increasingly important.

\subsubsection{The Effect of the Outer Pressure}

\begin{figure*}
	\includegraphics[width=0.49\hsize]{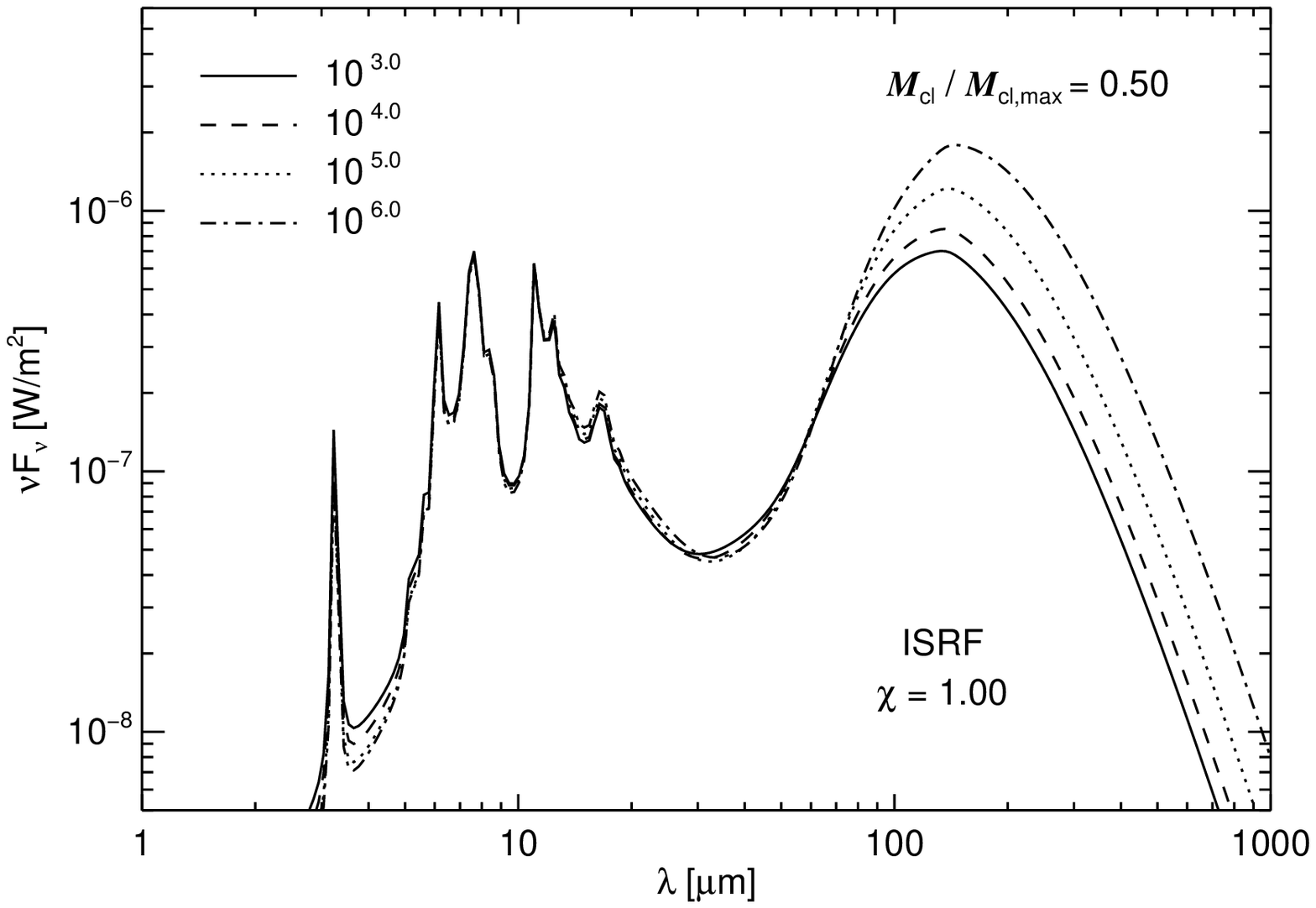}
	\hfill
	\includegraphics[width=0.49\hsize]{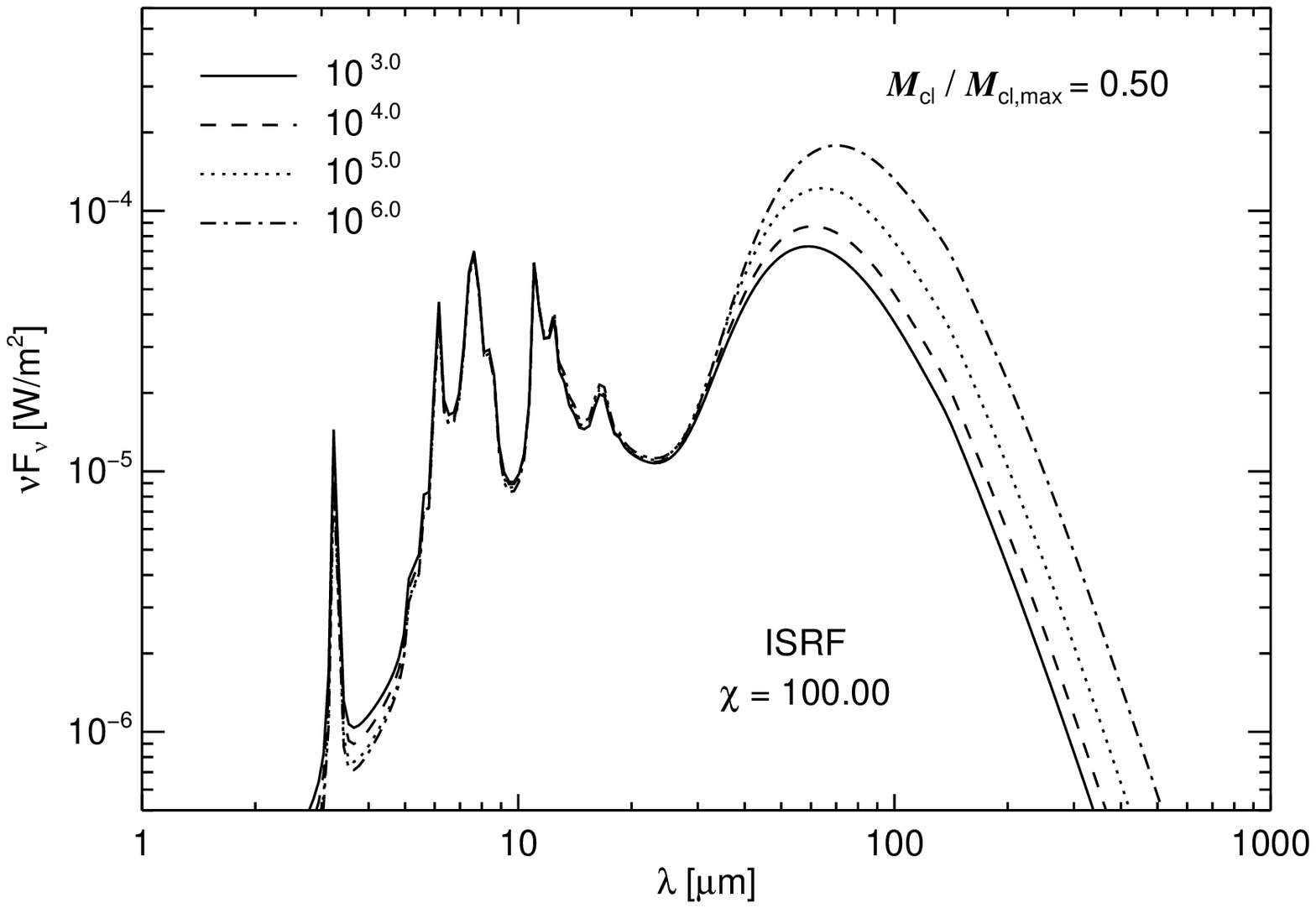}

	\caption{\label{figsedpressure} 
	\emph{Left panel:} The effect of the outer pressure on the SED's of isothermal clouds heated by the solar vicinity interstellar radiation field ($\chi=1.0$). The critical mass fraction is fixed to $M_{\rm cl}=0.5 M_{\rm cl,max}$. The pressure is taken to be $p_{\rm ext}/k=10^3$, $10^4$, $10^5$ and $10^6~\rm{\rm K/cm^3}$. The spectra are normalised to the emission at $11.3~\mu{\rm m}$ in respect to the spectrum for $p_{\rm ext}/k=10^4~{\rm K/cm^3}$. For comparison \emph{right panel}: The same clouds are heated by a local radiation field which is 100 times higher than the local radiation field ($\chi=100$).}
\end{figure*}
\begin{figure*}
	\includegraphics[width = 0.49\hsize]{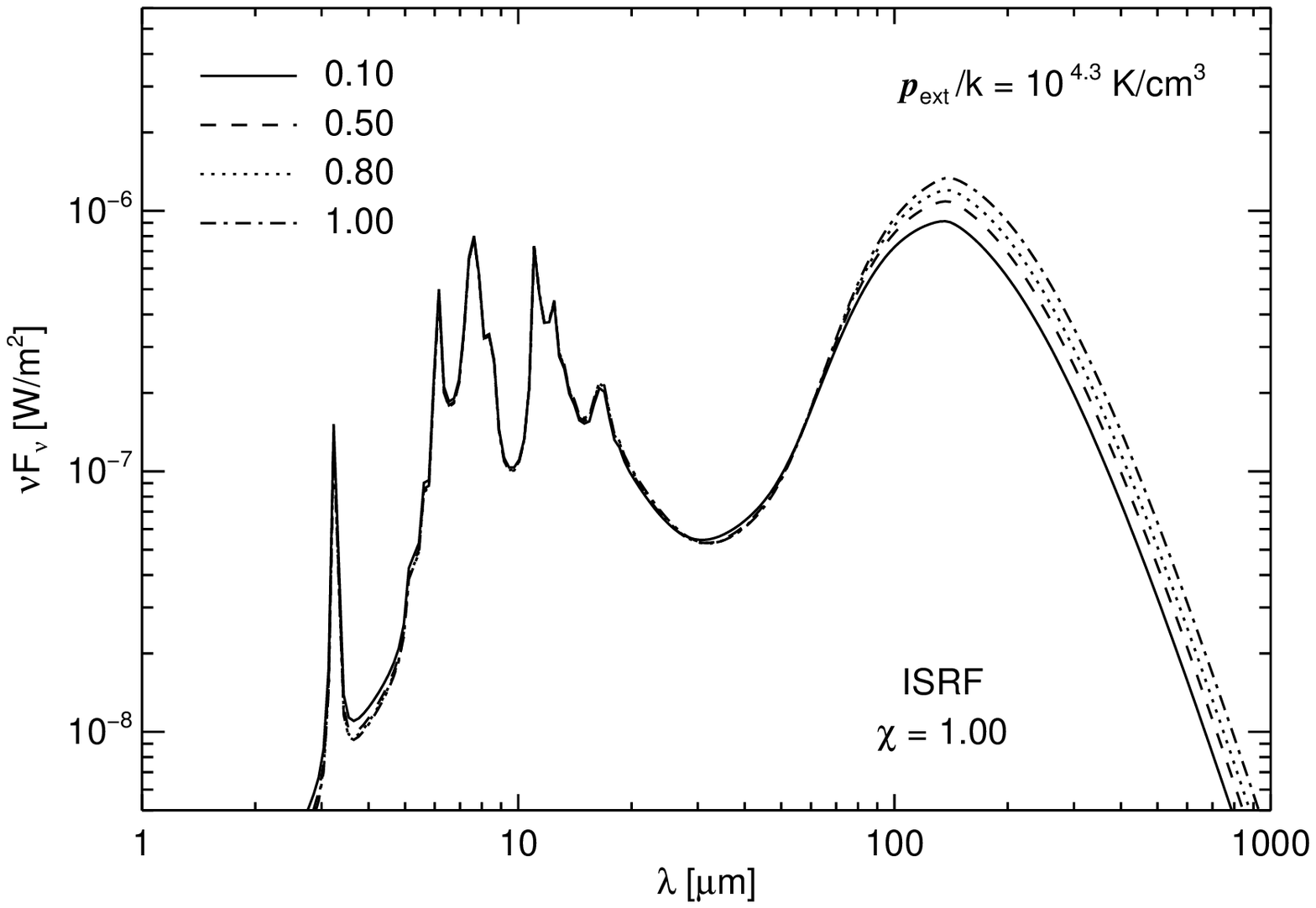}
	\hfill
	\includegraphics[width = 0.49\hsize]{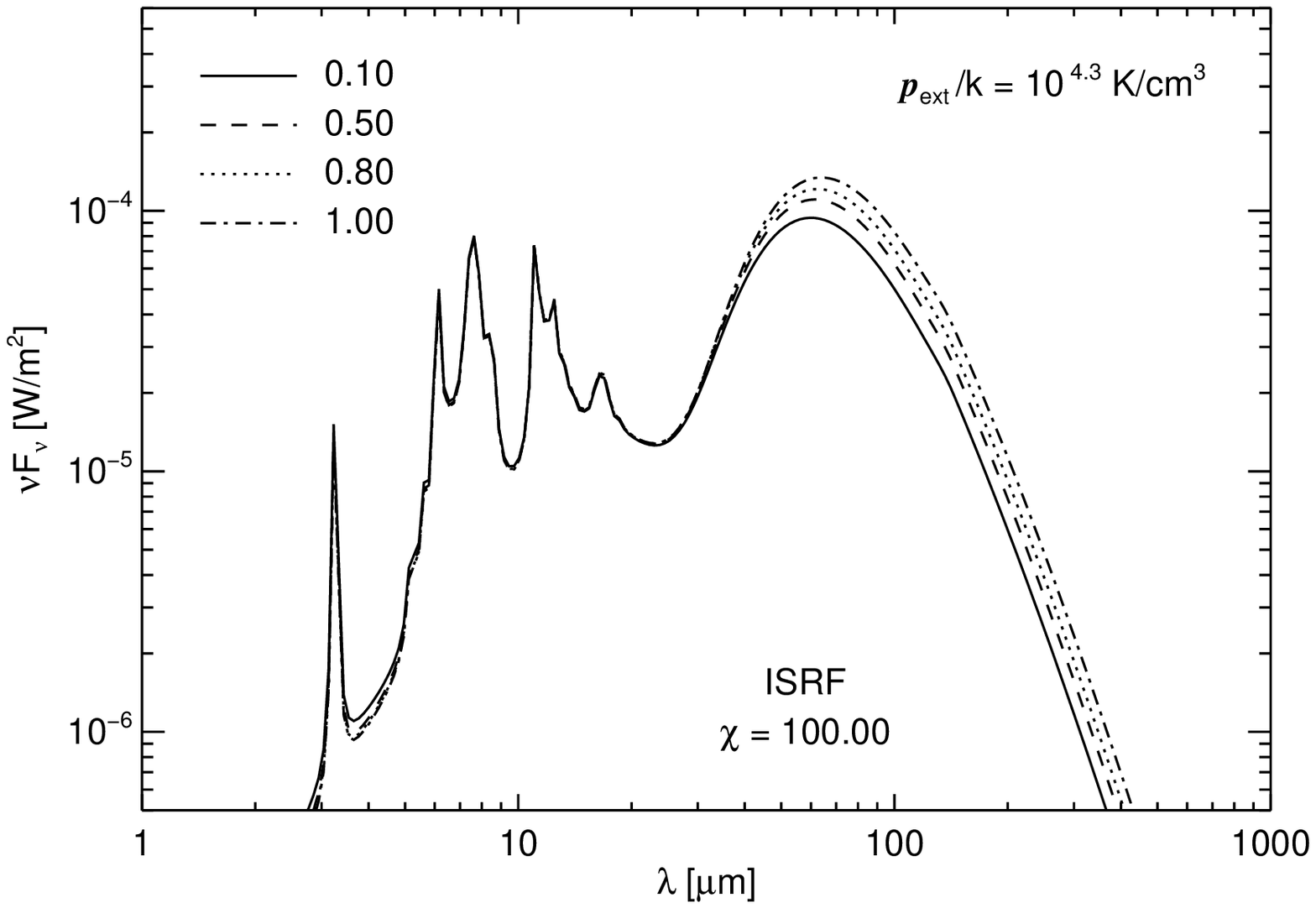}
	\caption{\label{figsedmass}Effect of the critical mass fraction on the SED of isothermal clouds. The outer pressure is taken to be $p_{\rm ext}/k=2\cdot 10^4~{\rm K/cm^3}$. The critical mass fraction is varied from 0.1 to 1.0 of the maximum stable cloud mass. The spectra are normalised to the emission at $11.3~\mu{\rm m}$ relative to the spectrum for $f=0.5$.}
\end{figure*}

As shown above, the clouds are more compact and optically thick in high pressure regions. One therefore expects more cold dust emission in high pressure regions if the clouds are heated by the same radiation field. To show the effect of the pressure on the SED we held the critical mass fraction fixed and only modified the outer pressure. The external radiation was chosen to be the ISRF with $\chi=1$ and $\chi=100$. The result is shown in Fig.~\ref{figsedpressure}.

The shape of the PAH-emission spectrum is mostly determined by the molecular heating associated with the absorption of single high energetic photons and is therefore rather insensitive to the emission spectrum. However, as can be seen the spectra become slightly less intense at short wavelengths in higher pressure regions because the energetic photons are strongly absorbed in the interior, resulting in less excitation of energetic molecular modes.

As the PAH molecules are predominantly heated by UV and optical light which is strongly absorbed inside the cloud their emission relative to the dust emission decreases towards more optical thick clouds. It also shows a shift towards longer wavelengths at higher pressures, as expected given the lower grain temperatures in the central cloud regions. Such dark clouds may therefore be important contributors to the sub-mm emission of starburst galaxies, which are characterized by higher pressures in the ISM, and in which the radiation of compact \ion{H}{2} regions would otherwise tend to dominate the warmer dust emission \citep{Dopita05}.

As seen in the figure, for both assumptions of the radiation field we obtain, with the exception of higher grain temperatures and proportional higher emission, quantitatively very similar results in temperature shift and in the ratio between dust and PAH-emission.

\subsubsection{The Effect of the Critical Mass Fraction}

A very similar effect in the dust temperature distribution is caused by increasing the critical mass fraction, as one also would also expect for simple homogeneous clouds. Here, the effect is complicated by the change of the radial density profile. We varied the critical mass fraction from 0.1 to 1.0 assuming an outer pressure $p_{\rm ext}/k=2\cdot 10^4~{\rm K/cm^3}$. The spectra derived for $\chi=1$ and $\chi=100$ are shown in Fig.~\ref{figsedmass}. 

\subsection{The Brightness Profiles}

Here we analyse different aspects on the brightness profiles of isothermal clouds. 
First we present profiles in the optical/NIR regime where the scattered light has a strong contribution to the SED. 
Then we will present brightness profiles of the thermal emission from dust grains and PAH molecules.
To provide an angular scale the clouds have been placed at a distance $D=1~{\rm kpc}$ from the observer.

\subsubsection{The Scattered Emission}

\label{sectprofsca}

\begin{figure*}[htb]
	\hspace{0.44cm}
	\includegraphics[width=0.42\hsize]{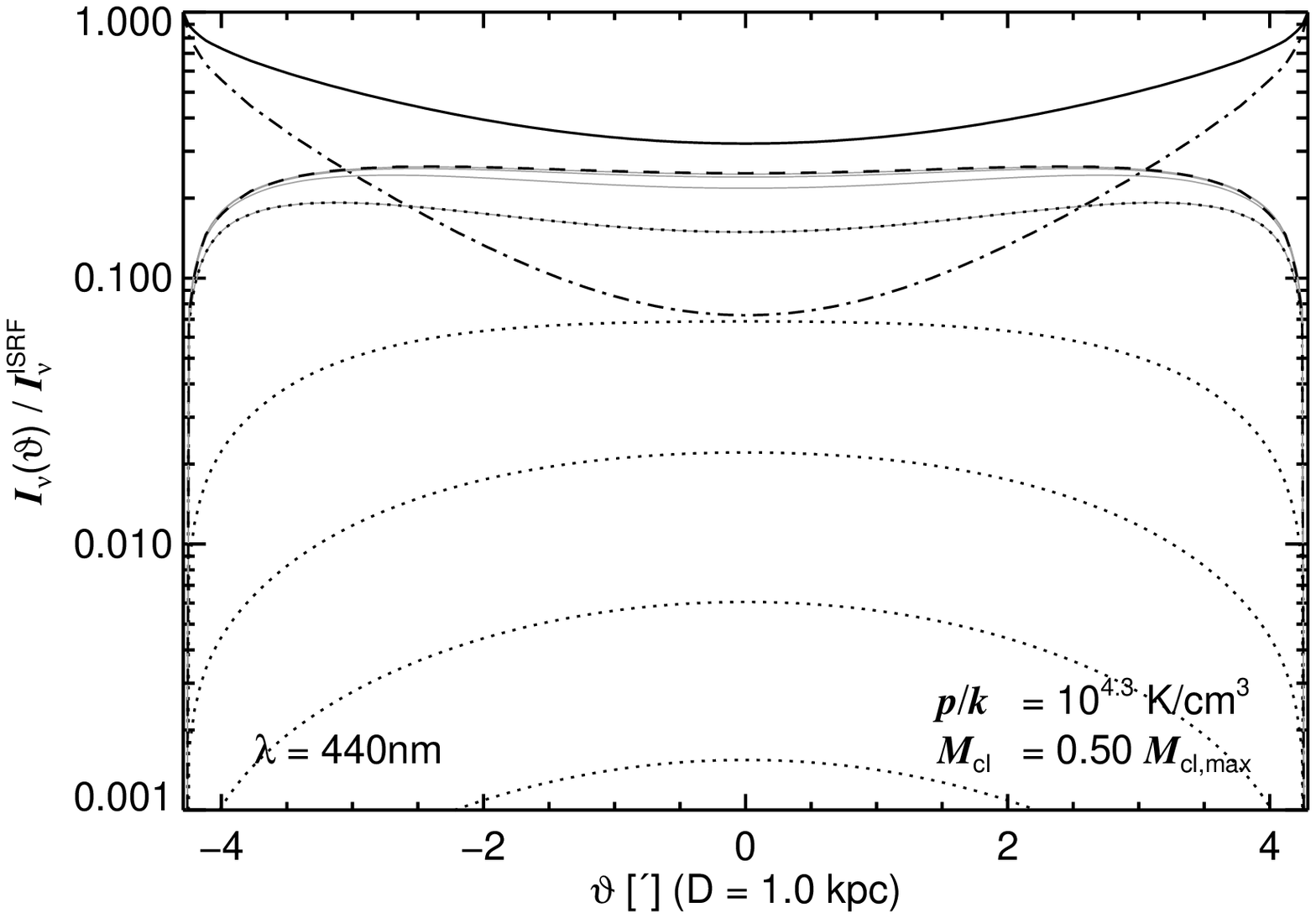}
	\hfill
	\includegraphics[width=0.42\hsize]{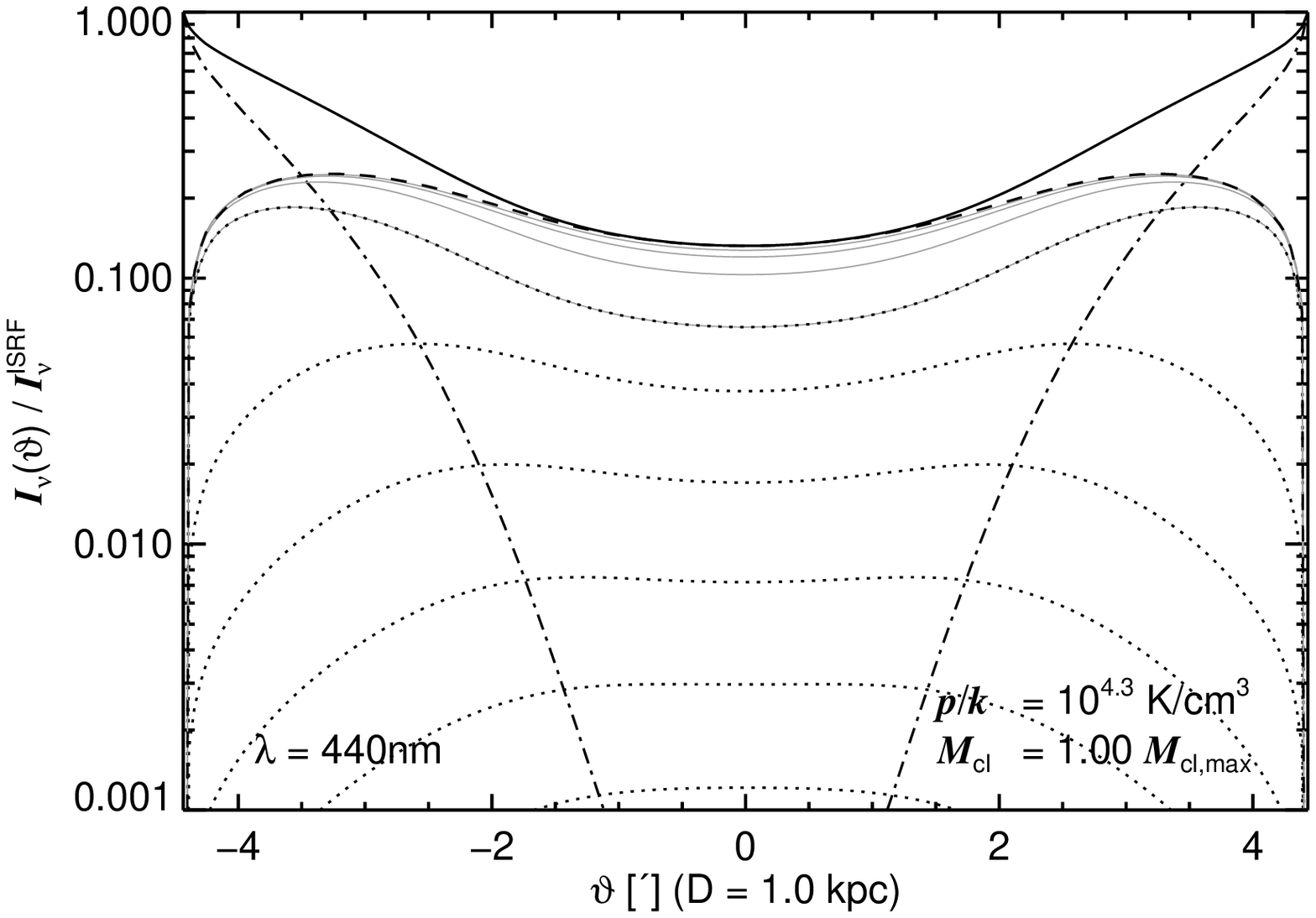}
	\hspace{0.5cm}

	\vspace{0.1cm}
	\hspace{0.44cm}
	\includegraphics[width=0.42\hsize]{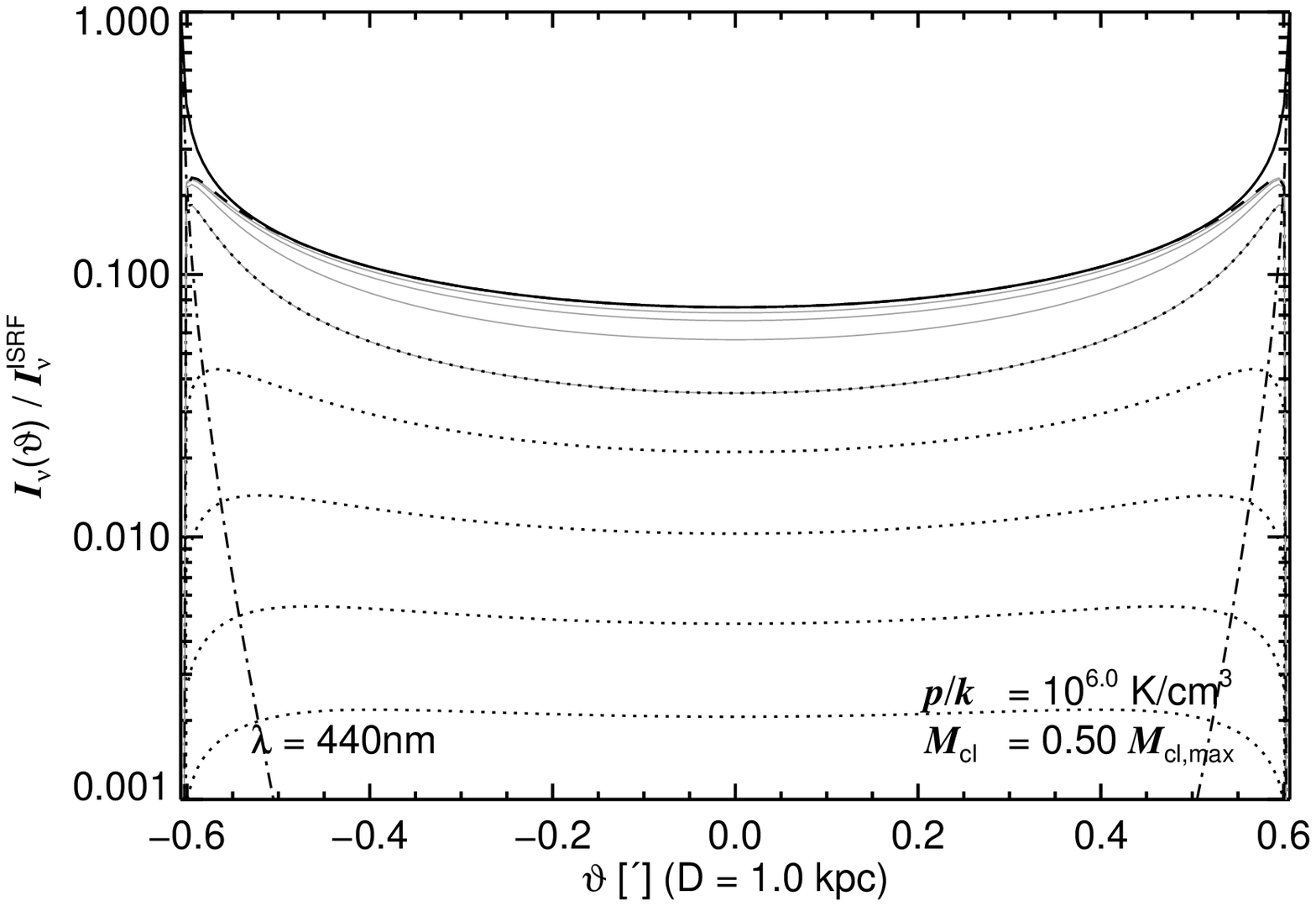}
	\hfill
	\includegraphics[width=0.42\hsize]{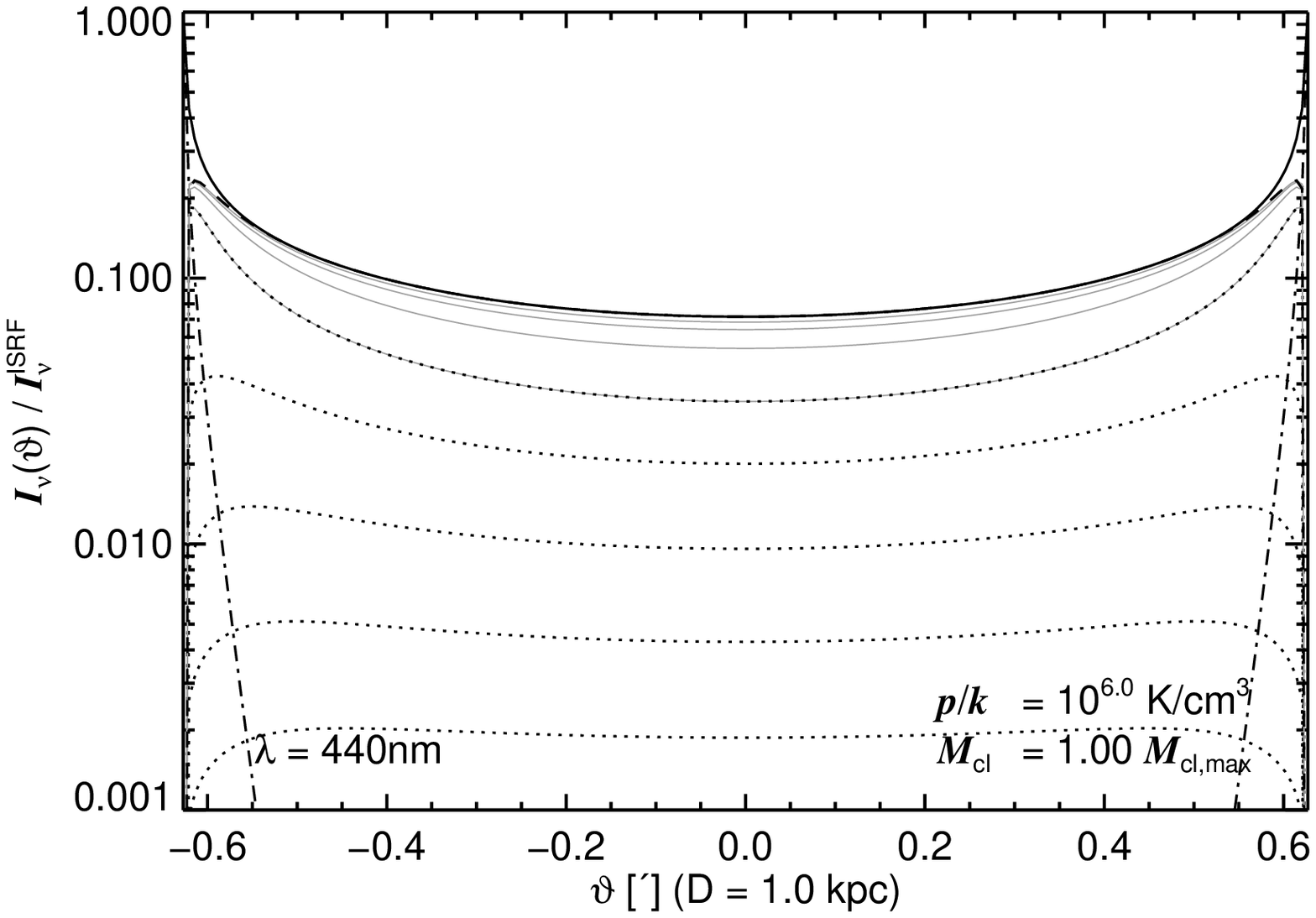}
	\hspace{0.5cm}

	\vspace{0.1cm}
	\hspace{0.44cm}
	\includegraphics[width=0.42\hsize]{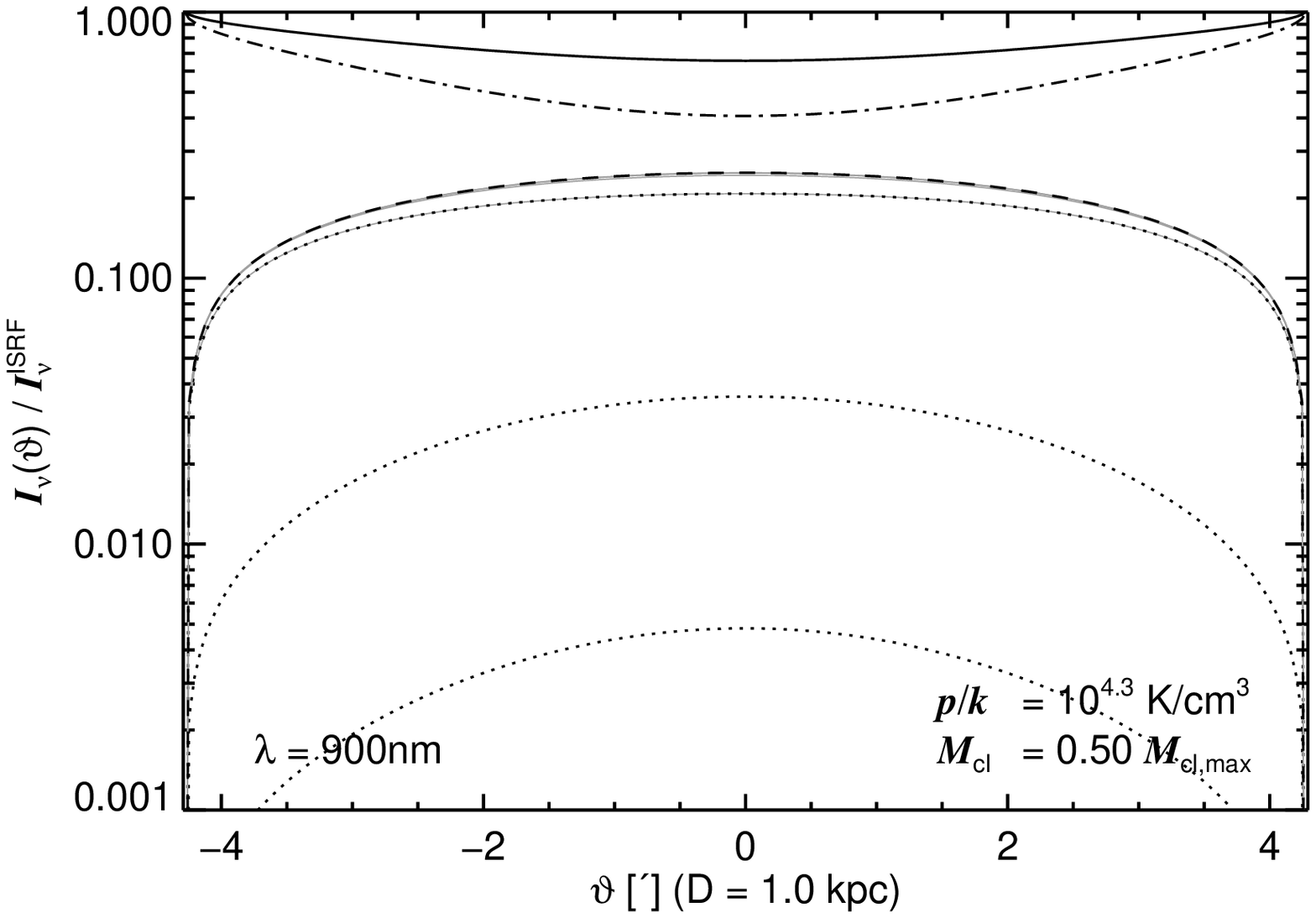}
	\hfill
	\includegraphics[width=0.42\hsize]{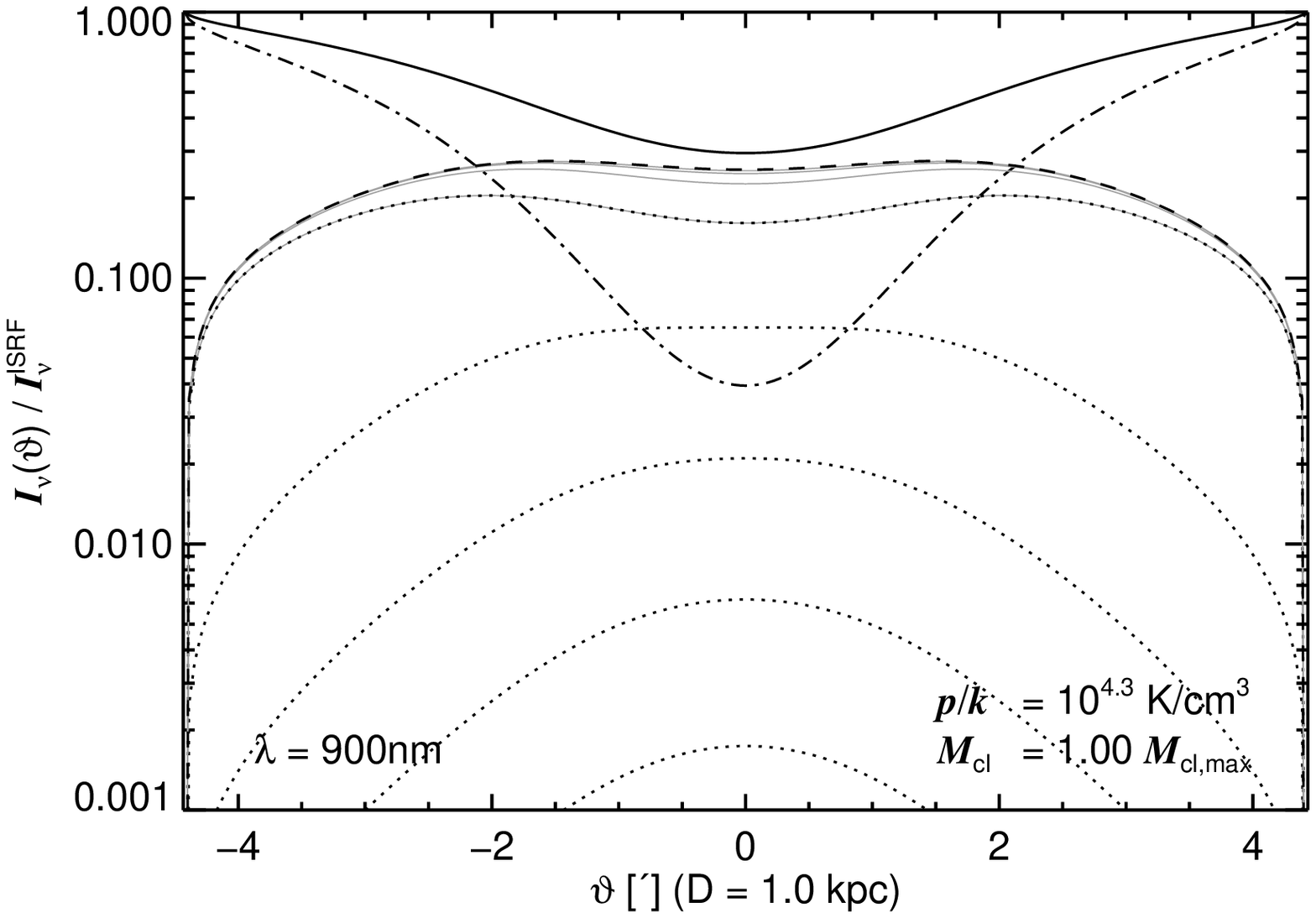}
	\hspace{0.5cm}

	\vspace{0.1cm}
	\hspace{0.44cm}
	\includegraphics[width=0.42\hsize]{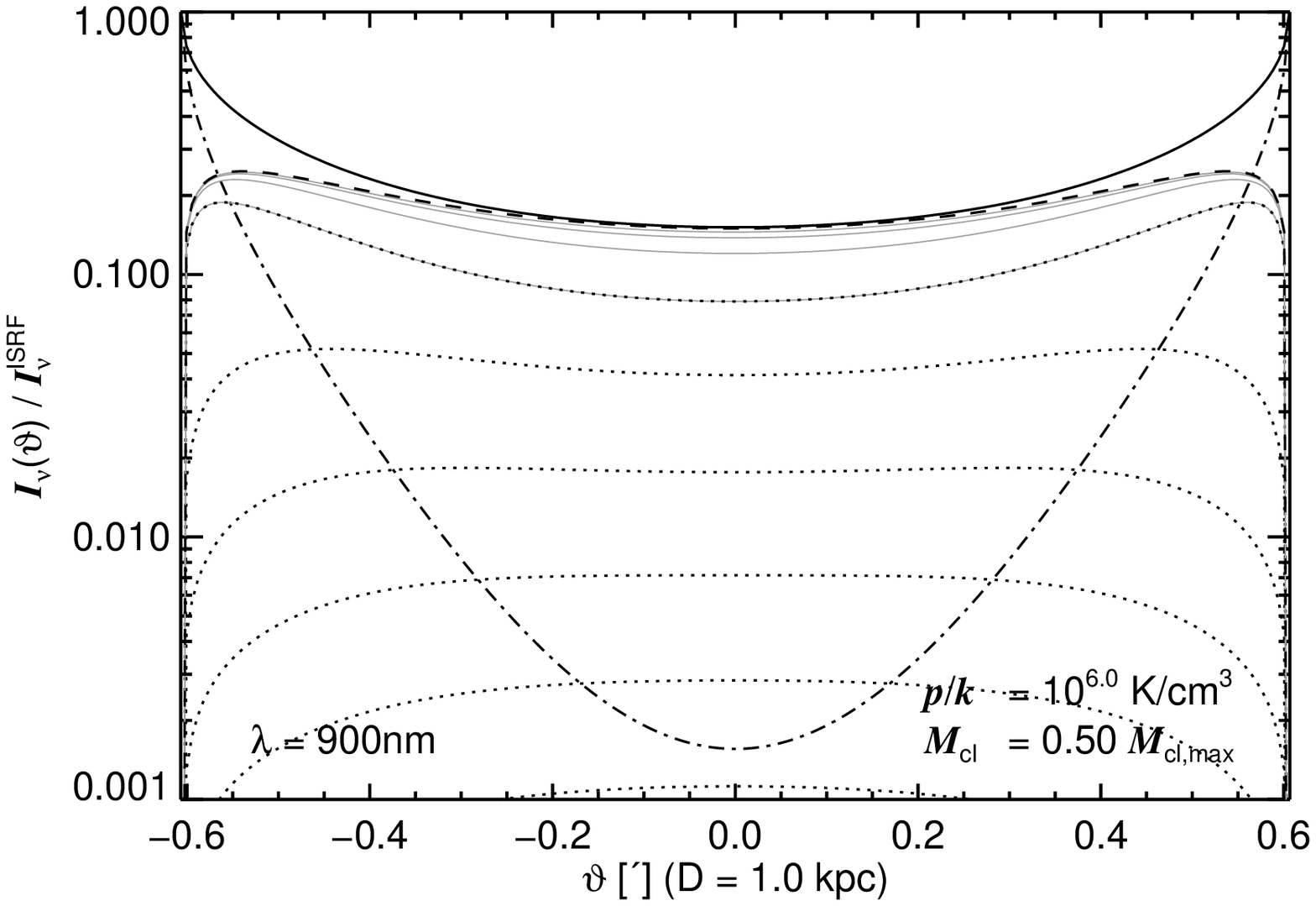}
	\hfill
	\includegraphics[width=0.42\hsize]{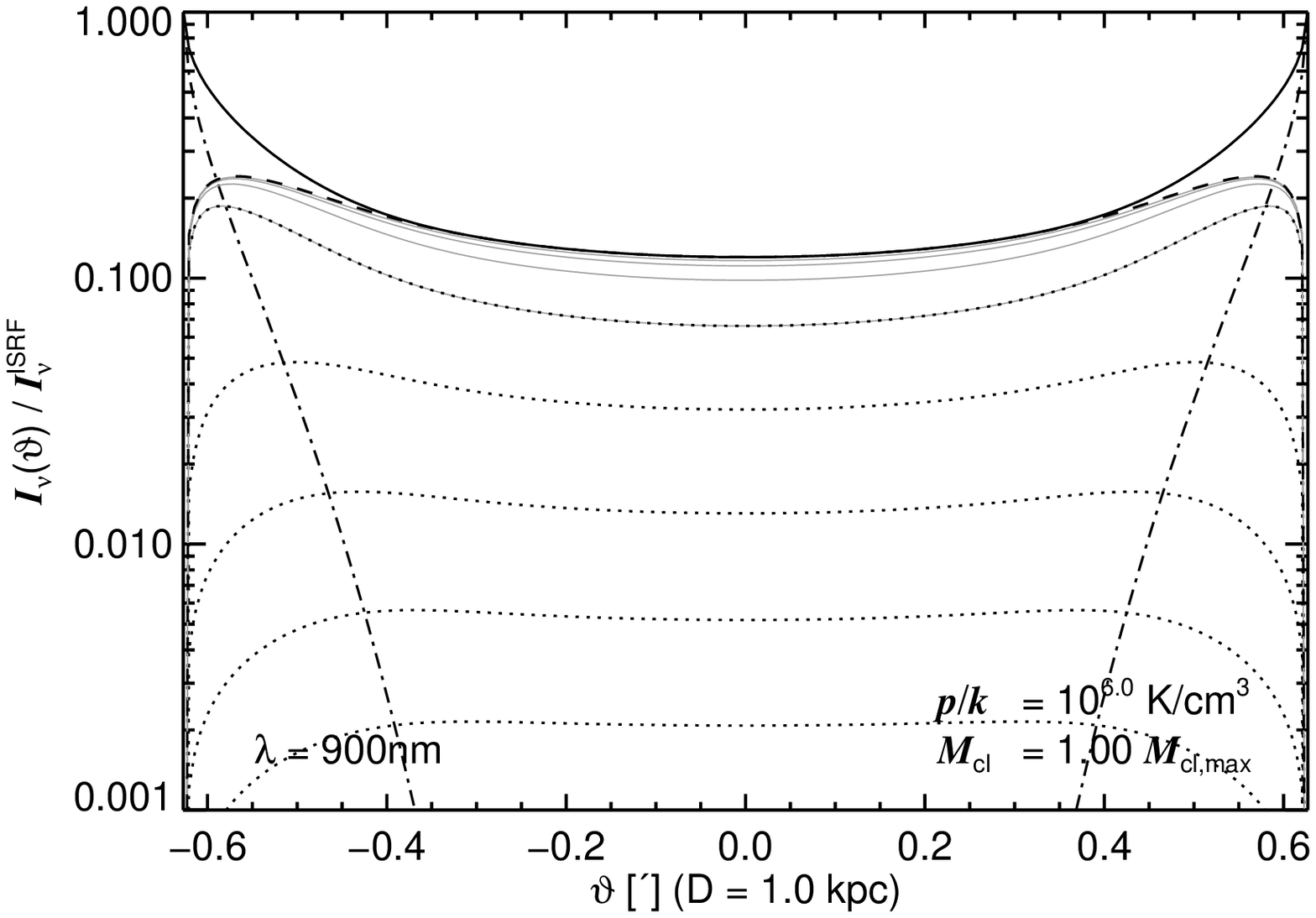}
	\hspace{0.5cm}
	
	\caption{\label{figprofsca}Brightness profiles at 440~nm (upper four panels) and 900~nm (lower four panels) for isothermal clouds located at a distance of 1~kpc heated by the ISRF. The pressure $p_{\rm ext}/k$ is chosen to be $2\cdot 10^4$ and $10^6~{\rm K/cm^3}$ while the mass fraction $f$ is assumed to be either $0.5$ or $1.0$.
Shown are the brightness profiles of the total flux (solid line), the scattered flux (dashed line), and the attenuated external flux (dashed-dotted line). The dotted lines are brightness profiles of photons scattered $s$-times before they escape from the dust cloud. The brightness profiles corresponding to the photons scattered at maximum $s$ times is shown for comparison as grey lines.}
\end{figure*}

\begin{figure*}[htp]

	\includegraphics[width=0.49\hsize]{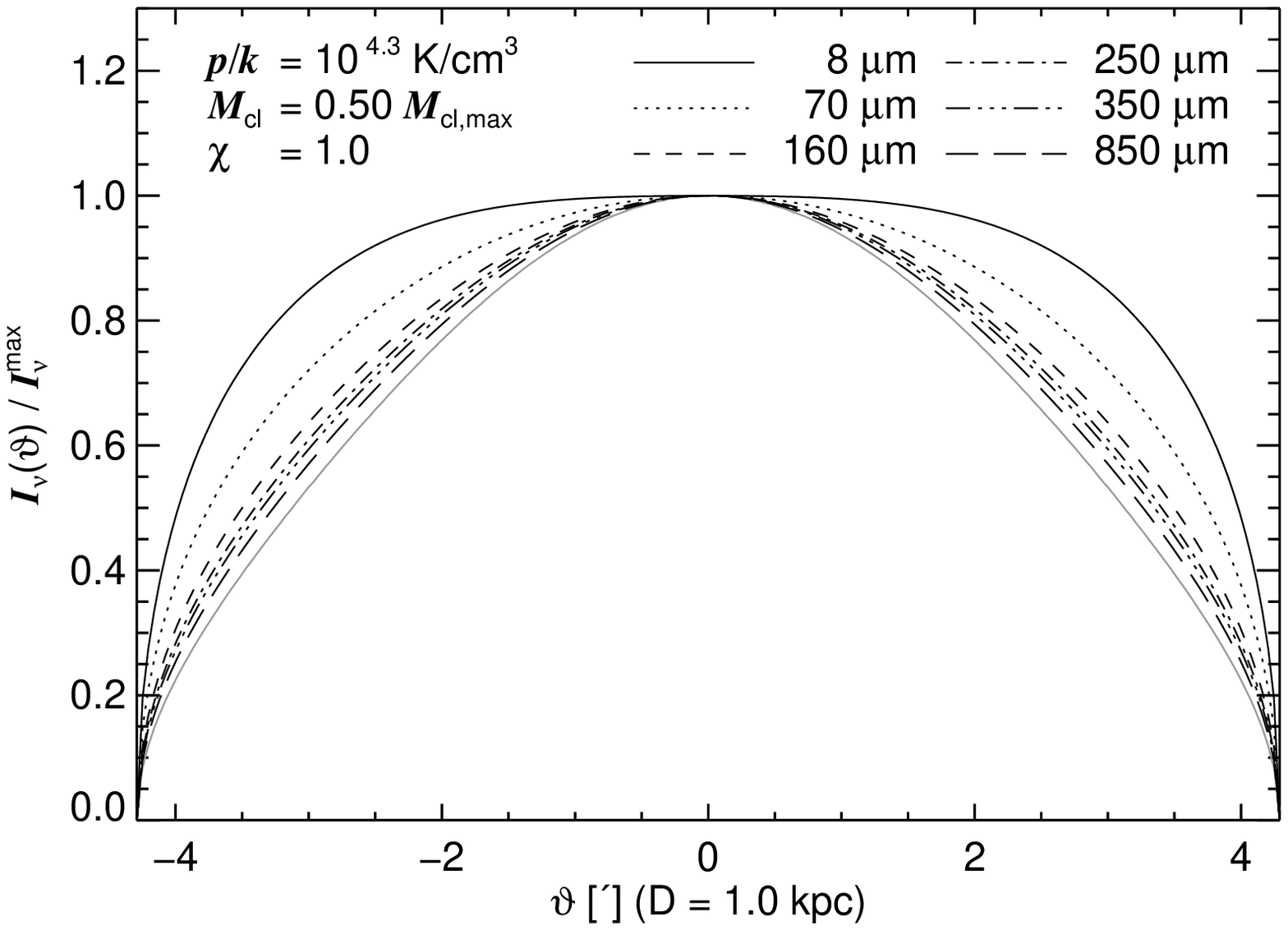}
	\hfill
	\includegraphics[width=0.49\hsize]{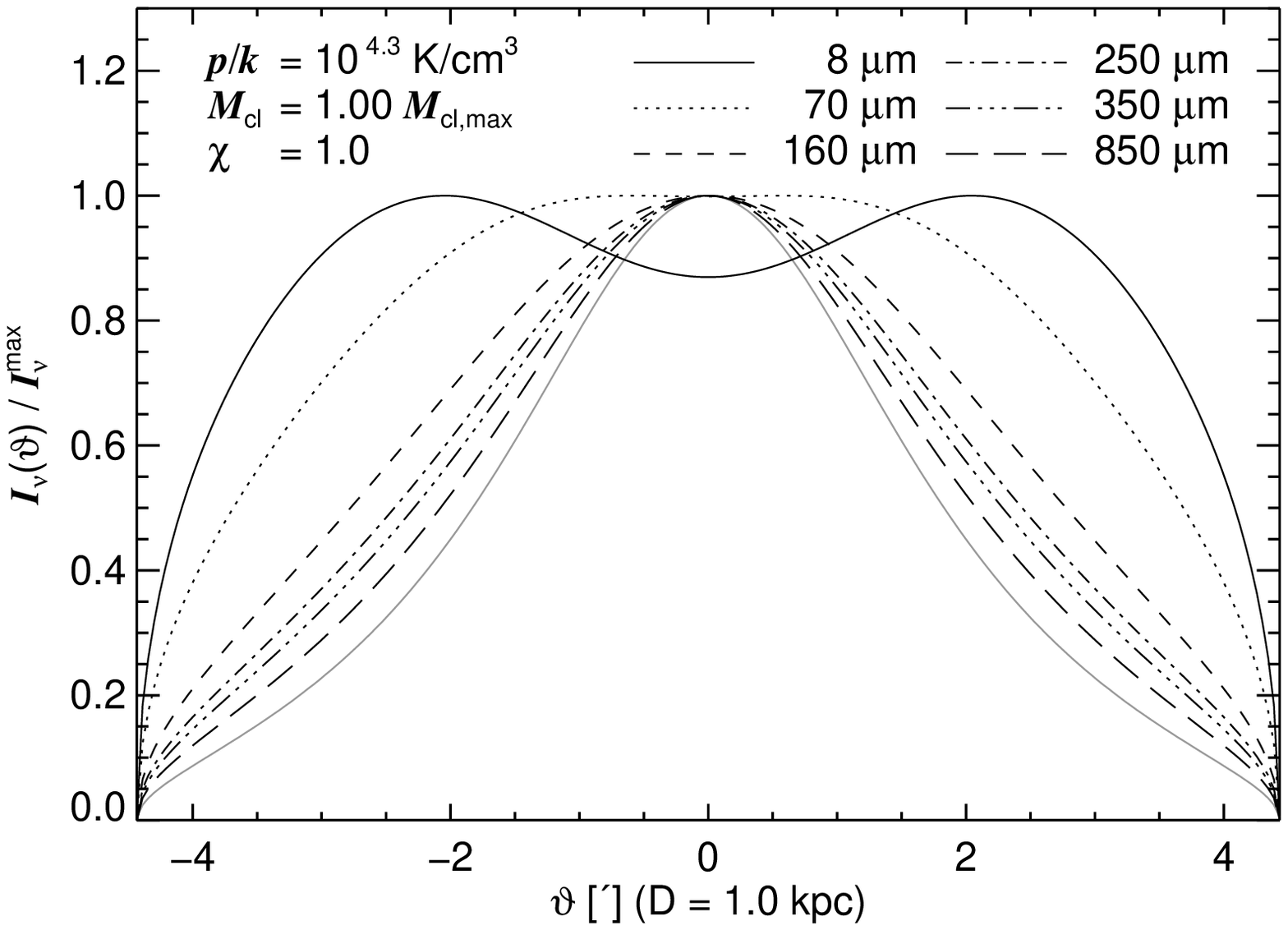}
	\hspace{0.5cm}

	\vspace{0.1cm}

	\includegraphics[width=0.49\hsize]{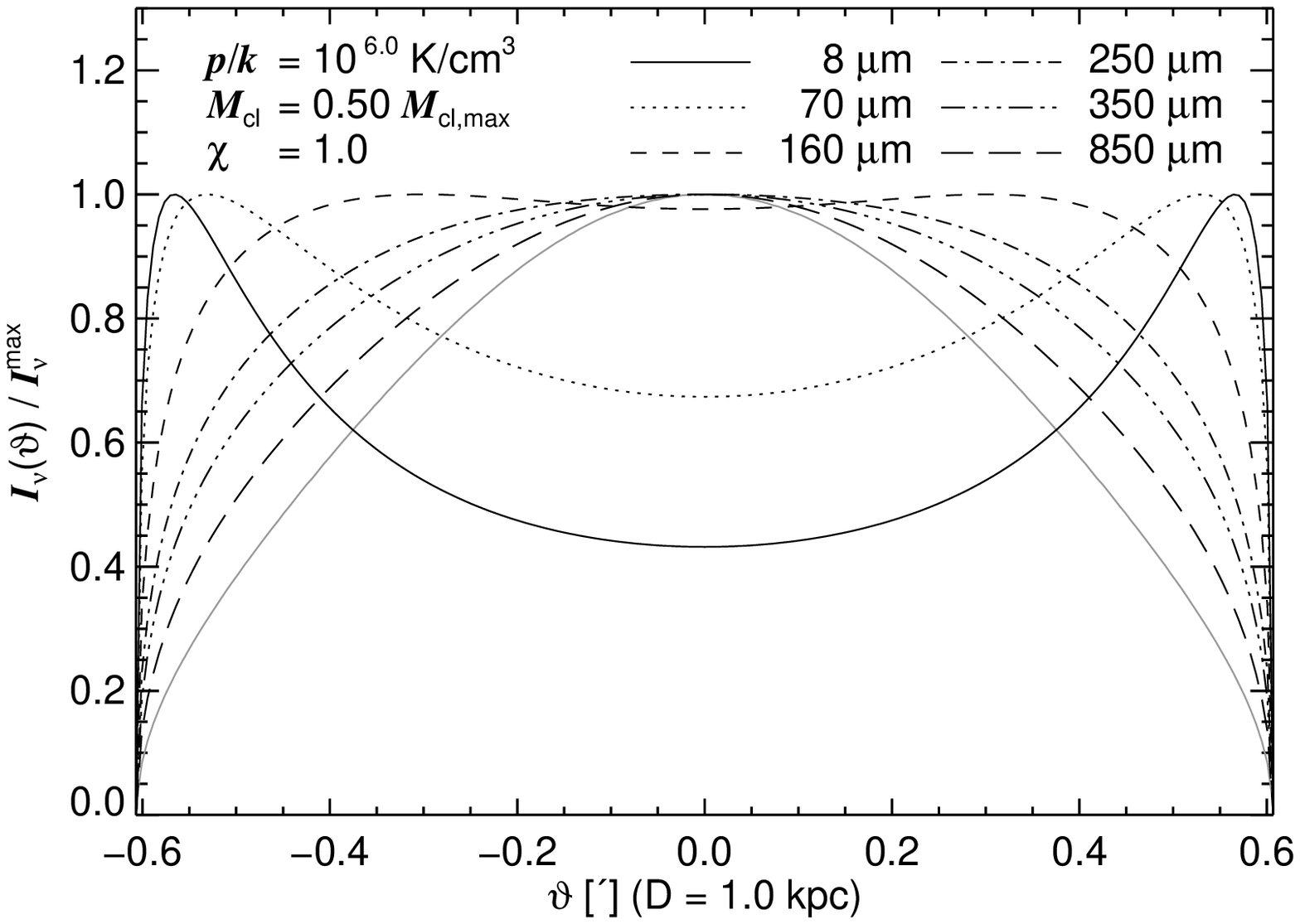}
	\hfill
	\includegraphics[width=0.49\hsize]{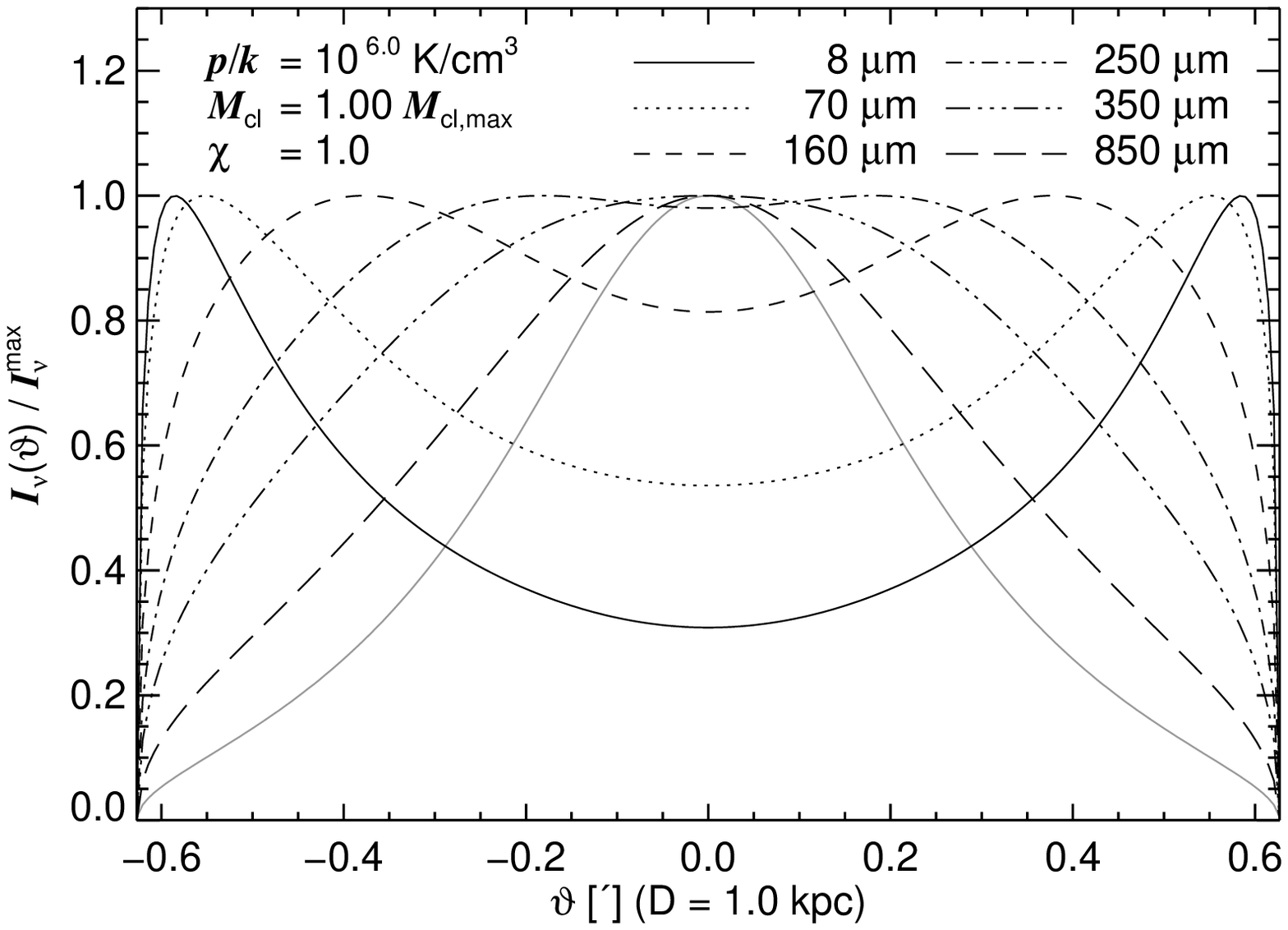}
	\hspace{0.5cm}
	\caption{\label{figirprofiles1} Normalized brightness profiles at several infrared wavelengths of isothermal clouds, which are located at a distance of 1~kpc and heated by the local ISRF ($\chi=1$). The profiles are compared with profiles of the column density (grey lines). The outer pressure is assumed to be $p_{\rm ext}/k=2\cdot 10^4~{\rm K/cm^3}$ (upper panels) and $p_{\rm ext}/k=10^6~{\rm K/cm^3}$ (lower panels). The mass fraction is taken to be 0.5 and 1.0 (left and right panels)} 
\end{figure*}
	
\begin{figure*}[htb]
	\includegraphics[width=0.49\hsize]{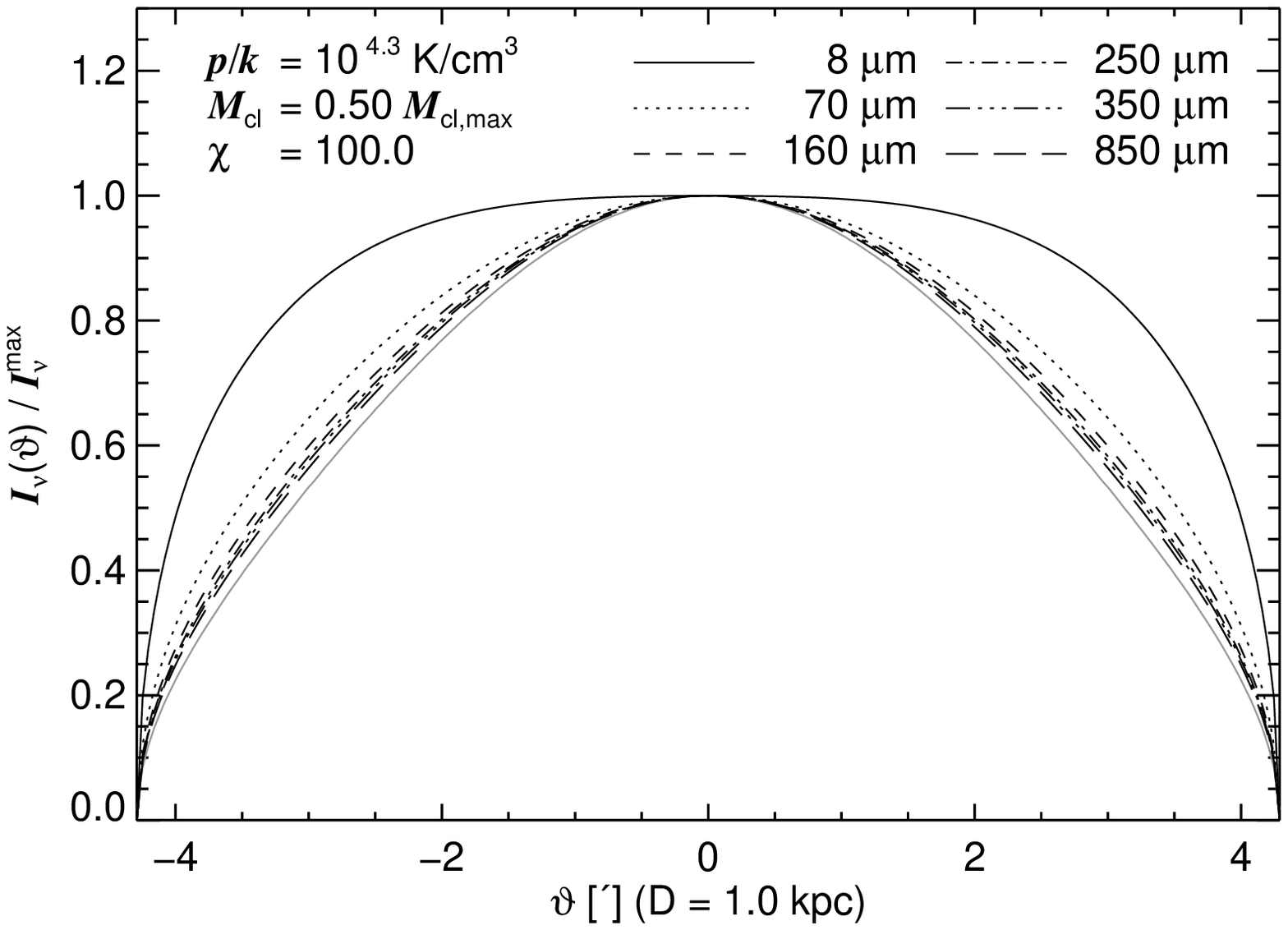}
	\hfill
	\includegraphics[width=0.49\hsize]{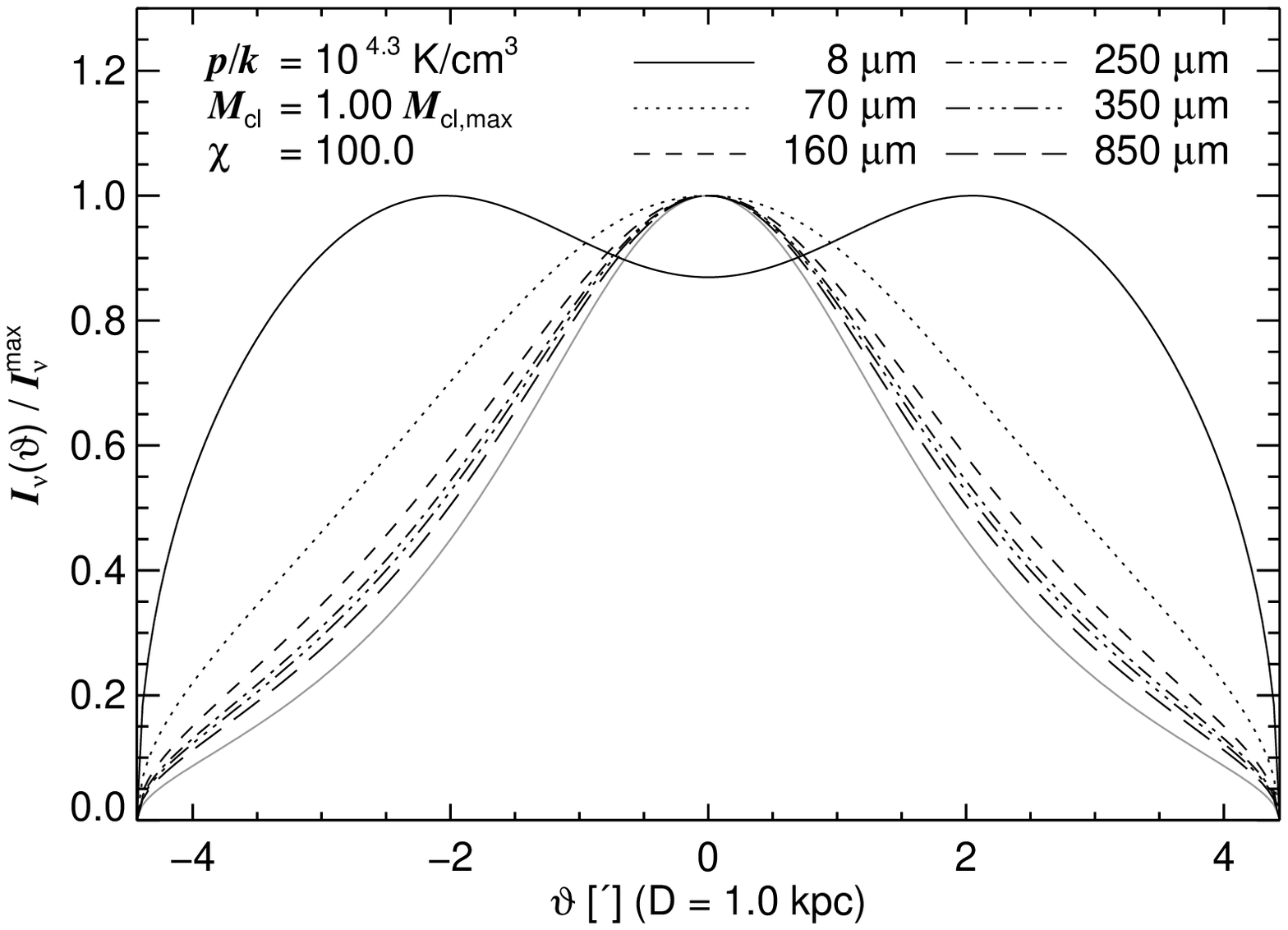}
	\hspace{0.5cm}

	\vspace{0.1cm}

	\includegraphics[width=0.49\hsize]{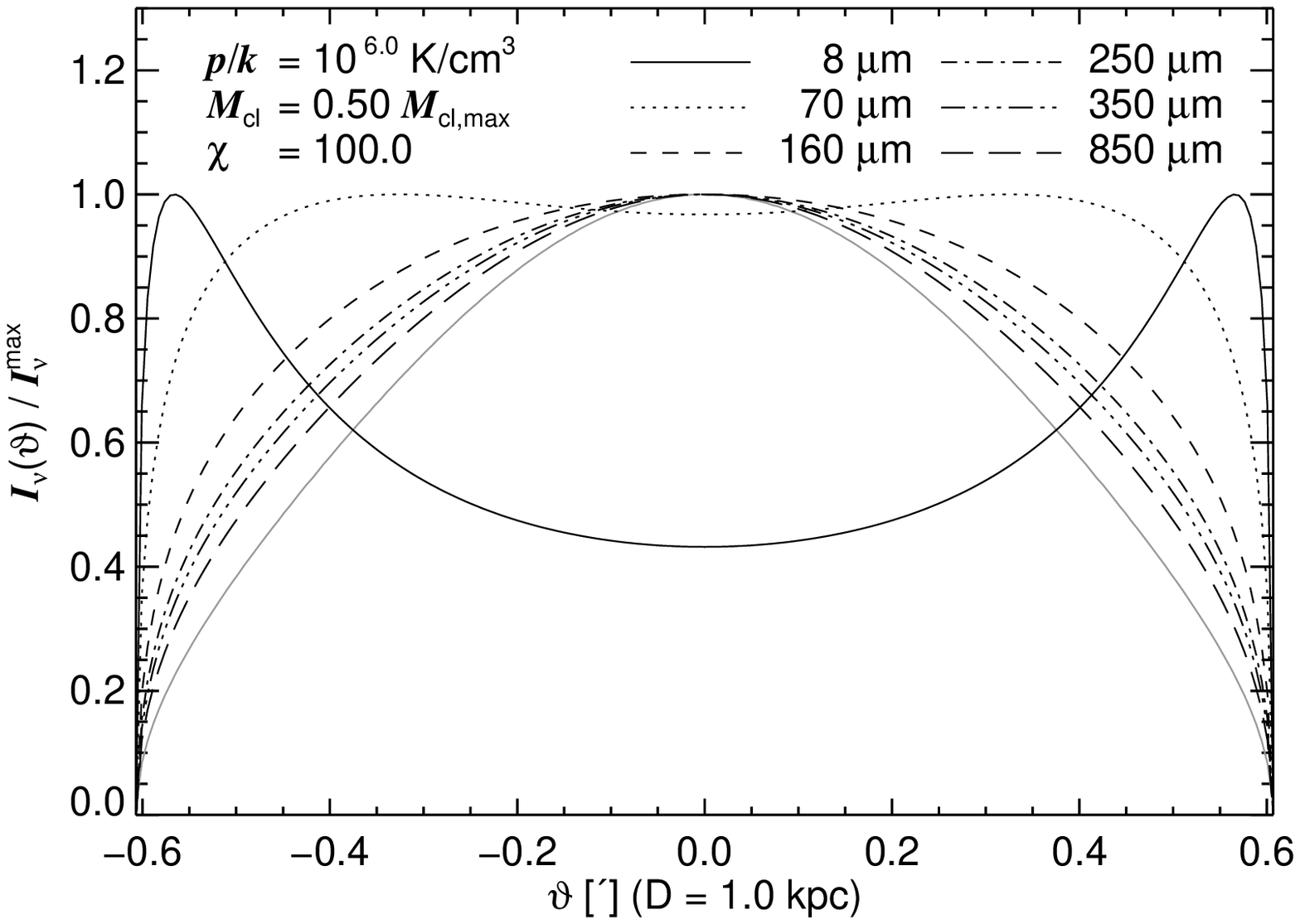}
	\hfill
	\includegraphics[width=0.49\hsize]{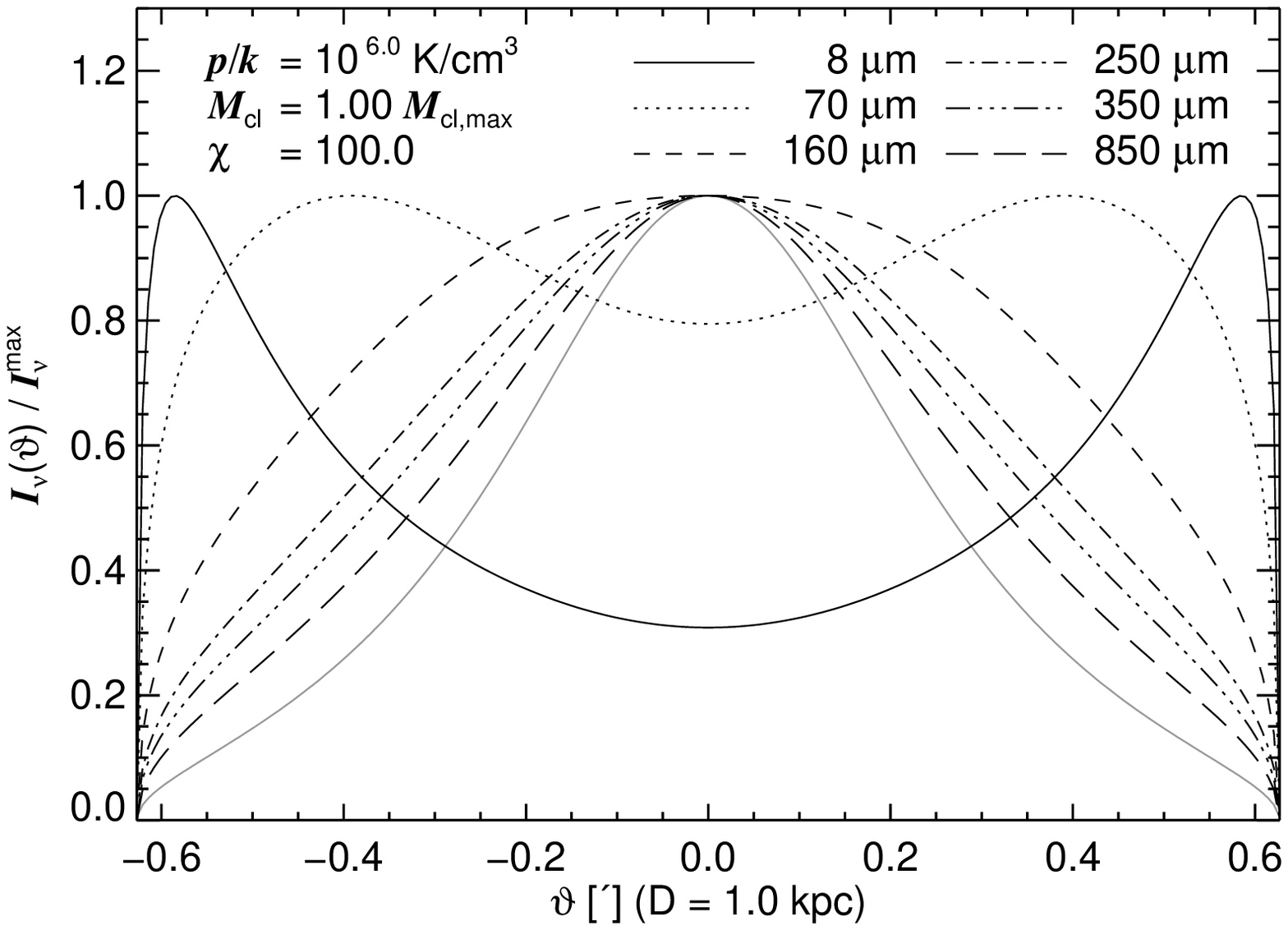}
	\hspace{0.5cm}
	
	\caption{\label{figirprofiles2} Same as Fig.~\ref{figirprofiles1} but for a 100 times stronger 
	radiation field.}
\end{figure*}

For comparison we have chosen wavelengths at 440 and 900~nm typical for the B and the I-filter. 
For highly opaque clouds the scattered emission seen by the observer arises predominantly from the outskirts and is therefore not very sensitive to the actual density structure. We see this in Fig.~\ref{figprofsca} for the considered clouds embedded in a pressure medium of $p_{\rm ext}/k=10^6~{\rm K/cm^3}$ where the profiles at 440~nm are almost identical. 
At longer wavelength the same clouds appear less optically thick so that the profiles show some dependence on the mass fraction $f$.  We see that the corresponding brightness profiles at 900~nm show qualitatively a similar shape as at 440~nm but the emission is less concentrated towards the cloud edges. While the maximum flux is similar strong at $f=0.5$ and $f=1$ the brightness towards the cloud centre decreases with mass fraction $f$ because of the higher probability that the scattered emission is also absorbed in the cloud centre. 

In the lower pressure region the actual brightness profile depends strongly on the actual density structure of the clouds. While the brightness profile at 440~nm of low massive clouds follow qualitatively the column density profile it developes a clear wing structure towards higher mass fraction which can be considered as a limb brightening effect. At longer wavelength the emission arises more closely from the cloud centre reducing this effect which at 900~nm has almost disappeared. 

Fig.~\ref{figprofsca} also shows the brightness profiles of photons scattered different times in the cloud before those photons escape. We see that the radial profiles become more centrally concentrated the higher the number of scattering events. The strong concentration in flux towards the outer part of the cloud in the single scattering case is therefore reduced when we consider the integrated multiple scattering profile.

\subsubsection{The Thermal emission}

\begin{deluxetable}{cccccccccc}

\tablewidth{22cm}
\rotate
\tablecolumns{9}
	
	\tablecaption{\label{tablefluxes}Theoretical Fluxes of the Thermal Emission Spectrum}
	\tablehead{
		 & \multicolumn{4}{c}{$p/k=2\cdot 10^4~{\rm K/cm^3}$} & \multicolumn{4}{c}{$p/k=1\cdot 10^6~{\rm K/cm^3}$} &\\
		 & \multicolumn{2}{c}{$f=0.5$} & \multicolumn{2}{c}{$f=1.0$} & \multicolumn{2}{c}{$f=0.5$} & \multicolumn{2}{c}{$f=1.0$} &  \\
		\colhead{$\lambda_{\rm ref}$} & \colhead{$\nu I_\nu^{(3)}(0)$\tablenotemark{a}} & \colhead{$\nu F_\nu^{(3)}$} & 
		\colhead{$\nu I_\nu^{(3)}(0)$} & \colhead{$\nu F_\nu^{(3)}$} & \colhead{$\nu I_\nu^{(3)}(0)$} & 
		\colhead{$\nu F_\nu^{(3)}$} & \colhead{$\nu I_\nu^{(3)}(0)$} & \colhead{$\nu F_\nu^{(3)}$} \\
		\colhead{$\mu{\rm m}$} & \colhead{$10^{-7} \rm W/m^{2}/sr$} & 
		\colhead{$10^{-7} \rm W/m^2$} & \colhead{$10^{-7} \rm W/m^{2}/sr$} & 
		\colhead{$10^{-7} \rm W/m^2$} & \colhead{$10^{-7} \rm W/m^{2}/sr$} & 
		\colhead{$10^{-7} \rm W/m^2$} & \colhead{$10^{-7} \rm W/m^{2}/sr$} & 
		\colhead{$10^{-7} \rm W/m^2$}
		}
		\startdata
	& \multicolumn{8}{c}{$\chi=1$}\\
		7.8727\tablenotemark{b} 	& 1.458 & 3.528 & 1.386 & 3.890 & 0.918 & 4.710 & 0.679  & 4.560\\
		23.843\tablenotemark{b} 	& 0.310 & 0.708 & 0.333 & 0.807 & 0.231 & 1.054 & 0.181 & 1.037\\
		72.556\tablenotemark{b} 	& 1.389 & 2.937& 1.686 &  3.415 & 1.097 & 4.367 & 0.886 & 4.170 \\
		156.893\tablenotemark{b} 	& 4.987 & 9.549 & 9.323 & 13.28 & 9.063 & 25.81 & 8.263 & 27.40 \\
		250\tablenotemark{c} 		& 2.280 & 4.217 & 5.141 & 6.360 & 6.261 & 15.08 & 7.594 & 17.97\\
		350\tablenotemark{c} 		& 0.864  & 1.568 & 2.139 & 2.466 & 2.910 & 6.447 & 4.373 & 8.313 \\
		850\tablenotemark{c} 		& 0.028 & 0.049 & 0.079 & 0.082 & 0.127  & 0.249 & 0.280 & 0.371\\

	& \multicolumn{8}{c}{$\chi=100$}\\
		7.8727 	& 146.1 & 353.6 & 138.8 & 389.7 & 91.94 & 471.7 &  68.02 & 456.7\\
		23.843 	& 55.17 & 126.3 & 56.86 & 141.1 & 37.09 & 172.9 &  28.28 & 167.3 \\
		72.556 	& 471.6  & 911.3 & 857.6 & 1250. &  815.0 & 2366. & 727.2  & 2489.\\
		156.893 	& 110.9  & 202.4 & 268.2 & 314.8 & 356.4 & 803.9 & 521.6  & 1021. \\
		250 		& 18.54  & 33.13 & 49.77 & 54.03 & 74.82  & 153.9 & 144.0 & 216.7\\
		350 		& 4.633 & 8.196 & 13.01 & 13.65 & 20.59 & 40.77 & 44.73 & 60.23\\
		850 		& 0.085 & 0.148 & 0.254 & 0.254 &  0.431 & 0.809 & 1.103 & 1.282
		\enddata

	\tablenotetext{a}{Brightness towards the cloud centre.}
	\tablenotetext{b}{Reference wavelengths of IRAC 4 and the three MIPS filters of the Spitzer satellite.}
	\tablenotetext{c}{Fluxes given for narrow band filter.}
\end{deluxetable}

The brightness profiles at IR and FIR wavelengths provide some insights about the radial variation of the radiation field and of the grain temperatures inside the clouds. 


For the purposes of displaying brightness profiles we have chosen six wavelengths covering the range from $5$ to $850~\mu{\rm m}$.
At shorter wavelengths we give the theoretical flux profiles for three different broad band filters of the Spitzer satellite, namely IRAC 4 ($8~{\mu\rm m}$), and MIPS 2 and 3 ($70$ and $160~{\mu\rm m}$). In the diffuse ISM they
characterize the PAH emission, the warm dust emission, and the cold dust emission. 
The three profiles at longer wavelengths, namely at 250, at 350, and at $850~\mu{\rm m}$, are given for narrow band filters. 
The profiles are normalized to the maximum brightness and compared to the profile of the column density. The results for the two assumed strengths of the radiation fields are shown in Fig.~\ref{figirprofiles1} and \ref{figirprofiles2}, respectively.

The brightness $\nu I_{\nu}^{(3)}(0)$ towards the cloud centre and the flux density $\nu F_{\nu}^{(3)}$ at the cloud radius are summarized in table~\ref{tablefluxes}. For completeness we also list the values for the MIPS 1 filter at $24~\mu{\rm m}$. 
We note that in our model the diffuse dust emission at $24~\mu{\rm m}$ is dominated by the PAH-emission. The flux can therefore not being used to identify warm dust emission.



As mentioned the PAH molecules in our model are predominantly heated by the UV/optical radiation which is strongly absorbed inside the cloud. In case of optically thick clouds ($p/k=10^6~{\rm K/cm^3}$) the PAH emission is strongest at the outskirts which causes a limb brightening effect as is seen for clouds in the high pressure medium.  The effect is less prominent for clouds embedded in the lower pressure region: A slight minimum in the brightness profile towards the cloud centre appears only for clouds close to the critical mass. The maximum of the brightness profile is 
furthermore not strongly related to the limb as it is located at a position roughly half of the projected radius. 
As seen in the figure the brightness profiles of the PAH emission are - as a result of its stochastic nature - almost insensitive to the radiation strength. 

The brightness profiles of the thermal emission from dust grains is a result of the decreasing radiation field inside the cloud and the temperature shift of the grains to colder grain temperatures. Naturally they depend on the strength of the external radiation field as a stronger radiation field produces warmer dust emission.

In case that the clouds are illuminated by the mean ISRF the brightness profiles at $70~\mu{\rm m}$ show qualitatively the same shape as the PAH emission although the limb brightening is smaller and the emission peak more closely located towards the cloud centre. 

In the low pressure region the brightness profiles at $170~{\rm nm}$ follow qualitatively more closely the profile of the column density with a prominent maximum towards the cloud centre for critically stable clouds. In the high pressure environment the profile produces a wing like shape because of the cold dust emission from the cloud centre.

In case of a 100 times stronger radiation field the diffuse dust emission has a peak around
$70~\mu{\rm m}$. For the corresponding brightness profiles at $70~\mu{\rm m}$ we obtain 
qualitatively the same shapes as for the lower radiation field at $170~\mu {\rm m}$. 

The brightness profiles at long wavelengths appear always broader in comparison with the profile of the column density. The difference is larger for shorter wavelengths and more prominent for more optically thick clouds. In general the effect decreases for higher radiation fields.

\section{Discussion}

The model we have presented here relates the SED from self-gravitating clouds with the physical properties of the diffuse in-homogeneous ISM. Even though it is based on a simplified structure of the compact clouds it might find application in helping us to understand the SED of whole galaxies. 

We have seen that clouds in the ISM of our galaxy are essentially optically thin unless they are close to the critical mass or are in a state of collapse. The effect will depend on how much dust mass is actually distributed in those clouds. Assuming a mass spectrum ${\rm d}n(M) \propto M^{-2}{\rm d}M$ for the clouds which is not only consistent with simple hierarchical models \citep{Fleck1996} but also for a turbulent medium \citep{Elmegreen2002} and therefore expected for the diffuse ISM most of the dust mass is distributed in optically thin clouds. If correct, the in-homogeneous density structure would have only a small effect on the global SED of our own galaxy or of normal galaxies in general. A simplified model will be presented in the following sub-section to quantify the effect.

The situation might be very different in star-forming regions or star-burst galaxies, where the pressure of the diffuse ISM might be significantly higher \citep{Dopita05} so that the stable clouds will be more compact. In those cases the in-homogeneous structure of the ISM might have an important effect on the SED, provided that the mass in dense self-gravitating clouds is significant.

Disk galaxies should be characterised by a radially-decreasing global pressure profile. 
Self-gravitating clouds in low pressure regions have to be much more massive to become gravitational unstable. They are less compact but they are also less optical thick. Therefore, the scale of both the dense clouds and the  ISM structure should decrease towards the galactic centre. Equally, stable clouds at larger scale height should be more extended and less optical thick in comparison with clouds located in the centre of the thin disc.



\subsection{Effect of the `clumpy' medium on the SED}

To analyse the effect of an in-homogeneous interstellar medium on the global SED we 
considered an idealised model where the dust is located in isothermal clouds of the same size and
where the clouds are heated by the same mean radiation field taken to be the ISRF.
We assumed that the clouds are embedded in a medium with $p_{\rm ext}/k=2\cdot 10^4~{\rm K/cm^3}$ and varyied the mass fraction $f$ from $0.1~\%$ up to $100~\%$ of the critical mass. By doing this we covered a range of different optical depths in $V$ towards the cloud centre from $0.08$ up to $3.57$ so that the cloud with smallest mass fraction can be considered to be optically thin.
We used this spectrum as reference.
The model implies that we have for each cloud size or mass fraction the same total dust mass but only distributed in a different number of clouds.

An interesting question is how much the colour $F_\lambda/F_{11.3\mu{\rm m}}$ might vary because
of the in-homogenous medium which increases both towards longer wavelength and higher mass fraction. The colour $F_{60\mu{\rm m}}/F_{11.3\mu{\rm m}}$ is almost insensitive on the cloud mass showing a variation of less than $4\%{}$. The increase of the colour $F_{100\mu{\rm m}}/F_{11.3\mu{\rm m}}$ on the other hand can be up to $\sim 40\%{}$. At even longer wavelengths the difference in colour can be larger than a factor of two if the clouds are close to gravitational instability.

\begin{figure}
	\includegraphics[width = \hsize]{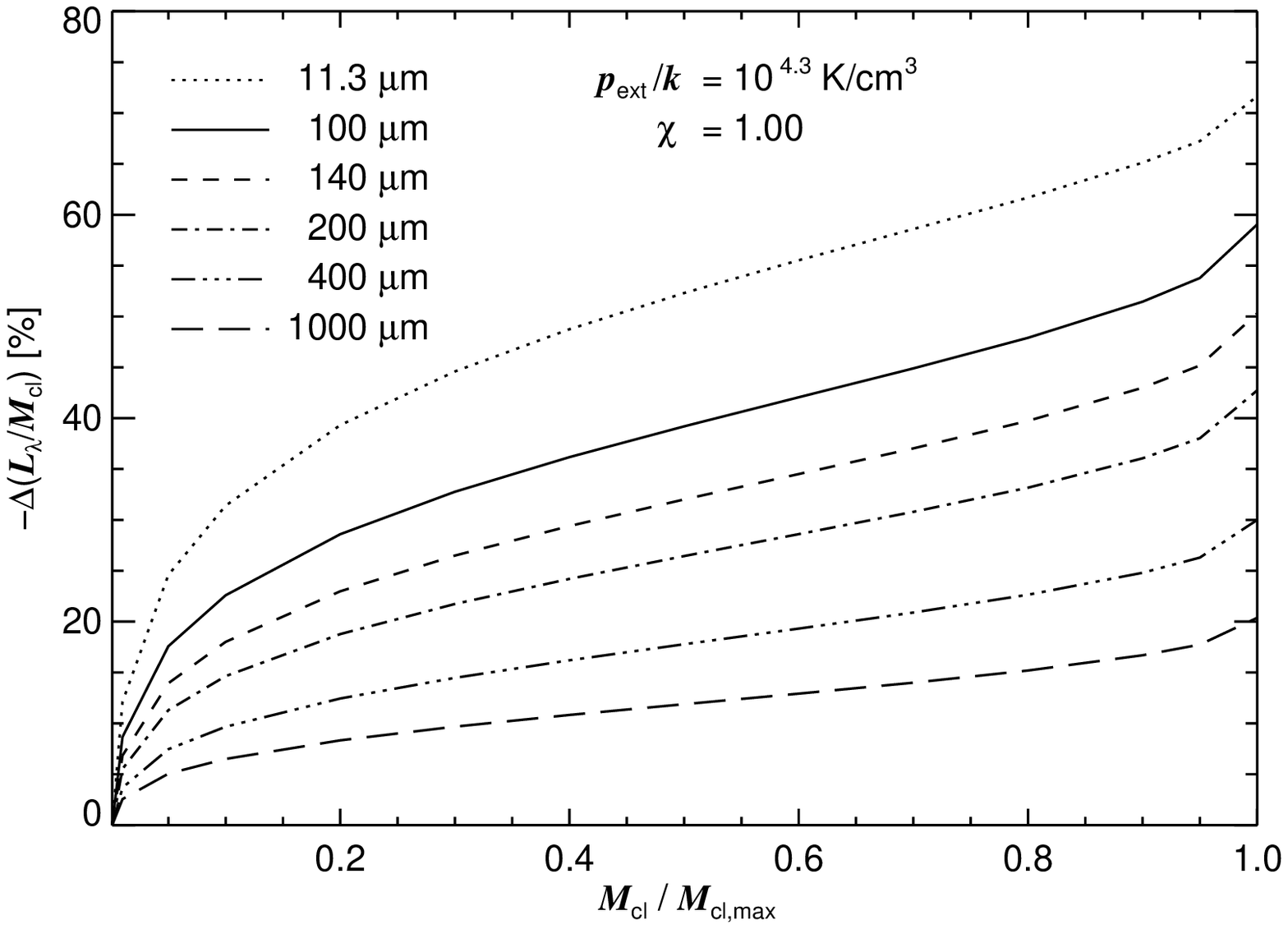}
	\caption{\label{figlump40} Variation of the normalised luminosity $L_\lambda/M_{\rm cl}$ relative to a cloud 
		with $f=0.001$ at several wavelengths as function of mass fraction. The outer pressure is taken to be
		 $2\cdot 10^4~{\rm K/cm^3}$. Shown are the negative values.}
\end{figure}

Since grains in the dense clouds are only heated by a strongly attenuated photon field, the conversion efficiency of the illuminating flux into thermal emission is decreased in respect to the emission of less compact clouds. 
To obtain an estimate of the decrease of the total emission from the dust clouds we normalised the total IR-emission $L_\lambda=4\pi r_{\rm cl}^2 F_\lambda$ by the cloud mass $M_{\rm cl}$.  
The result is shown in Fig.~\ref{figlump40}. 

As can be seen, the `clumpy' density structure leads to a lower total flux in the re-emitted spectrum at all wavelengths, an effect that becomes, as expected, stronger with mass fraction. Because of the colder grain temperatures and the resulting shift of the dust emission spectrum towards longer wavelengths the flux at longer wavelengths are in general less affected. The strongest decrease of the total flux we obtain for the PAH emission and the warm dust emission at $60~\mu{\rm m}$. Their flux is at most $70~\%$ lower compared to a homogeneous medium. At $1000~{\mu\rm m}$ the decrease of the total flux is only $\sim 20\%$.
 
\subsection{Limitations of the Dust Model}

One main assumption in this paper made to obtain the SED of isothermal clouds is that the grain properties inside the clouds are the same as in the diffuse ISM. In compact clouds it is generally thought that dust growth occurs because of dust accretion and adsorption of atoms and molecules from the gas phase. 

Grain growth would have the effect that the extinction curve becomes less steep in the optical and UV/FUV as smallest grains become less abundant. A measure of the rise of the extinction curve is given by the absolute-to-relative extinction $R_{\rm V}=A_{\rm V}/E(B-V)$ which is higher for flatter curves. 
\cite{Whittet2001} observed an increase of $R_{\rm V}$ in the Taurus dark clouds for $A_{\rm V}>3$. 
In addition the $3~\mu{\rm m}$ water ice feature started to appear, in agreement with the expectation of the formation of ice mantles on the grain surfaces in dense gas.

Based on the model of isothermal clouds we can estimate the corresponding critical density for a change of the grain properties using the central density as the main indicator.
In the local diffuse ISM with $p_{\rm ext}/k\approx 2\cdot 10^4~{\rm K/cm^3}$ the cloud mass would need to be close to the critical value ($f \gtrsim 0.7$) to produce a total extinction as large as $A_V>3$ (Fig.~\ref{figcolumn}). 
The kinetic temperatures inside a sample of translucent clouds observed with {\em FUSE} (Far Ultraviolet Spectroscopic Explorer) lie in the total range $40$ to $100$~K with an average value $68\pm15~{\rm K}$ \citep{Rachford2002}. The value is in agreement with the kinetic temperatures of $77\pm 17~{\rm K}$ derived for 61 lines of sights with $log N_2>18.0$  observed with the {\rm Copernicus} satellite \citep{Savage1977}. Following Rachford et al. the mean kinetic temperature drops to $55\pm 8~{\rm K}$ after using only a sub sample of sight lines with $\log N({\rm H_2}>20.4)$, consistent with their own sample.
From Fig.~\ref{figcentraldensity} we conclude that processes responsible for a flattening of the extinction curve occurs at a central density of $n_{\rm H}\sim 1000~{\rm cm^{-3}}$. 

How the grain growth process affects the optical thickness of the cloud is not certain. It is believed that
the extinction in the NIR is rather insensitive to the actual grain size distribution.
For the effect in the optical one might consider
the extinction curve towards $\rho$~Oph~A. The curve is not only unusually flat with $R_V=4.34$ 
\citep{Cardelli1989} but shows also with $N_{\rm H}/E(B-V)=9.3\cdot10^{21}~{\rm cm^{-2}mag^{-1}}$ 
\citep{Diplas1994} a significant higher `gas-to-dust-ratio' relative to the mean value of $5.8\cdot 10^{21}\,{\rm cm^{-2}}{\rm mag}^{-1}$ found for the diffuse ISM \citep{Bohlin1978} which is also used in this work. 
If correct and the situation can be generalised the grain growth process leads to clouds which are less optically thick. 
The effect should be more dramatic in the UV/FUV which is mostly responsible for the grain heating in the diffuse ISM. 

Measurements of the extinction curve through translucent clouds furthermore show a broader feature at 2200~\AA{} \citep{Boulanger1994,Rachford2002}. \citet{Boulanger1994} found for their extinction curves of different sight-lines towards the Chameleon cloud complex that the
feature becomes not only broader but also weaker if the clouds are more optically thick. This indicates that the carriers inside dense clouds become less abundant presumably because they stick to larger grains. If our dust model is correct and the 2200~\AA{}-feature is related to the PAH molecules their emission is expected to be less strong.

Grain growth processes in clouds are also indicated by infrared observations \citep{Laureijs1989,Bernard1992,Bernard1993,Rawlings2005}. Based on their simplified model of the radial density structure and on the assumption that the clouds are illuminated by an isotropic interstellar radiation field \citet{Bernard1992} concluded that the observed brightness profiles of clouds with a total attenuation as low as 2 to 4 cannot be purely explained by radiative transfer effects. \citet{Bernard1993} used the dust composition as a free parameter to fit the observed brightness profile of dense clouds. For comparison with observations they also allowed to have the cloud illuminated by a nearby young star. Applying this method they obtained for dense clouds an under-abundance of PAH and very small grains in the central region. 

One of their infrared sources is the dark cloud G~300.2--16.8 which is thought to be mainly heated by the ISRF \citep{Laureijs1989,Bernard1993}. 
Local variations of the dust properties have also been suggested by \citet{Rawlings2005} to explain the strong variations of the IR-spectra taken towards three different sight-lines
\citep{Lemke1998} that show a change of the PAH emission by a factor $\sim 2$, much stronger than predicted also by our model calculations.
Following \citet{Rawlings2005} the maximum extinction through the cloud is $A_V\sim 3.5~{\rm mag}$ so that a larger grain population in the region of high column density would be in agreement with the findings of \citet{Whittet2001}.

The effect of grain growth on the SED should be even more evident in Bok Globules which are characterised by very low temperatures around $10~{\rm K}$. Their densities are therefore above $n_{\rm H}\sim 10^3~{\rm cm^{-3}}$, our expected critical value for grain growth in clouds. 

Several observations of scattered and emitted light indicate a larger grain population in Globules relative to the diffuse interstellar medium \citep{FitzGerald1976,Lehtinen1996,Lehtinen1998}. \citet{FitzGerald1976} concluded from modelling the scattered optical light from the Thumbprint Nebula that the grain population should produce a rather flat extinction curve with $R_{\rm V}=5.7\pm 1.1$. This result has been verified by modelling the scattered light in the NIR  in J,H,K by \citet{Lehtinen1996}. 

A population of larger grains seems to be supported by observations made at 100 and at 
$200~\mu{\rm m}$ \citep{Lehtinen1998} as
the fluxes seen towards the cloud centre gave a colour temperature
significantly smaller than theoretical predictions presented by \citet{Bernard1992}.

\subsubsection{PAH emission spectrum}

\label{sect_pah}
The strength of the PAH-emission features depends on the ionisation state of the PAH molecules. The model of \citet{LiDraine2001} differentiates between non charged and charged states. In general the ionisation states will vary inside the clouds. In particular, one would expect that the probability of charged PAH molecules decreases inwards with the decrease in the strength of the UV-field. Our calculations assume the same probabilities of ionised and non ionised molecules taken to be the value for the CNM. Therefore, the model could be improved by adding an explicit solution of the PAH charge state, and this would then enable the PAH spectrum to be used as a probe of the UV field within the dark cloud.

\subsection{Comparison with Dense-Core-Models}

The model presented here can be compared with
the work of \citet{Evans2001} and of \citet{Stamatellos2003} who were modeling the emission 
in the FIR and submm-regime of highly dense clouds to obtain better estimates for the dust mass. 
As in our model, the calculations were (at least partly) based upon an isothermal 
self-gravitating sphere. However, they applied the model also to unstable clouds which show higher overpressures and can be more optically thick relative to the critical stable clouds.
One of the main differences to our calculations, apart from the fact that \citet{Evans2001} did not consider the complication through scattered light, lies in the method used to derive the thermal emission from dust grains.

In their models, the dust emission at different depths inside the cloud is obtained assuming 
that the dust emission can be described by an effective grain temperature
where the cooling rate is equal to the heating rate, both derived using the mean optical dust properties.  
Although this assumption 
fullfills the energy conservation, it does not necessarily reflect the actual dust temperatures and, therefore, the correct 
shape of the dust emission spectrum. We have seen that this might only be
true in the innermost core region of highly optical thick clouds. 
Otherwise the simplification will underestimate the dust emission at short wavelengths 
where it is dominated by the emission from small particles as they cool in general less efficiently and
might show strong temperature fluctuations.
In the FIR or submm-regime a physical model of the grain temperatures might be less important as the emission is dominated by large grains.
However, this aspect should be analyzed more quantitatively to improve the mass estimate for highly compact and dense clouds.

The dust composition and the optical properties in the dense medium are major uncertainties.
It is possible that these overshadow the uncertainties which arise by using a rather simplified
dust model.
In both works mentioned above, the calculations are based on the mean dust opacities provided by \citet{Ossenkopf1994} 
(OH-model), which were derived on theoretical considerations of grain growth in a dense medium and may, therefore, be more appropriate for the dust properties in dense compact clouds.
As the data are given only for wavelengths larger then $1~\mu{\rm m}$, extrapolations
to the optical and UV light were made, independent on the actual grain properties assumed by \citet{Ossenkopf1994}.  
\citet{Stamatellos2003} combined the OH-model with the model of \citet{Mathis1977} (MRN-model) which was originally 
obtained by fitting the mean extinction curve of the interstellar medium\footnote{However, it is not mentioned why the 
strong absorption peak at 2200~\AA{} is missing although the calculations were based on the optical properties of 
\citet{DraineLee1984} where the bump would be produced by small graphite particles.}. 
The MRN-model predicts an overabundance
of small dust particles to explain the rise in the UV and the optical while,
as discussed in the previous section, a flat extinction curve is expected. Although it is not mentioned,
a flatter curve was probably assumed 
in the work of \citet{Evans2001} as indicated by the lower effective dust temperature at the outskirts of the clouds (14 K instead of 17 K as stated by \citet{Stamatellos2003}). For comparison, in our model the equilibrium grain temperatures (shown in Fig.~\ref{figgraintemp}) vary between different sizes and compositions. 
If we consider only the silicate and graphite particles the temperatures at the edge of highly optical thick clouds range 
from 12 to ${18~{\rm K}}$.

The absorption properties in the optical determine not only the effective temperature in the outskirts of the highly optically thick clouds but also how deep the optical emission can penetrate inside the cloud. The stronger temperature gradient in case of \citet{Stamatellos2003} should produce a steeper
brightness profile at FIR and submm-wavelengths if compared with \citet{Evans2001}.
It is true that in highly dense clouds most of the optical emission is absorbed at the cloud outskirts so 
that the dust in the core region is predominantly heated by IR photons.
However, most power of the diffuse ISRF lies, even if we take the DIRBE measurements of the IR emission over the entire sky into account which has been neglected in our calculations but has been considered in the two other works, in the optical. As consequence, the emission from the outskirts should affect to a certain degree not only the spectral shape but also, as has been discussed in the previous section, the position of the emission peak. On the other
hand, it is indeed expected that the dependence of the SED beyond the emission peak on the optical absorption is less strong and weakens towards more optically thick clouds.

To study the  dependence of the optical properties on the dust emission spectrum, \citet{Stamatellos2003} considered different scattering properties of the dust grains. The authors found only a weak dependence of 10 to 20\%. The conclusion
would be in agreement with our calculations if we had based them also on an effective dust temperature: As we have 
seen in Sect.~\ref{sectheating} the dust scattered light makes only a minor contribution to the dust heating at the cloud outskirts. Where it is important 
the heating by scattered light is less than $50~\%$ of the total heating rate. As the effective temperature depends only 
as $T_{\rm eff}\propto h(r)^{1/6}$ on the heating rate $h(r)$ the effective temperature would not vary by more than a factor
$\sim 2^{-1/6}$ if we would have ignored the scattered light completely. Because in the Rayleigh-Jeans part the emission 
spectrum is proportional to the grain temperature, only a weak temperature dependence beyond the emission peak would arize.

\subsection{The Clouds Structure }

The shape of the cloud will have certainly an efffect on the radiative transfer problem. As mentioned, a following paper
will focus on the SED from cylindrical cloud structures where this aspect will, at least to a certain degree, be considered.

Apart from the actual overall shape, the SED of interstellar clouds will be affected by the in-homogeneous 
substructure 
caused by non-thermal motion which is probably related to magneto-hydrodynamic turbulence inside the cloud \citep{Curry2000}. The turbulent motion will provide an additional pressure support term, so such clouds might be more stable against gravitational collapse. As discussed for the ISM the turbulent motion will create a broad density distribution of the local density which is approximately described by a simple log-normal function. 

The effect of the turbulent substructure might be considered by using an idealistic structure of a turbulent medium superimposed on the simple radial structure discussed here. The turbulent motion might even lead to a broadening so that the radial density profile of turbulent clouds might be less steep. The maximum structure is related to the cloud size and the width of the density distribution of the local density to the strength of the turbulence. 

Because of the fractal structure of high density clouds the light will penetrate to deeper cloud regions so that grains in those regions become warmer than in clouds without turbulence. In addition turbulence produces density enhancements which are optically thick where grains show colder temperature. It will be important to determine how large these effects actually are. 

Our model still does not explain all the observational properties, particularly those of dense clouds. In the case of Bok Globules or also in the case of dense cores in molecular clouds it has been found that the non-thermal contribution to the line width increases for more optically thick lines indicating an increase of the turbulent motion from the cloud centre outwards \citet{Caselli1995}. This observation is generally explained by the strong dissipation of magnetic  motion in high density gas. This leads to a core-envelope structure of the cloud where the core region is mainly supported by thermal pressure while the envelope is supported mainly by magnetic and turbulent pressure. Based on this observation  \citet{Curry2000} suggested a model where the the core and the envelope are described by two different polytropes where the in the two regions the pressure is related to the density via $p\propto \rho^{\gamma}$ with different values for $\gamma=1+1/n$ where $n$ is the polytrope index. This model is able to produce the high density contrast observed for a number of clouds which is larger than the maximum value of 14.04 of the Bonner-Ebert sphere. \citet{McKee1999} discussed a model for molecular clouds using multipressure polytropes where the different pressure components in the cloud, thermal, turbulent, and magnetic pressure, are described by different polytropes. It produces arbitrarily large density contrasts for stable clouds.

On the other hand,  it has been shown based on extinction measurements in the NIR that the profiles of the projected column density of rather spherical Globules are well described by the density structure of the isothermal sphere \citep{Alves2001,Kandori2005}. The small sample of 10 Globules studied by Kandori et al. indicate that most of the Globules are at the critical state to gravitational collapse. The derived external pressure, however, shows strong variation from $2.1\times 10^4$ up to $1.8\times 10^5~{\rm K/cm^3}$  with a mean of $5.7\times 10^4~{\rm K/cm^2}$ which is almost 3 times larger than the value assumed here which should be typical for the ISM of our galaxy. It might well be that in certain cases the simplification using an isothermal cloud model has to be compensated by a higher external pressure. Following Kandori et al. the model of an isothermal sphere is also able to mimic in the limit of the observational uncertainties the density profiles of clouds which undergo the gravitational collapse.

\section{Summary}

We have analysed in some detail the SED from spherical, self gravitating, isothermal stable clouds illuminated by an isotropic ISRF.
The radiative transfer through those clouds is, for given radiation field and dust properties inside the cloud, determined by the pressure of the ambient medium and by the critical mass fraction $f=M_{\rm cl}/M_{\rm cl,max}$ of the cloud mass $M_{\rm cl}$ and the critical mass $M_{\rm cl, max}$ where the cloud becomes gravitational unstable. 
The dust properties in the clouds are assumed to be consistent with the ones in the diffuse ISM of our galaxy. We derived the SED for different assumptions of the critical mass fraction $f$, outer pressure $p_{\rm ext}$ and strength of the external ISRF. Our special attention was the SED from clouds embedded in the WNM with a pressure of $p_{\rm ext}/k=2\cdot 10^4~{\rm K/cm^3}$ which were compared with SEDs from clouds in higher pressure regions.
We find that:
\begin{itemize}
	\item The grain temperatures in stable isothermal clouds embedded in the WNM are generally higher than 10~K. Lower temperatures are expected for collapsing clouds or for clouds in higher pressure regions.
	\item The ratio of the dust emission relative to the PAH-emission rises towards higher critical mass fraction and outer pressure.
	\item As clouds are generally more optically thick in higher pressure regions more light is transferred to thermal emission.
\end{itemize}
We further analysed the brightness profiles of the scattered emission at 440 and at 900~nm and the thermal emission from dust grains and PAH-molecules in the IR/FIR and in the submm/mm regime. We find that:
\begin{itemize}
	\item For stable isothermal clouds embedded in the WNM the scattered emission at 440~nm developes for increasing mass fraction a broad wing-like profile as seen for Bok-Globules in our galaxy. At longer wavelength the scattered emission is brighter towards the cloud centre. In high pressure regions the scattered emission is more concentred to the limb of the clouds.
	\item For stable isothermal clouds in the WNM the PAH brightness profile shows a weak wing-like profile if the clouds are critical to collapse. In high density regions the profile shows a strong limb-brightening effect.
	\item The brightness profiles of the dust emission is generally broader in respect to the profile of the column density. The effect increases towards shorter wavelengths and is more prominent in high pressure regions.
\end{itemize}

\begin{acknowledgements}
M. A. Dopita acknowledges the support of both the Australian National  University and of the Australian Research Council through his ARC Australian Federation Fellowship. Both authors acknowledge financial support for this research through an ARC Discovery project grant DP0208445.
J\"org Fischera acknowledges helpful discussions he had with Ute Liesenfeld,  Endrik Kruegel, and 
Richard Tuffs.
\end{acknowledgements}

\begin{appendix}

\section*{The Physical parameters of isothermal clouds}\label{Appendix}

Here we want to justify our statements that both the density profile and (for a given outer pressure) the column density to the centre of gravitational stable isothermal clouds can be characterized by the critical mass fraction $M_{\rm cl}/M_{\rm cl,max}$ where $M_{\rm cl}$ is the cloud mass and $M_{\rm cl,max}$ the maximum cloud mass critical stable against gravitational collapse. In addition we provide the accurate expression for the cloud size.

The Ansatz of self gravitating isothermal spherical clouds leads to the famous Lane-Emden equation, a differential equation of the unit free potential $\omega=\Phi/K$ where $\Phi$ is the gravitational potential and where $K=kT_{\rm cl}/\mu m_{\rm u}$. $k$, $T_{\rm cl}$, $\mu$, and $m_{\rm u}$ are the Boltzmann-constant, the cloud temperature, the mean molecular weight, and the atomic mass unit. The Lane-Emden equation is given by
\begin{equation}
	\frac{{\rm d}^2 \omega}{{\rm d}z^2}+\frac{2}{z}\frac{{\rm d}\omega}{{\rm d}z} = e^{-\omega}
\end{equation}
where $z = Ar $ with $r$ as the radius and $A^2 = 4\pi G\rho_c/K$. $\rho_c$ is the density in the cloud centre and $G$ the gravitational constant. The radial density and the pressure are simply given by the exponentials:
\begin{equation}
	\label{eqpressure}
	\rho(z) = \rho_c\,e^{-\omega(z)},\quad p(z) = p_c\,e^{-\omega(z)}.
\end{equation}
where $p_c = K \rho_c$.

\begin{figure}
  \includegraphics[width=0.49\hsize]{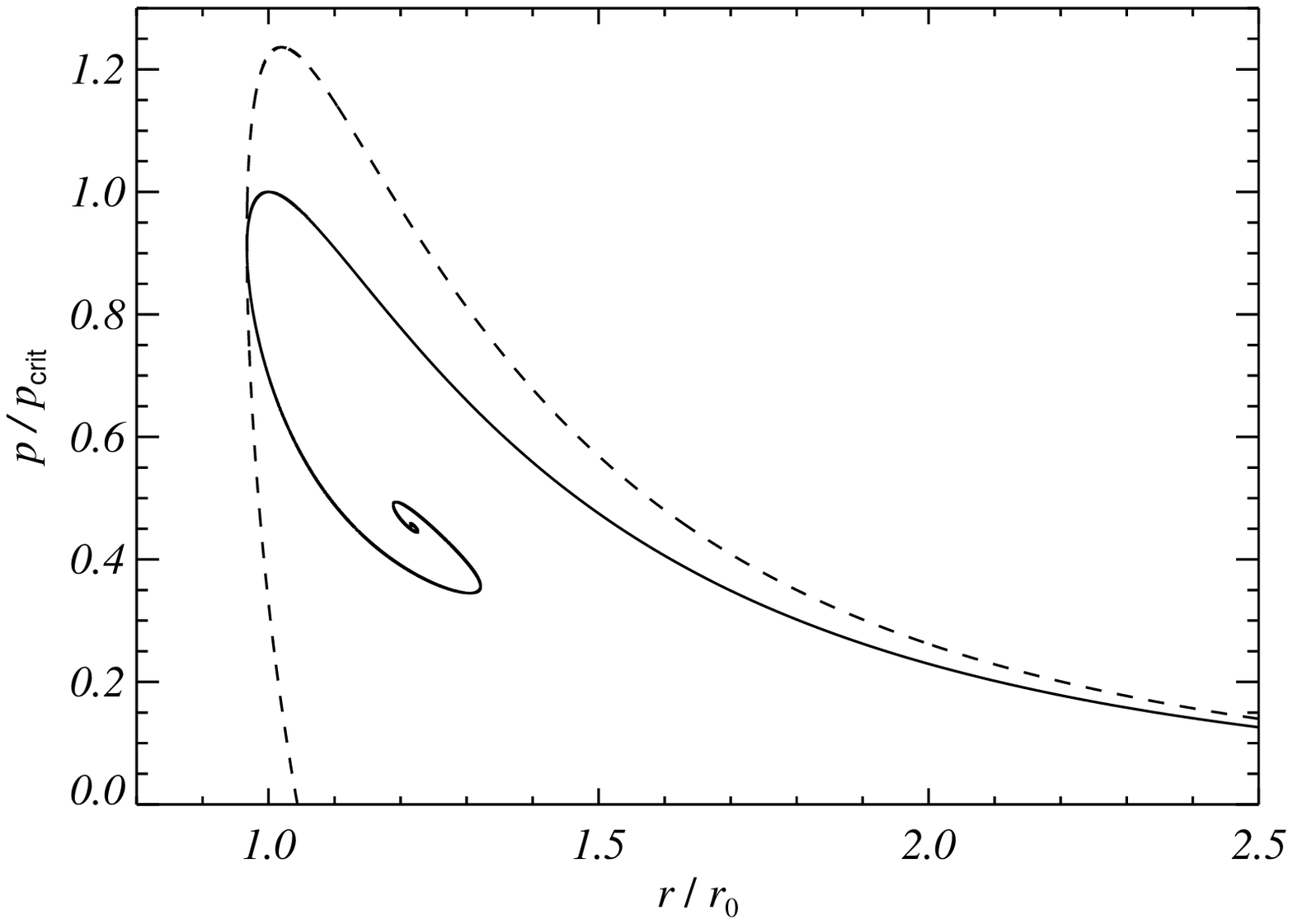}
	\hfill
  \includegraphics[width=0.503\hsize]{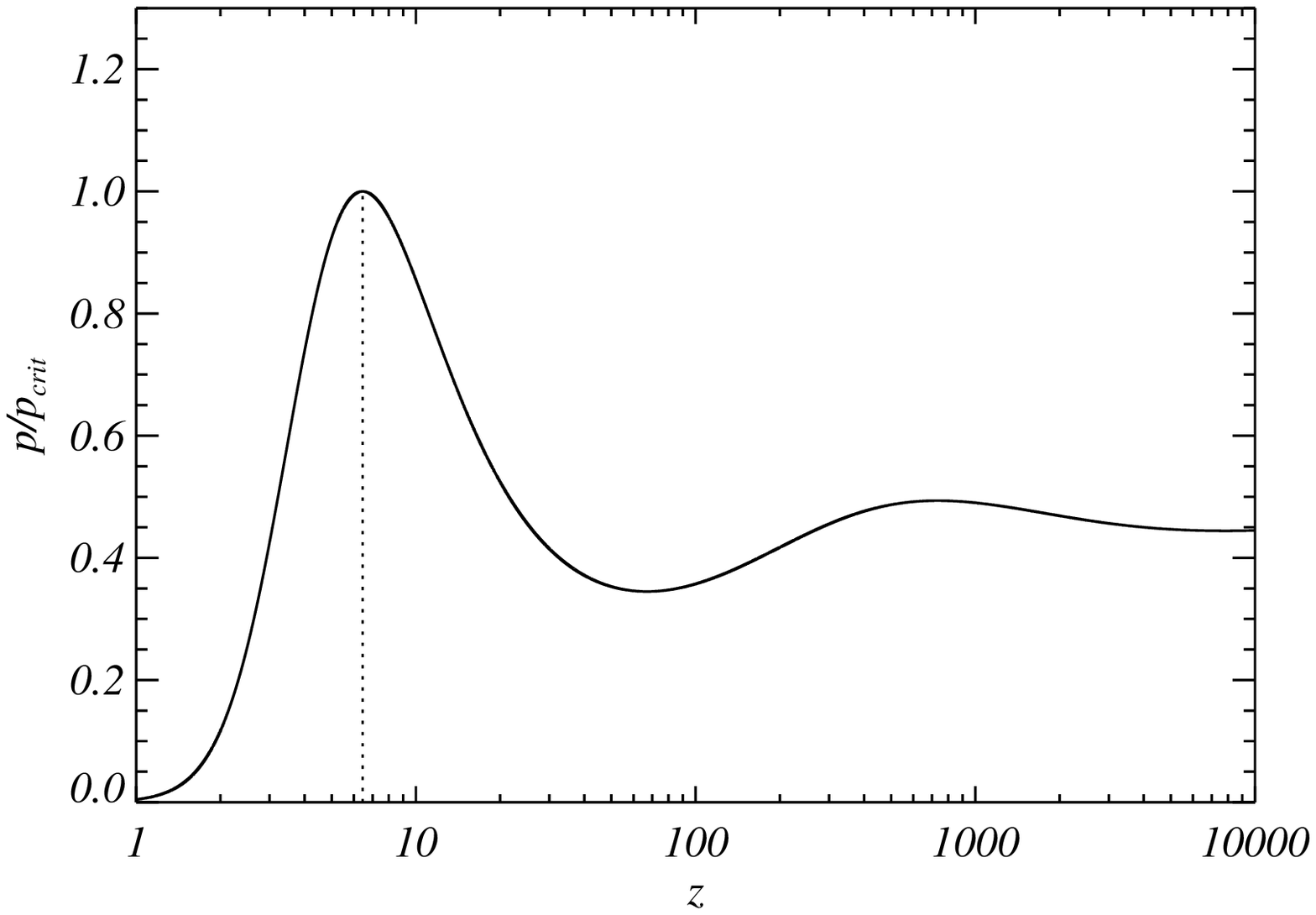}
  \caption{\label{figpressure}The normalized pressure $p/p_{\rm crit}$ at the outer boundary of the isothermal sphere as function of the normaized radius $r/r_0$ where $r_0=0.411 M_{\rm cl} G/K$ (left panel) and as function of $z$ (right panel). $G$ is the gravitational constant and $K=kT_{\rm cl}/(\mu m_{\rm u})$ where $m_{\rm u}$ and $\mu$ are the atomic mass unit and the mean molecular weight, respectively. The dashed line in the left panel is derived using a simple application of the virial theorem while the solid curve is obtained by solving the Lane-Emden equation. The dotted curve in the right panel marks the boundary where the clouds become gravitationally unstable which defines $z_{\rm max}$. The solutions considered in this paper of gravitational stable clouds lie to the left.}
\end{figure}

First we will consider the pressure at the outer radius of a cloud of certain mass $M_{\rm cl}$. We express the constant $A$ through the mass $M_{\rm cl}$ of the cloud:
\begin{equation}
	\label{eq_aconst}
  A(z_{\rm cl}) = \frac{K}{M_{\rm cl}G}\int_0^{z_{\rm cl}} {\rm d}z'\, z'^2\,e^{-\omega(z')}.
\end{equation}
If we replace $\rho_c$ by $A^2 K/(4\pi G)$ and the constant $A$ through Eq.~\ref{eq_aconst} we obtain for the pressure of a cloud with mass $M_{\rm cl}$ the following variation as function of the parameter $z_{\rm cl}$:
\begin{equation}
	p(z_{\rm cl}) = \frac{K^4}{{4\pi G^3 M_{\rm cl}^2}}e^{-\omega(z_{\rm cl})}\left(\int_0^{z_{\rm cl}}{\rm d}z'\,z'^2\,e^{-\omega(z')}\right)^2.
\end{equation}
The variation of the outer pressure as function of $z_{\rm cl}$ is shown in Fig.~\ref{figpressure}.
For small values the pressure increases as function of $z_{\rm cl}$ up to a maximum at $z_{\rm max}\approx 6.451$. At $z_{\rm cl}>z_{\rm max}$ the function shows smooth variations with declining amplitude around a certain pressure value. The solutions for the isothermal clouds which are stable against gravitational collapse lie in the region $z_{\rm cl} \le z_{\rm max}$. The maximum pressure is related to a maximum stable cloud mass $M_{\rm cl, max}$.
\begin{equation}
	\label{eq_mcrit}
	M_{\rm cl,max}(z_{\rm max}) = \frac{K^2}{\sqrt{4\pi G^3 p_{\rm ext}}}e^{-\omega(z_{\rm max})/2}\int_0^{z_{\rm max}}{\rm d}z'\,z'^2\,e^{-\omega(z')}.
\end{equation}
In this paper we have chosen to measure the cloud mass $M_{\rm cl}$ in units of the critical mass of a cloud with the same 
temperature. If we replace the cloud mass by $f=M_{\rm cl}/M_{\rm cl,max}$ the pressure at the outer cloud radius is given by:
\begin{equation}
	\frac{p(z_{\rm cl})}{p_{\rm ext}} = \frac{1}{f^2}\frac{e^{-\omega(z_{\rm cl})}}{e^{-\omega(z_{\rm max})}}\left(
	\frac{\int_0^{z_{\rm cl}} {\rm d}z\,z^2\,e^{-\omega(z)}}
	{\int_0^{z_{\rm max}}{\rm d}z\,z^2\,e^{-\omega(z)}}\right)^2.
\end{equation}
In the region of stable clouds which are in pressure equlibrium with their external medium ($p_{\rm ext}=p(z_{\rm cl})$) 
the value $z_{\rm cl}$ can be considered as a pure function of $f$. 

Often the model is applied also to clouds which are in a state of collapse or show in general a higher over-pressure 
than for critical stable clouds. In this situation it is not clear how well the pressure $p(z_{\rm cl})$ at the cloud edge reflects the external pressure $p_{\rm ext}$. From Fig.~\ref{figpressure} we see that the pressure at the outskirst would drop below the external pressure if the cloud becomes unstable.
However, the isothermal cloud model seems to be useful as an approximate solution of the density structure of non stable clouds as it resembles the power law density profile close to $\rho(r)\propto r^{-2}$ at the outskirts and a flat profile in the centre as observed for pre-stellar cores.

The radial density and pressure profile is directly obtained by the solution of the Lane-Emden equation:
\begin{equation}
	\frac{\rho(z)}{\rho(z_{\rm cl})} = \frac{p(z)}{p(z_{\rm cl})}=\frac{e^{-\omega(z)}}{e^{-\omega(z_{\rm cl})}}.
\end{equation}
Because the transformation from $z$ to $r$ is linear the same applies to the density profile 
$\rho(r)/\rho(r_{\rm cl})=p(r)/p(r_{\rm cl})$.  As $\omega(z=0)=0$ the central density is given by:
\begin{equation}
	\label{eqcentraldensity}
	\rho_{\rm c} = \frac{p(z_{\rm cl})\,e^{\omega(z_{\rm cl})}}{K}.
\end{equation}

An equally simple relation can be found for the cloud radius:
\begin{equation}
	r_{\rm cl} = z\frac{K}{\sqrt{4\pi G p(z_{\rm cl})}}e^{-\omega(z_{\rm cl})/2}.
\end{equation}

As last parameter of the isothermal clouds we consider the column density, given by 
\begin{eqnarray}
  N_{\rm H} &=& \int {\rm d}r\,n_{\rm H}(r) = \left(m_{\rm u}\sum_i \frac{n_i}{n_{\rm H}}\mu_i\right)^{-1}\int_0^{r_{\rm cl}}{\rm d}r\,\rho(r)
\end{eqnarray}
where $n_i$ is the number density of element $i$ and $\mu_i$ its mean molecular weight. Using the solution
of the isothermal cloud for its radius, density, and pressure one finds:
\begin{equation}
	N_{\rm H}(z_{\rm cl}) = \left(m_{\rm u}\sum_i \frac{n_i}{n_{\rm H}}\mu_i\right)^{-1} 
	\sqrt{\frac{p({z_{\rm cl}})}{4\pi G}}
		\int_0^{z_{\rm cl}}{\rm d}z\,e^{-\omega(z)}\,e^{\omega(z_{\rm cl})/2.}
\end{equation}
For given parameter $z_{\rm cl}$ the column density is proportional to the pressure 
$\sqrt{p(z_{\rm cl})}$ at the edge of the cloud. 

\end{appendix}

\end{document}